%% file: main.tex
\RequirePackage{ifpdf}
\documentclass[12pt]{article}
\input{gdefn.tex}

\usepackage[title]{appendix}
\usepackage{amsmath}
\usepackage{amssymb}
\usepackage{multirow}

\begin{document}

\parindent=12pt

\begin{center}

{\Large \bf TsT vs. LCR and Gravity Dual of Non-relativistic Fluid}
\end{center}

\baselineskip=18pt

\bigskip

\centerline{Suvankar Dutta$^a$, Taniya Mandal$^b$ and Sanhita Parihar$^a$}

\bigskip

\centerline{\large $^a$\it Indian Institute of Science Education and Research
Bhopal}
\centerline{\large  \it Bhopal bypass, Bhopal 462066, India}

\centerline{\large $^b$\it National Institute of Theoretical and Computational Sciences}
		\centerline{\large\it School of Physics and Mandelstam Institute for Theoretical Physics}
\centerline{\large\it University of the Witwatersrand, Wits, 2050, South Africa.}
\bigskip

\centerline{E-mail: suvankar@iiserb.ac.in, taniya.mandal@wits.ac.za, sanhita18@iiserb.ac.in}

\vskip .6cm
\medskip

\vspace*{4.0ex}

\centerline{\bf Abstract} \bigskip

\noindent 
We discuss two different approaches to synthesise holographic non-relativistic fluid from its relativistic counterpart. In the first approach we obtain the non-relativistic fluid by light-cone reduction of a relativistic conformal fluid. In the second approach we consider the bulk dual of the relativistic fluid, uplift the solution to 10 dimensions and perform TsT transformations on the bulk solution to change the asymptotic structure. Reducing the TsT transformed geometry over $S^5$ we find an effective 5 dimensional locally boosted solution. We then use the bulk-boundary dictionary to compute the non-relativistic constitutive relations. We show that the non-relativistic fluids obtained by these two methods are equivalent up to second order in derivative expansion. Our results also provide explicit expressions for different constitutive relations and transports of holographic $U(1)$ charged non-relativistic fluids (both parity odd and even) up to second order in derivative expansion.

\vfill\eject

\tableofcontents

%%%%%%%%%%%%%%%%%%%%%%%%%%%%%%%%%%%%%%%%%%
%%%%%%%%%%%%%%%%%%%%%%%%%%%%%%%%%%%%%%%%%%
%%%%%%%%%%%%%%%%%%%%%%%%%%%%%%%%%%%%%%%%%%
\section{Introduction}\label{sec:intro}

Finding holographic descriptions of non-relativistic systems is an interesting area of research as they are realised in low energy experiments and various condensed matter systems. One possible way to construct such descriptions is to take non-relativistic limits on both sides of the AdS/CFT duality \cite{Maldacena:1997re, Gubser:1998bc, Witten:1998qj}.

There are various distinct ways to derive non-relativistic systems from a given relativistic theory. Light-Cone reduction (LCR) is one of such techniques. The technique is based on the fact that a relativistic system respects Poincar\'e invariance whereas a system is called non-relativistic if it respects Galilean symmetry. It is well known that the Galilean algebra in $d$ space dimensions can be obtained by reducing the Poincar\'e algebra $so(d+1,1)$ in $(d+ 1,1)$ dimensions ($d+1$ space directions, one time direction) on the light-cone directions $x^\pm \sim x^0 \pm x^{d+1}$, where $x^0$ is the relativistic time and $x^{d+1}$ is a spatial coordinate. The light-cone reduced theory respects the Galilean symmetry while evolving along the light-cone time $x^+$. In the same way if we consider conformal algebra $so(d+2,2)$ in $(d+1, 1)$ dimensions, under LCR it boils down to Schr\"odinger algebra \cite{PhysRevD.5.377,Mehen:1999nd,Nishida:2007pj,Sakaguchi:2008ku,Kovtun:2008qy,Duval:2008jg,Rangamani:2009zz}  in $d$ space dimensions.

In order to provide a holographic description of a theory one has to first check the isometries on both sides. A conformal theory in $(d+1, 1)$ dimensions is dual to a gravity theory in $AdS_{d+3}$ whose metric is given by,
\begin{equation}\label{eq:adsd3}
    ds^2 = -r^2 \eta_{\mu\nu}dx^\mu dx^\nu + \frac{dr^2}{r^2}
\end{equation}
where $\eta$ is a $d+1,1$ dimensional Minkowski metric and $\mu,\nu = {0, \cdots, d+1}$. The isometry group of $AdS_{d+3}$ is $SO(d+2,2)$ which is the conformal group of the boundary theory. In a similar spirit, the holographic dual of a theory with Schr\"odinger isometry was first proposed in \cite{Son:2008ye,Balasubramanian:2008dm,Goldberger:2008vg,Barbon:2008bg}. The bulk metric is given by\footnote{In \cite{Balasubramanian:2008dm} the bulk metric was proposed for Galilean symmetry algebra with an arbitrary scaling $z$. $z=2$ corresponds to Schr\"odinger algebra with an extra generator corresponds to special conformal transformation in time direction. Here we consider $z=2$ case.}
\begin{equation}\label{eq:schst}
    ds^2 = r^2 \lb -2 dx^+ dx^- -r^2 (dx^+)^2 + d \vec{x}^2\rb + \frac{dr^2}{r^2}
\end{equation}
where $\vec x$ is a $d$ dimensional vector. This spacetime is known as Schr\"odinger spacetime, denoted by Sch$_{d+3}$ and has the Schr\"odinger algebra as  a global symmetry algebra \cite{Adams:2008zk,Schafer-Nameki:2009dsc}. Although the Schr\"odinger algebra of the boundary theory can be derived from the conformal algebra by LCR, the Schr\"odinger metric (\ref{eq:schst}) can not be obtained from the $AdS_{d+3}$ metric (\ref{eq:adsd3}) by LCR. However, there is a way to get the Sch$_{d+3}$ spacetime from the $AdS$ geometry using the solution generating technique \cite{Herzog:2008wg,Rangamani:2008gi,Adams:2009dm,Brattan:2010bw}, know as \emph{TsT} transformation \cite{Lunin:2005jy} (or equivalently \emph{null-Melvin twist} \cite{Adams:2008wt, Herzog:2008wg,Alishahiha:2003ru,Gimon:2003xk}). One can start with a $AdS_5$ geometry, uplift the metric to ten dimensions to embed in a type IIB solution. Identify two isometry directions in ten dimensions : $x^-$ and $\psi$, say. The TsT transformation consists of three steps. A T-duality along $x^-$ direction, followed by a shift in $\psi$ : $\psi \ra \psi + x^-$ and finally a further T-duality along $x^-$ direction. As a result the reduced five dimensional spacetime is given by (\ref{eq:schst}).

\begin{figure}[htp]
    \centering
    \includegraphics[width=16cm]{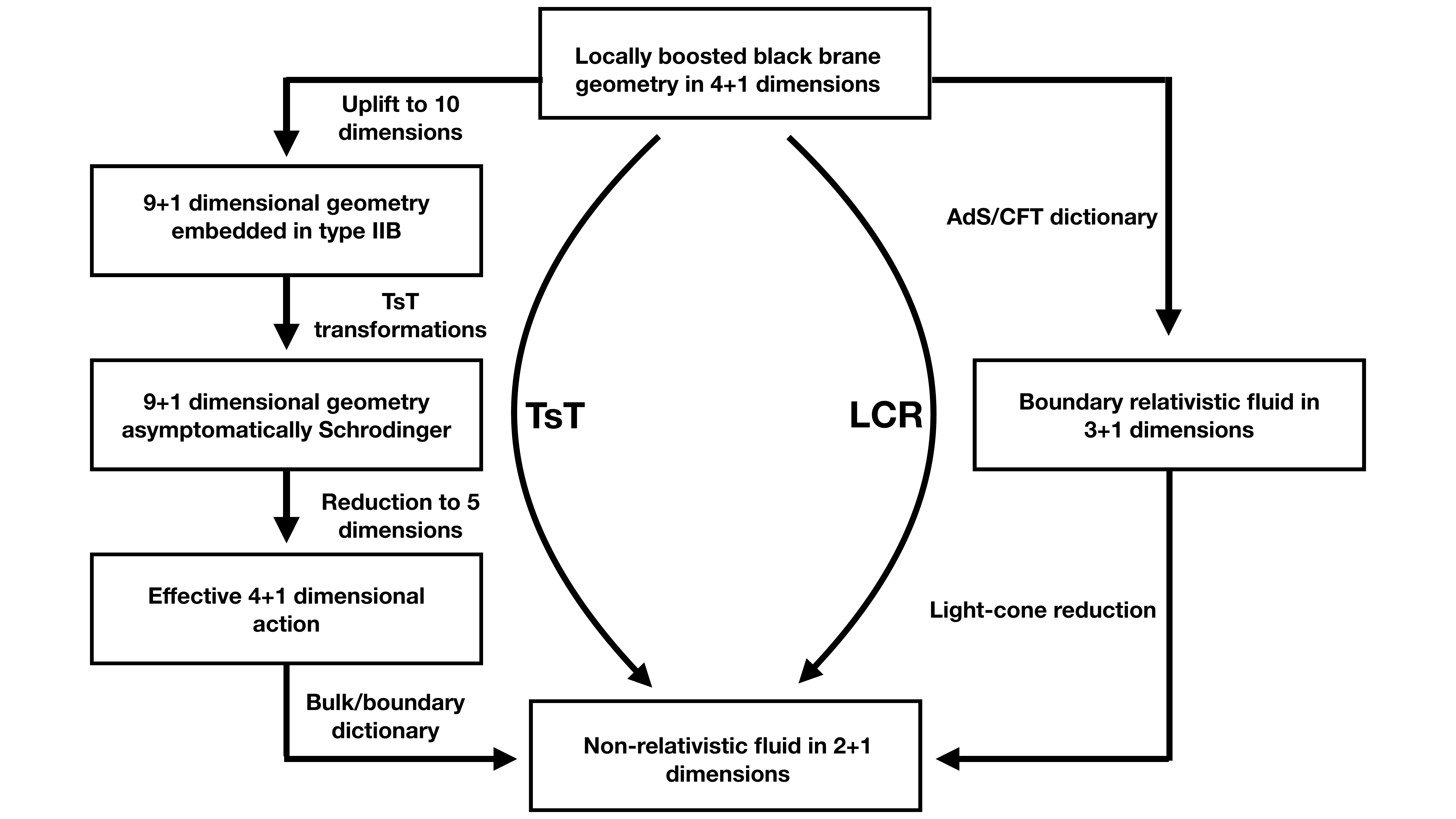}
    \caption{Two different paths to obtain second order non-relativistic fluids.}
    \label{fig:LCR vs TsT}
\end{figure}

LCR on the boundary and TsT in the bulk are therefore consistent at the level of isometry, in a sense that the asymptotic symmetry of a TsT transformed bulk spacetime matches with the corresponding Schr\"odinger symmetry of the boundary theory. Our goal is to understand the consistency between the LCR and TsT transformation beyond the geometry, in particular at the hydrodynamic scale, which is considered to be a low energy or long wavelength departure from local thermal equilibrium. Our starting point is a hydrodynamic system with conformal invariance and its gravity dual. Since hydrodynamics is an effective description, its evolution is governed by the set of conservation equations (known as constitutive equations). The holographic description of a hydrodynamic system is given by locally boosted black brane geometry \cite{Bhattacharyya:2007vjd}. In the LCR method, we reduce the relativistic constitutive relations over the light-cone directions to obtain a set of equations consistent with Schr\"odinger isometry and thus we get a \emph{non-relativistic} hydrodynamic system with stress tensor, energy current, charge current etc. In the TsT method, we uplift the locally boosted black brane geometry to ten dimensions, perform the TsT transformation and get a reduced locally boosted five dimensional Schr\"odinger spacetime. Then we use the bulk-boundary dictionary of \cite{Ross:2009ar} to write down the conserved quantities (stress tensor, energy current, charge current etc) of the boundary non-relativistic system from the reduced effective five dimensional effective action. The question is whether these two reduced systems are identical. We describe the problem with help of the following commutative diagram in fig. \ref{fig:LCR vs TsT}. 

A partial answer to this question was given by \cite{Dutta:2018xtr}. It was shown that both the paths (LCR and TsT) commute up to first order in derivative expansion. The first order calculations were somewhat trivial and it was not clear whether the commutativity would hold at higher orders. In this paper we address this question more rigorously. We consider second order $U(1)$ charged conformal hydrodynamics with both parity-odd and parity-even sectors and its gravity dual \cite{Banerjee:2008th}. We show that the non-relativistic fluid obtained via LCR and TsT transformation are \emph{identical} order by order in derivative expansion up to second order in derivative expansion. As a byproduct of our calculation we gave a complete holographic description of non-relativistic charged fluid with Schr\"odinger isometry up to second order in derivative expansion.

\section{Locally boosted black brane geometry and the relativistic fluid}\label{sec:relflu}

We start with Einstein-Maxwell theory in $AdS_5$ with a Chern-Simons term
\begin{eqnarray}\label{eq:EMCS}
S=\int d^{5}x\big(R-2\Lambda-\mathcal{G}_{a b}\mathcal{G}^{a b}-\frac{4\kappa}{3}\epsilon^{a b c d e}\mathcal{C}_{a}\mathcal{G}_{b c}\mathcal{G}_{d e}\big)
\end{eqnarray}
where $\Lambda = - 6/L^2$ is the five dimensional cosmological constant, $\cG$ field strength of the $U(1)$ gauge field. The Einstein's and the Maxwell's equations obtained from this action admit the following solutions
\begin{eqnarray}\label{eq:qbbsol0}
\begin{split}
ds^{2} & = -2u_{\mu}dx^{\mu}dr-r^{2}f(r,m,q)u_{\mu}u_{\nu}dx^{\mu}dx^{\nu}+r^{2}P_{\mu \nu}dx^{\mu}dx^{\nu}\\
\mathcal{C}_{\mu} & = \frac{\sqrt{3}q}{2 r^{2}}u_{\mu}
\end{split}
\end{eqnarray}
where
\begin{equation}
     x^\mu = \{v, x, y, z\}
\end{equation}
are boundary coordinates, $u^\mu$ is a constant four velocity whose different components are given by
\begin{eqnarray}\label{eq:metdef}
  u_v  = -\gamma, \quad u_i = \gamma \beta_i, \quad \gamma = \frac{1}{\sqrt{1-\vec \beta^2}}
\end{eqnarray}
such that $u^2 = - 1$ and
\begin{equation}
    f(r,m, q) = 1-\frac{m}{r^{4}} +\frac{q^2}{r^6}. 
\end{equation}
$P_{\mu\nu} $ is the projection operator given by
\begin{eqnarray}\label{eq:Pdef}
P_{\mu\nu} = \eta_{\mu\nu}+u_\mu u_\nu, \quad \eta_{\mu\nu} = \text{diag}\{-1, 1,1,1\}.
\end{eqnarray}
The solutions (\ref{eq:qbbsol0}) are parametrised by five constant parameters : three boosts $\vec \beta$, one mass parameter $m$ and one charge parameter $q$ and represent an electrically charged black hole with horizon at $r = R$, given by $f(R,m,q)=0$. In the rest frame ($\vec \beta = 0$) the metric becomes the standard Reissner–Nordstr\"om  metric written in Eddington–Finkelstein coordinates. We have taken the $AdS$ radius $L=1$.

We now consider the black hole parameters $\vec \beta$, $m$ and $q$ to be slowly varying functions of boundary coordinates $x^\mu$
\begin{equation*}
    m \ra m(x^\mu) , \quad \vec\beta \ra \vec\beta(x^\mu), \quad q \ra q(x^\mu).
\end{equation*}
With such replacement the metric and gauge field (\ref{eq:qbbsol0}) do not solve Einstein's as well as Maxwell's equations any more. In order to find the solutions with local parameters we have to supplement the metric and gauge field in (\ref{eq:qbbsol0}) with extra terms proportional to the derivatives of the parameters. Assuming $m$, $q$ and $\vec \beta$ are slowly varying functions of $x^\mu$ , one can solve Einstein's equations order by order in derivative expansion. Up to second order in derivative expansion the solutions are given by \cite{Banerjee:2008th},
\begin{eqnarray}\label{uc bulk dual}
\begin{split}
ds^{2} &= -2u_{\mu}dx^{\mu}dr-r^{2}f(r, m, q)u_{\mu}u_{\nu}dx^{\mu}dx^{\nu}+r^{2}P_{\mu \nu}dx^{\mu}dx^{\nu} \\
& + \frac{2r^{2}}{R}F_{2}\lb\frac{r}{R},\frac{m}{R^{4}}\rb\sigma_{\mu\nu}dx^{\mu}dx^{\nu} 
+\frac{2}{3}r u_{\mu}u_{\nu}(\partial u)dx^{\mu}dx^{\nu}-2r u_{\mu}(u^{\lambda
}\partial_{\lambda} u_{\nu})dx^{\mu}dx^{\nu} \\
& -2u_{\mu}\Big(\frac{\sqrt{3}\kappa q^{3}}{m r^{4}}l_{\nu} + \frac{6 q r^{2}}{R^{7}} P^{\lambda}_{\nu}\cD_{\lambda}qF_{1}\lb \frac{r}{R},\frac{m}{R^{4}}\rb\Big)dx^{\mu}dx^{\nu} + r^{2}\alpha_{\mu\nu}(r) dx^{\mu}dx^{\nu} \\
&+ 3 h(r) u_{\mu} dx^{\mu} dr +r^{2}h(r)P_{\mu\nu}dx^{\mu}dx^{\nu}-12 r^{2}j_{\alpha}P^{\alpha}_{\nu}u_{\mu}dx^{\mu}dx^{\nu} + \frac{k(r)}{r^{2}}u_{\mu}u_{\nu}dx^{\mu}dx^{\nu}
\end{split}
\end{eqnarray}
and
\begin{eqnarray}\label{eq:bulkgaugefield}
   \mathcal{C}_{\mu} = \frac{\sqrt{3}q}{2 r^{2}}u_{\mu}+\frac{\sqrt{3}w(r)}{2r^{2}}u_{\mu}+\frac{3 \kappa q^{2}}{2m r^{2}}l_{\mu}-\frac{\sqrt{3}r^{5}}{2 R^{8}}P^{\lambda}_{\mu}\mathcal{D}_{\lambda}q F_{1}^{(1,0)}\lb\frac{r}{R},\frac{m}{R^{4}}\rb + \frac{\sqrt{3}r^{5}}{2}g_{\mu}(r).
\end{eqnarray}
$\sigma^{\mu\nu}$, $l^\mu$ and the Weyl covariant derivative $\cD_\mu$ are given by
\begin{equation}
    \sigma^{\mu\nu} = P^{\mu\alpha}P^{\nu\beta} \partial_{(\alpha}u_{\beta)} - \frac13 P^{\mu\nu}\partial\cdot u, \quad l^\mu = \epsilon^{\nu\lambda\sigma\mu} u_\nu \partial_\lambda u_\sigma, \quad P^\lambda_\mu \cD_\lambda q = P^\lambda_\mu \partial_\lambda q + 3 (u\cdot \partial u_\mu)q.
\end{equation}
The functions $F_1$ and $F_2$ are given by
\begin{eqnarray}
   \begin{split}
   F_{1}\lb\frac{r}{R},\frac{m}{R^{4}}\rb  =& \frac{1}{3}(1-\frac{m}{r^{4}}+\frac{q^{2}}{r^{6}})\int_{\frac{r}{R}}^{\infty}dp\frac{1}{(1-\frac{m}{R^{4}p^{4}}+\frac{q^{2}}{R^{6}p^{6}})^{2}}\Big(\frac{1}{p^{8}}-\frac{3}{4 p^{7}}(1+\frac{R^{4}}{m})\Big)\\
   F_{2}\lb \frac{r}{R},\frac{m}{R^{4}}\rb  =& \int_{\frac{r}{R}}^{\infty}dp\frac{p(p^{2}+p+1)}{(p+1)(p^{4}+p^{2}-\frac{m}{R^{4}}+1)}.
   \end{split}
\end{eqnarray}
The parameters $m$, $q$ and $R$ are not independent at the leading order, they are related by
\begin{equation}\label{eq:qmR}
    q^2 = R^{2}(m-R^{4}).
\end{equation}
The exact form of various scalar, vector and tensor functions ($h(r)$, $k(r)$,$w(r)$, $j_\mu(r)$, $g_\mu(r)$ and $\alpha_{\mu\nu}(r)$) can be found in \cite{Banerjee:2008th}. We have also presented these functions in light cone coordinates in appendix \ref{bulkg}.

The metric and gauge field given by Eqn.(\ref{uc bulk dual}) and (\ref{eq:bulkgaugefield}) provide holographic description of $U(1)$ charged conformal fluid in flat $3+1$ dimensions. One can use the AdS/CFT dictionary to compute the constitutive relations (stress-energy tensor and $U(1)$ current) of the relativistic fluid up to second order in derivative expansion. They are given by
\begin{eqnarray}\label{eq:st2ndorder}
\begin{split}
	T^{\mu\nu}  = & (E+P)u^{\mu}u^{\nu}+P \eta^{\mu\nu}-2\eta \sigma^{\mu\nu}+ \mathcal{N}_{1} u^{\lambda}\mathcal{D}_{\lambda}\sigma^{\mu\nu} + \mathcal{N}_{2} (\omega^{\mu}\hspace{0.2pt}_{\lambda}\sigma^{\lambda\nu}+\omega^{\nu}\hspace{0.2pt}_{\lambda}\sigma^{\lambda\mu}) \\
	& + \mathcal{N}_{3} (\sigma^{\mu}\hspace{0.2pt}_{\lambda}\sigma^{\lambda\nu}-\frac{1}{3}P^{\mu\nu}\sigma^{\alpha\beta}\sigma_{\alpha\beta}) + \mathcal{N}_{4} (\omega^{\mu}\hspace{0.2pt}_{\lambda}\omega^{\lambda\nu}+\frac{1}{3}P^{\mu\nu}\omega^{\alpha\beta}\omega_{\alpha\beta}) \\ & + \mathcal{N}_{5}q^{-1}\Pi^{\mu\nu\alpha\beta}\mathcal{D}_{\alpha}\mathcal{D}_{\beta}q  + \mathcal{N}_{6}q^{-2}\Pi^{\mu\nu\alpha\beta}\mathcal{D}_{\alpha}q\mathcal{D}_{\beta}q + \mathcal{N}_{7}\Pi^{\mu\nu\alpha\beta}(\mathcal{D}_{\alpha}l_{\beta}+\mathcal{D}_{\beta}l_{\alpha})  \\
    &+\mathcal{N}_{8}q^{-1}\Pi^{\mu\nu\alpha\beta}l_{\alpha}\mathcal{D}_{\beta}q+\mathcal{N}_{9}q^{-1}\epsilon^{\alpha\beta\lambda(\mu}\sigma^{\nu)}_{\lambda}u_{\alpha}\mathcal{D}_{\beta}q
	\end{split}
	\end{eqnarray}
		and
	\begin{eqnarray}\label{eq:Jrel}
	\begin{split}
	   J^{\mu} = & 4\sqrt{3}q u^{\mu}-4\sqrt{3}\lambda_{r} P^{\mu\nu}\cD_{\nu}q + \xi_{r} l^{\mu} + \gamma_1 P^{\mu\nu}\mathcal{D}_{\lambda}\sigma^{\lambda}_{\nu} + \gamma_2 P^{\mu\nu}\mathcal{D}_{\lambda}\omega^{\lambda}_{\nu} + \gamma_3 l^{\lambda}\sigma_{\lambda}^{\mu}\\& + \gamma_4 q^{-1}\sigma^{\mu\lambda}\mathcal{D}_{\lambda}q + \gamma_5 q^{-1}\omega_{\mu\lambda}\mathcal{D}_{\lambda}q.
  \end{split}
	\end{eqnarray}
The tensors $\omega^{\mu\nu}$ and $\Pi^{\alpha\beta\mu\nu}$ is given by
\begin{eqnarray}
        \omega^{\mu\nu}&=&\frac{1}{2}P^{\mu\alpha}P^{\nu\beta}(\partial_{\alpha}u_{\beta}-\partial_{\beta}u_{\alpha})\\
	\Pi^{\alpha\beta\mu\nu}&=&\frac{1}{2}\lb P^{\alpha\mu}P^{\beta\nu}+P^{\alpha\nu}P^{\beta\mu}-\frac{2}{3}P^{\alpha\beta}P^{\mu\nu}\rb.
\end{eqnarray}
One can independently construct this stress tensor by demanding Weyl covariance \cite{Erdmenger:2008rm}. The transport coefficients for such fluid in general depend on the temperature and charge density. The holographic values of different independent transport coefficients appear in (\ref{eq:st2ndorder}) and (\ref{eq:Jrel}) are given in table \ref{eq:reltransports}. $m_0$ and $q_0$ appearing in different transports are leading (unperturbed) values of $m$ and $q$. Also we have set the leading value of the horizon radius $R$ to be one.
\begin{table}[H]
    \begin{tabular}{| c|c |}
    \hline
       $\eta$  & $R^3$ \\
         \hline
         $\lambda_{r}$  & $\frac{m+R^4}{4m R}$\\
         \hline
         $\xi_{r}$ & $\frac{12\kappa q^2}{m}$\\
         \hline
         \end{tabular}
         \quad 
         \begin{tabular}{| c|c |}
         \hline
         $\mathcal{N}_{1}$ & $ 2 + \frac{m_{0}}{\sqrt{4m_{0}-3}}\log\lb\frac{3-\sqrt{4m_{0}-3}}{3+\sqrt{4m_{0}-3}}\rb$  \\
         \hline
         $\mathcal{N}_{2}$  & $\mathcal{N}_{1}-2$\\
         \hline
         $\mathcal{N}_{3}$  & 2\\
         \hline
         $\mathcal{N}_{4}$  & $-\frac{4(m_{0}-1)}{m_{0}}(12(m_{0}-1)\kappa^{2}-m_{0})$\\
         \hline
         $\mathcal{N}_{5}$  & $-\frac{(m_{0}-1)}{2  m_{0}}$ \\
         \hline
         $\mathcal{N}_{6}$  & $\frac{1}{2 n^{2}}(m_{0}-1)(3\log(2)-1)$\\
         \hline
         $\mathcal{N}_{7}$  & $\frac{\sqrt{3}\kappa(m_{0}-1)^{\frac{3}{2}}}{m_{0}}$\\
         \hline
         $\mathcal{N}_{8}$, $\mathcal{N}_{9}$  & 0\\
         \hline
         \end{tabular}
         \quad 
         \begin{tabular}{| c|c |}
         \hline
         $\gamma_1$  & $\frac{3\sqrt{3}q_{0}}{m_{0}}$\\
         \hline
         $\gamma_2$  & $\frac{2\sqrt{3} q_{0}^{3}\kappa^{2}}{m_{0}^{2}}$\\
         \hline
         $\gamma_3$  & $-\frac{12q_{0}^{2}\kappa }{m_{0}^{2}}$\\
         \hline
         $\gamma_4$  & $2\sqrt{3}q_{0}\log2$\\
         \hline
         $\gamma_5$  & $- \frac{\sqrt{3}q_{0}(m_{0}^{2}-48\kappa^{2}q_{0}^{2}+3)}{2m_{0}^{2}}$\\
         \hline
    \end{tabular}
    \caption{Relativistic transports.}
    \label{eq:reltransports}
\end{table}

\section{Light-cone reduction and non-relativistic second order fluid }\label{lcr}

As we have already discussed, a non-relativistic fluid can be obtained by LCR of relativistic fluid. The idea follows from the fact that LCR of conformal algebra renders Schr\"odinger algebra in one lower dimension. We  start  with  relativistic  constitutive  equations  and reduce these equations along a light-cone direction. As a result, the non-relativistic quantities like energy density, pressure, stress tensor, charge current etc.  are given in terms of different components of relativistic stress tensor and $U(1)$ current.

The conservation equations for a $U(1)$ charged relativistic fluid are given by
	\begin{eqnarray}
	\partial_{\mu}T^{\mu\nu}=0,\hspace{1cm}\partial_{\mu}J^{\mu}=0.
	\end{eqnarray}
We are considering the fluid in a flat background and in absence of any external electric or magnetic fields.  The flat metric in ($d+1, 1$) dimensions is given by
\begin{equation}
    ds^2 = -(dx^0)^2 + (dx^{d+1})^2 + \sum_{i=1}^d (dx^i)^2.
\end{equation}
In the light-cone frame $x^\pm$, defined by
\begin{eqnarray}	    
    x^{+} = \frac{x^{0}+x^{d+1}}{\sqrt{2}}, \hspace{1cm}x^{-} = \frac{x^{0}-x^{d+1}}{\sqrt{2}}	\end{eqnarray}
the metric takes the form
\begin{equation}
    ds^2 = -2 dx^+dx^- + \sum_{i=1}^d (dx^i)^2.
\end{equation}
It turns out that \cite{Rangamani:2008gi, Brattan:2010bw, Banerjee:2014mka} under LCR the  relativistic constitutive equations boil  down to the non-relativistic constitutive equations in one lower dimension. These equations describe a non-relativistic fluid in one lower dimension.

To obtain the Schr\"odinger algebra one has to look for the subalgebra of the conformal algebra where all the generators commute with $P_-$ (translation along $x^-$). Therefore we consider only those solutions to the relativistic equations that do not depend on $x^-$, i.e. $x^-$ is an isometry direction. To make the reduced theory consistent with discrete light-cone quantisation we reduce the theory along $x^-$ direction, and consider $x^+$ to be the non-relativistic time. 

In the light-cone frame ($x^\mu = \{x^+, x^-, x^i\}$) the non-zero components of metric and partial derivatives are given by
\begin{equation}
\begin{split}
    g^{+-} & = g_{+-} = -1, \quad g^{ij} = g_{ij} = 1 \\
    \partial_{\mu} & = \{\partial_+, \partial_-, \partial_i\}, \quad \partial^{\mu} = \{- \partial_-, - \partial_+, \partial_i\}.
    \end{split}
\end{equation}
After reduction, the relativistic constitutive equations are given by
\begin{eqnarray}
	\partial_{+}T^{+ -}+\partial_{i}T^{i -}=0,\hspace{0.3cm}\partial_{+}T^{+ +}+\partial_{i}T^{i +}=0,\hspace{0.3cm}\partial_{+}T^{+ j}+\partial_{i}T^{i j}=0,\hspace{0.3cm}\partial_{+}J^{+}+\partial_{i}J^{i}=0.\nonumber
\end{eqnarray}
These equations can be interpreted as dynamical equations governing the motion of non-relativistic fluid if we identify different components of energy-momentum tensor and charge current with non-relativistic quantities as follows
\begin{eqnarray}\label{eq:rel-nonreldic}
	T^{++}=\rho,\quad T^{+i}=\rho v^{i},\quad T^{+-}=\epsilon+\frac{1}{2}\rho v^{2},\quad T^{ij}=t^{ij},\quad Q=J^{+},\quad j^{i}=J^{i}.
\end{eqnarray}
A relativistic fluid in four space time dimensions is described in terms of its normalized four velocity $u^{\mu}$, temperature $T$ and charge density $q$. On the other hand the non-relativistic fluid in one lower dimension is described in terms of its mass density $\rho$, pressure $p$, spatial velocity $v^{i}$ and charge density $Q$. Hence the number of fluid variables in both the theories are the same and a mapping between them can be found. Since fluid dynamics admits derivative expansion, the relations between the non-relativistic and relativistic fluids can be obtained order by order in derivative expansion.

LCR of conformal uncharged and charged fluid up to first order in derivative expansion have been derived in \cite{Rangamani:2008gi,Brattan:2010bw,Banerjee:2014mka,Dutta:2018xtr}. In a recent paper \cite{Dutta:2020gcg} LCR of conformal uncharged second order fluid has been considered thoroughly. It was shown that the Rivlin-Ericksen fluid \cite{RE} is a subclass of such non-relativistic fluid. In this paper we further extend those calculations. We light-cone reduce a $(3+1)$ dimensional second order Weyl invariant $U(1)$ charged fluid (\ref{eq:st2ndorder}, \ref{eq:Jrel}) and obtain the corresponding second order charged non-relativistic fluid.

To perform the reduction we need to consider two sets of constraints relations. The first set of relations is derived from the conservation laws of the first order corrected stress tensor and charge current. We list the independent first order data in table \ref{lcr constraint1}. We have categorise the non-relativistic fluid data in scalar, vector and tensor with respect to the $SO(2)$ symmetry group, as the non-relativistic fluid has an $SO(2)$ symmetry (rotation about $z$ axis). 
\begin{table}[h!]
\begin{center}
\begin{tabular}{|c|c|c|c|}
    \hline
     & Data & Constraint & Independent data\\
     \hline
        Scalar & $\partial_{+}T$, $\partial_{+}\phi$, & $\partial_{\mu}T^{\mu +}=0$,  & $\partial_{i}v^{i}$, $\epsilon^{ij}\tilde{\omega}_{ij}$\\
       & $\epsilon^{ij}\omega_{ij}$, $\partial_{+}u^{+}$, $\partial_{i}u^{i}$ & $\partial_{\mu}T^{\mu -}=0$, $\partial_{\mu}J^{\mu }=0$ & \\
        \hline
         Vector & $\partial_{i}T$, $\partial_{i}\phi$, $\partial_{i}u^{+}$, $\partial_{+}u^{i}$ & $\partial_{\mu}T^{\mu i}=0$ & $\partial_{i}\tau$, $\partial_{i}\nu$, $\partial_{i}\mu_m$ \\
& $\epsilon^{ik}\partial_{k}T$,$ \epsilon^{ik}\partial_{k}\phi$, $\epsilon^{ik}\partial_{k}u^{+}$ &  &$\epsilon^{ik}\partial_{k}\tau$, $\epsilon^{ik}\partial_{k}\nu$, $\epsilon^{ik}\partial_{k}\mu_m$ \\
        \hline 
        Tensor & $\sigma^{ij}$ &  & $\tilde{\sigma}^{ij}$ \\
        \hline 
    \end{tabular}
    \caption{Independent non-relativistic fluid data at first order.}
    \label{lcr constraint1}
\end{center}
\end{table}

The second set of constraints can be derived by taking derivatives of the zeroth order conservation laws. The independent second order data are given in table \ref{lcr constraint2}.
\begin{table}
\begin{center}
    \begin{tabular}{|c|c|c|c|}
    \hline
     & Data & Constraint & Independent data\\
     \hline
        \multirow{4}{3em}{Scalar} & $\partial_{+}^{2}T$, $\partial_{+}^{2}\phi$, &$\partial_{+}\partial_{\mu}T^{\mu +}=0$,  & $\partial_{i}^{2}\tau $, $\partial_{i}^{2}\nu$\\
         & $\partial_{+}^{2}u^{+}$, $\partial_{i}^{2}T$, $\partial_{i}^{2}\phi$, & $\partial_{+}\partial_{\mu}T^{\mu -} = 0$, & $\partial_{i}^{2}\mu_m$\\
      & $\partial_{i}^{2}u^{+}$,$\partial_{i}^{2}u^{+}$ & $\partial_{i}\partial_{\mu}T^{\mu i}=0$, & \\
        &  & $\partial_{+}\partial_{\mu}J^{\mu }=0$ & \\
        \hline
         Vector & $\partial_{j}\sigma^{ij}$, $\partial_{j}\omega^{ij}$,  & $\partial_{+}\partial_{\mu}T^{\mu i}=0$, & $\partial_{j}\tilde{\sigma}^{ij}$, $\partial_{j}\tilde{\omega}^{ij}$ \\
         &$\partial_{+}\partial_{i}u^{+}$, $\partial_{+}^{2}u^{i}$ & $\partial_{i}\partial_{\mu}T^{\mu +}=0$, & $\epsilon^{il}\partial_{k}\tilde{\sigma}^{l k}$,$\epsilon^{il}\partial_{k}\tilde{\omega}^{l k}$ \\
         &$\partial_{+}\partial_{i}T$, $\partial_{+}\partial_{+}\phi$& $\partial_{i}\partial_{\mu}T^{\mu -}=0$  & \\
         &$\epsilon^{il}\partial_{k}\tilde{\sigma}^{l k}$,$\epsilon^{il}\partial_{k}\omega^{l k}$  & $\partial_{i}\partial_{\mu}J^{\mu }=0$ &\\
        \hline 
        Tensor & $\partial^{i}\partial^{j }T$, $\partial^{ i}\partial^{j }\phi$,  &  & $\partial^{\langle i}\partial^{j \rangle }\tau $, $\partial^{\langle i}\partial^{j \rangle }\nu $ \\
         & $\partial^{ i}\partial^{j } u^+$, \  $\partial_{+}\sigma^{ij}$,   & $\partial_{i}\partial_{\mu}T^{\mu j}=0$ & $\partial^{\langle i}\partial^{j \rangle }\mu_m$, \\
         & $\epsilon^{ik}\partial^{k}\partial^{j} u^{+}$, $\epsilon^{ik}\partial^{k}\partial^{j}T$  &  & $\epsilon^{\langle ik}\partial^{k}\partial^{j\rangle }\mu_m$, \ $\epsilon^{\langle ik}\partial^{k}\partial^{j\rangle }\tau $, \\
         &$\epsilon^{ik}\partial^{k}\partial^{j}\phi$   &  & $\epsilon^{\langle ik}\partial^{k}\partial^{j\rangle }\nu $ \\
        \hline 
    \end{tabular}
\caption{Independent non-relativistic fluid data at second order.}
\label{lcr constraint2}
\end{center}
\end{table}
$\tilde \sigma_{ij}$ and $\tilde\omega_{ij}$ are given by
\begin{equation}
    \tilde \sigma^{ij} = \partial^{i}v^j + \partial^{j}v^i -\delta^{ij} \partial_{k} v^k, \quad \tilde \omega^{ij} = \partial^{i}v^j - \partial^{j}v^i.
\end{equation}
The angle brackets denote symmetric traceless combinations
\begin{eqnarray}
		A ^{\langle i}B^{j\rangle}&=&(A^{i}B^{j}+A^{j}B^{j}-\delta^{i j}A^{k}B_{k}).
\end{eqnarray}
The non-relativistic temperature and $U(1)$ chemical potential are defined as,
\begin{eqnarray}\label{eq:Ttaumuphi}
		\tau=\frac{T}{u^{+}},\hspace{0.5cm} \mu=\frac{\phi}{u^{+}}.
\end{eqnarray}
The relativistic variables satisfy Euler relation : $E+P - q \phi = TS$. Using the relation between relativistic fluid variables and non-relativistic fluid variable at leading order \cite{Dutta:2018xtr, Rangamani:2008gi, Brattan:2010bw}
\begin{equation}
    \rho = (E+P)(u^+)^2, \quad p = P, \quad \epsilon = \frac{E-P}{2}, \quad Q = 4\sqrt{3} q u^+
\end{equation}
we see that the non-relativistic variables satisfy
\begin{equation}\label{eq:nonreleuler}
    \epsilon+p-\rho \rho_{m}-\mu \lb \frac{Q}{4\sqrt{3}} \rb =\tau s
\end{equation}
where $\rho_{m}$ is given by,
\begin{equation} \label{eq:rhom}
	 \rho_{m}=-\frac{1}{2 (u^{+})^{2}}.
\end{equation}
and $s = u^+ S$ is the entropy density of the non-relativistic fluid. This equation is interpreted as the Euler relation for the non-relativistic fluid with an extra fluid variable $\rho_m$, which can be considered as chemical potential associated with particle number conservation \cite{Banerjee:2014mka}. In \cite{Dutta:2018xtr} a new basis has been introduced in terms of  reduced chemical potential $\nu$ and redefined mass chemical potential $\mu_{m}$ given by,
\begin{eqnarray} \label{eq:numumumup}
      \nu=\frac{\mu}{\tau},\hspace{0.5cm}\partial_{i}\mu_{m}=(u^{+})^{3}\partial_{i}\rho_{m} \ \text{i.e.} \ \mu_m = u^+.
\end{eqnarray}
In this paper we use the variables $\{\mu_m, \nu, \tau\}$ to express the non-relativistic constitutive relations.

In tables \ref{lcr constraint1} and \ref{lcr constraint2} all the non-relativistic symmetric tensors are traceless. After using the constraint relations we have three independent second order symmetric traceless tensors in the parity even sector and three in the parity odd sector. In addition to these, we have composite symmetric traceless tensors constructed from the independent first order vectors. We have seven such vectors in parity even sector and seven in parity odd sector\footnote{In higher space dimensions there can be more composite terms like $\sigma^{\langle i k}\sigma_{}^{k j\rangle}$ which identically vanishes in two space dimensions}. Thus, in total we have 10 parity even and 10 parity odd second order independent symmetric traceless tensors. The light-cone reduced fluid stress tensor is expected to depend on these independent data.

After a careful computation and using the set of constraints we finally write down the stress tensor in a simplified form given by,
\begin{eqnarray}\label{eq:nonrelst2nd}
\begin{split}
    t^{i j} & = \rho v^{i} v^{j} + p \delta^{ij}  - n_{1}\tilde{\sigma}^{i j} + n_{a+1} \partial^{\langle i}\partial^{j\rangle} X_a + c_{1} \tilde{\sigma}^{\langle i k}\tilde{\omega}^{k j\rangle}  + \mathbf{c}_{ab} \partial^{\langle i} X_a \partial^{j\rangle} X_b \\ 
    & \quad + g_{a}\epsilon^{\langle i k}\partial^{k}\partial^{j \rangle} X_a + \mathbf{g}_{ab} \epsilon^{\langle i k} \partial_{k} X_a \partial^{j\rangle} X_b+g_{10} \epsilon^{\langle i k}\tilde{\sigma}^{k l}\tilde{\omega}^{l j\rangle}.
\end{split}
\end{eqnarray}
The indices $a,b$ run over $1, 2$ and $3$, $X_a = \{\mu_m, \nu, \tau\}$ denotes an array of non-relativistic parameters, $\mathbf{c}_{ab}$ and $\mathbf{g}_{ab}$ are $3\times 3$ transport matrices given by
\begin{equation}\label{eq:cglcr}
    \mathbf{c} = \begin{pmatrix}
c_2 & c_3/2 & c_4/2\\
c_3/2 & c_5 & c_6/2 \\
c_4/2 & c_6/2 & c_7
\end{pmatrix},  \quad \mathbf{g} = \begin{pmatrix}
g_4 & g_5 & g_6\\
g_5 & g_7 & g_8 \\
g_6 & g_8 & g_9
\end{pmatrix}.
\end{equation}
The other non-relativistic fluid variables : mass density $\rho$, pressure $p$, velocity $v^i$, energy density $\epsilon$, energy current $e_i$ and non-relativistic transports are given in appendix \ref{lcrgen}. 

The non-relativistic charge current which comes with eleven parity even and eleven parity odd transports are given by,
\begin{eqnarray}\label{eq:nonrelJi}
    \begin{split}
       \mathcal{J}^{i} & =  Q v^{i} + \mathbf{q}_a \partial^{i} X_a + \tilde{\mathbf{q}}_a \epsilon^{ik}\partial_{k}X_a +{n}^{\mathcal{J}}_{1}\partial_{k}\tilde{\sigma}^{ik}+{n}^{\mathcal{J}}_{2}\partial_{k}\tilde{\omega}^{ik}+{n}^{\mathcal{J}}_{3}\epsilon^{il}\partial_{k}\tilde{\sigma}^{lk}+{n}^{\mathcal{J}}_{4}\epsilon^{il}\partial_{k}\tilde{\omega}^{lk} \\
       & \quad + \mathbf{c}^{\cJ}_a \tilde{\sigma}^{ik} \partial_{k} X_a + \tilde{\mathbf{c}}^{\cJ}_a \tilde{\omega}^{ik} \partial_{k} X_a + + \bar{\mathbf{c}}^{\cJ}_a (\partial_k v^k) \partial^{i} X_a + \mathbf{g}^{\cJ}_a \epsilon^{il}\tilde{\sigma}^{lk}\partial_{k} X_a + \tilde{\mathbf{g}}^{\cJ}_a \epsilon^{il}\tilde{\omega}^{lk}\partial_{k} X_a \\
     & \quad + \bar{\mathbf{g}}^{\cJ}_a (\partial_k v^k) \epsilon^{il} \partial_{l} X_a
    \end{split}
\end{eqnarray}
where 
\begin{eqnarray}\label{eq:non-relarray}
    \begin{split}
\mathbf{q}_a & = \{\lambda_q, \sigma_q, \kappa_q\}, \quad  \tilde{\mathbf{q}}_a = \{\tilde{ \lambda}_q, \tilde{\sigma}_q, \tilde{\kappa}_q \}, \quad  \mathbf{c}^\cJ_a = \{c_1^{\cJ}, c_7^{\cJ}, c_4^{\cJ}\}, \quad  \tilde{\mathbf{c}}^\cJ_a = \{c_2^{\cJ}, c_8^{\cJ}, c_5^{\cJ}\},\\
\bar{\mathbf{c}}^\cJ_a & = \{c_3^{\cJ}, c_9^{\cJ}, c_6^{\cJ}\}, \quad  \mathbf{g}^\cJ_a = \{g_1^{\cJ}, g_7^{\cJ}, g_4^{\cJ}\}, \quad \tilde{ \mathbf{g}}^\cJ_a = \{g_2^{\cJ}, g_8^{\cJ}, g_5^{\cJ}\}, \quad  \bar{\mathbf{g}}^\cJ_a = \{g_3^{\cJ}, g_9^{\cJ}, g_6^{\cJ}\} 
    \end{split}
\end{eqnarray}
are different transport arrays. The non-relativistic charge density $Q$ and different charge transports are given in appendix \ref{lcrgen}. 

As a consistency check we see that in the limit $q\ra 0$ all the constitutive relations reduce to those for uncharged fluid found in \cite{Dutta:2020gcg}. In \cite{Dutta:2020gcg} the stress tensor at second order was written in terms of a quantity $\mathcal{B}^{ij} = \partial^{i}a^j + \partial^j a^i + 2  (\partial^iv^k) (\partial^j v_k)$. The acceleration $a^i$ for charged fluid is given by,
\begin{equation}\label{eq:aidef}
   a^{i}=-\frac{\partial^{i}\mu_{m}}{u^{+}}-\frac{\partial^{i}\tau}{\tau}-\frac{\mu q}{(\tau s +\mu q)}\frac{\partial_{i}\nu}{\nu}.
\end{equation}
In order to match our results with \cite{Dutta:2020gcg} one has to replace $\mathcal{B}^{ij}$  using the above expression of $a^{i}$ (with $q=0$) and trade $\partial^i \mu_m$ with $a^i$ using the above relation.

Our next goal is to construct a holographic description of the non-relativistic fluid studied in this section.

\section{Holographic description of second order non-relativistic fluid} \label{sec:holography}

TsT transformation is a solution generating technique in string theory which has been used to generate a black hole solution with asymptotically Schrodinger isometry \cite{Herzog:2008wg,Maldacena:2008wh,Adams:2008wt,Yamada:2008if,Bobev:2009zf,Bobev:2009mw,Imeroni:2009cs,Kim:2010tf,Banerjee:2011jb}. The idea behind is based on the fact that supergravity has more symmetry than the string theory. If we perform TsT transformation on a string theory solution then it generates new solution of string theory, which is guaranteed to be a solution of supergravity theory. To perform the TsT transformation on a five dimensional geometry we first need to uplift the solution to ten dimensions to embed in string theory \cite{Herzog:2008wg,Imeroni:2009cs,Banerjee:2011jb}. This uplift can be performed such that the 10 dimensional metric is the direct sum of the five dimensional metric ($AdS$ part) and a five sphere which is written as a fibration over $CP^{2}$. Such ten dimensional uplift of the five dimensional metric will be a solution of type IIB string theory. 

We write the five dimensional metric and the gauge field in the following form,
\begin{eqnarray}\label{eq:5dmet}
\begin{split}
         ds^{2} &= - 2u_{\mu}dx^{\mu}dr+ S u_{\mu}dx^{\mu}dr+ S_{\mu \nu}dx^{\mu}dx^{\nu}\\
         \cC & = \cC_a dx^a = \cC_\mu dx^\mu.
\end{split}
\end{eqnarray}
The exact form of $S$ and $S_{\mu\nu}$ depend on the solution we are considering. Since we are interested in the bulk dual of second order charged fluid (\ref{uc bulk dual} and \ref{eq:bulkgaugefield}), $S$ and $S_{\mu\nu}$ are given by
\begin{eqnarray}\label{eq:SSmuC}
   \begin{split}
       S & =3 h(r)\\
        S_{\mu\nu} & = -r^{2}f(r, m, q)u_{\mu}u_{\nu}+r^{2}P_{\mu \nu} + \frac{2r^{2}}{R}F_{2}\lb\frac{r}{R},\frac{m}{R^{4}}\rb\sigma_{\mu\nu} 
+\frac{2}{3}r u_{\mu}u_{\nu}(\partial u)-2r u_{\mu}(u^{\lambda
}\partial_{\lambda} u_{\nu}) \\
& \quad - 2 u_{\mu}\Big(\frac{\sqrt{3} \kappa q^{3}}{m r^{4}} l_{\nu} + \frac{6 q r^{2}}{R^{7}} P^{\lambda}_{\nu} \cD_{\lambda}qF_{1}\lb \frac{r}{R},\frac{m}{R^{4}}\rb\Big) + r^{2}\alpha_{\mu\nu}(r)+r^{2}h(r)P_{\mu\nu}\\
& \quad -12 r^{2}j_{\alpha}P^{\alpha}_{\nu}u_{\mu} + \frac{k(r)}{r^{2}}u_{\mu}u_{\nu}  \\
        \cC_\mu & = \frac{\sqrt{3}q}{2 r^{2}}u_{\mu}+\frac{\sqrt{3}w_{2}(r)}{2r^{2}}u_{\mu}+\frac{3 \kappa q^{2}}{2m r^{2}}l_{\mu}-\frac{\sqrt{3}r^{5}}{2 R^{8}}P^{\lambda}_{\mu}\mathcal{D}_{\lambda}q F_{1}^{(1,0)}\lb\frac{r}{R},\frac{m}{R^{4}}\rb + \frac{\sqrt{3}r^{5}}{2}g_{\mu}(r).
    \end{split}
\end{eqnarray}
To uplift the five dimensional metric to ten dimensions we introduce the five dimensional Sasaki-Einstein manifold,	\begin{eqnarray}
	ds_{SE}^{2} = \lb d\psi+\mathcal{P}-\frac{2}{\sqrt{3}}\mathcal{C}\rb^{2}+ds_{CP^{2}}^{2}
	\end{eqnarray}
	where,
\begin{eqnarray}
 \begin{split}
	\mathcal{P} & = \frac{1}{3}(d\chi_{1}+d\chi_{2})-\sin^{2}\alpha(d\chi_{2}\sin^{2}\beta+d\chi_{1}\cos^{2}\beta)\\
	ds_{CP^{2}}^{2} & = d\alpha^{2}+\sin^{2}\alpha d\beta^{2}+\sin^{2}\alpha \cos^{2}\alpha(\cos^{2}\beta d\chi_{1}+\sin^{2}\beta d\chi_{2})^{2}\\
 & \quad + \sin^{2}\alpha\sin^{2}\beta\cos^{2}\beta(d\chi_{1}-d\chi_{2})^{2}.
 \end{split}
\end{eqnarray}
We uplift the five dimensional metric (\ref{eq:5dmet}) to ten dimensions as a fibration over $CP^2$. The ten dimensional metric is given by
\begin{eqnarray}\label{eq:10dmet}
		ds_{10}^{2}&=& -2u_{\mu}dx^{\mu}dr+ S u_{\mu}dx^{\mu}dr+ S_{\mu \nu}dx^{\mu}dx^{\nu}+\lb d\psi+\mathcal{P}-\frac{2}{\sqrt{3}}\mathcal{C}\rb^{2}+ds_{CP^{2}}^{2}
\end{eqnarray}
supported by a five form field strength
\begin{eqnarray}\label{eq:f5}
		\mathcal{F}_{5} = 2(1+\star_{10})\lB \lb d\psi +\mathcal{P}-\frac{2}{\sqrt{3}}\mathcal{C}\rb \wedge J_{2} -\frac{1}{\sqrt{3}}\star_{5}\mathcal{G}\rB \wedge J_{2},\hspace{0.4cm}J_{2}=\frac{1}{2}d\mathcal{P},
\end{eqnarray}
a two form field strength $\cG = d\cC$ and a dilaton $\Phi =0$.
		
To perform the TsT transformation on the ten dimensional solutions (\ref{eq:10dmet}) and (\ref{eq:f5}) we need to identify two isometry directions. One such isometry direction is $\psi$. To identify the other isometry direction we introduce the bulk light-cone coordinates $x^\pm$
\begin{eqnarray}
\begin{split}
		x^{+} & = (v+z), \\
		x^{+} & = \frac{1}{2}(v-z)
  \end{split}
\end{eqnarray}
and choose $x^{-}$ to be the second isometry direction. In the light-cone frame the ten dimensional metric can be written as,
		\begin{eqnarray}
		ds_{10}^{2} = A_{1}(dx^{-}+K_{1})^{2}+ds_{4}^{2}+\lb d\psi+\mathcal{P}-\frac{2}{\sqrt{3}}\mathcal{C}\rb^{2}+ds_{CP^{2}}^{2}
		\end{eqnarray}
where,
		\begin{eqnarray}
  \begin{split}
		A_{1} & = S_{--}, \quad 
		K_{1} = \frac{1}{S_{--}}\lb S_{+-}dx^{+}-u_{-}dr+S_{-i}dx^{i}+\frac{S}{2}u_{-}dr \rb, \\
		ds_{4}^{2} & = - 2u_{\bar{a}}dx^{\bar{a}}dr+S u_{\bar{a}}dx^{\bar{a}}dr+S_{\bar{a}\bar{b}}dx^{\bar{a}\bar{b}}-A_{1}K_{1}^{2} .
  \end{split}
		\end{eqnarray}
Here $i, j$ are running over $\{x, y\}$ and $\bar a, \ \bar b$ are running over $\{x^+, x, y\}$. Different components of $S : S_{+-}, S_{-i}, S_{\bar a\bar b}$ can be computed using (\ref{eq:SSmuC}) and the related expressions given in appendix \ref{bulkg}.
  
The TsT transformations correspond to performing T-duality along $\psi$ direction, followed by a shift\footnote{Here we set the twist parameter to 1.} along $x^-$,  i.e., $x^- \ra x^- - \psi$. Finally we perform T-duality back along the $\psi$ direction \cite{Mazzucato:2008tr,Banerjee:2011jb}. The TsT transformed solutions are given by \cite{Banerjee:2011jb}
\begin{eqnarray}\label{eq:tstgeo}
  \begin{split}
		d\hat{s}_{10}^{2} & = \mathcal{M}A_{1}(dx^{-}+K_{1})^{2}+\mathcal{M}\big(d\psi+\mathcal{P}-\frac{2}{\sqrt{3}}\mathcal{C}\big)^{2}+ds_{8}^{2}, \quad e^{2\hat{\phi}} = \mathcal{M},\\
		\quad \mathcal{F}_{3} & = g \wedge d\mathcal{P},\quad g = \frac{1}{2} d\mathcal{C}_{-}, \quad 
		\hat{B}_{2} = \mathcal{A}\wedge(d\psi+\mathcal{P}-\frac{2}{\sqrt{3}}\mathcal{C}), \quad 
		\hat{\mathcal{F}}_{5} = \mathcal{F}_{5}+B_{2}\wedge \mathcal{F}_{3}
  \end{split}
\end{eqnarray}
where,
		\begin{eqnarray}
		\mathcal{M} = (1+A_{1})^{-1}\quad \text{and} \quad \mathcal{A} = -\mathcal{M} A_{1}(dx^{-}+K_{1}).
		\end{eqnarray}
The TsT transformed ten dimensional fields can be truncated over $S^5$ directions \cite{Maldacena:2008wh,Adams:2009dm}. After truncation the effective five dimensional solutions consist of a metric, massive vector coming from $\hat B_2$, a scalar and a  massless vector field. The massless vector field was present in ten dimensions even before the TsT transformation and it remains unaltered after TsT and reduction. We also get an one form in five dimensions but this one form is not an independent excitation. The reduced five dimensional metric in Einstein frame and other fields are given by \cite{Adams:2009dm,Herzog:2008wg},
\begin{eqnarray}
  \begin{split}
		d\hat{s}_{E}^{2} & = e^{\frac{4\phi}{3}}A_{1}(dx^{-} + K_{1})^{2} + e^{\frac{-2\phi}{3}}ds_{4}^{2} \\
		 \hat{\mathcal{A}} & = -\mathcal{M} A_{1}(dx^{-}+K_{1}), \quad g = \frac{1}{2}d\mathcal{C_{-}}, \quad
		e^{2\phi} =\mathcal{M}.
  \end{split}
\end{eqnarray}
It can be verified that this reduced 5 dimensional geometry and fields are solution of  five dimensional effective action given by \cite{Brattan:2010bw},
\begin{eqnarray}
		S&=&\frac{1}{16 \pi G_{5}}\int dx^{5}\sqrt{-g}\Big(R-\frac{4}{3}(\partial_{a}\phi)(\partial^{a}\phi)-\frac{1}{4}e^{\frac{-8\phi}{3}}\mathcal{F}_{a b} \mathcal{F}^{a b}-4\mathcal{A}_{a}\mathcal{A}^{a}-4e^{\frac{2\phi}{3}}(e^{2\phi}-4)\nonumber\\&&-\frac{1}{3}e^{\frac{4\phi}{3}}\mathcal{G}_{a b}\mathcal{G}^{ab}-4 e^{2\phi}g_{a}g^{a}-e^{\frac{-4\phi}{3}}\mathcal{A}_{a}\mathcal{G}_{bc}\mathcal{A}^{a}\mathcal{G}^{bc}\nonumber\\&&-\frac{1}{2}e^{\frac{-2\phi}{3}}\big(-\frac{2}{\sqrt{3}}\mathcal{G}_{ab}-4 g_{a}\mathcal{A}_{b}\big)\big(-\frac{2}{\sqrt{3}}\mathcal{G}^{ab}-4 g^{a}\mathcal{A}^{b}\big)+\frac{4\kappa}{3}\mathcal{G}_{a b}\mathcal{G}_{c d}\mathcal{C}_{e}\varepsilon^{a b c d e}\Big) .
\end{eqnarray}

To find the boundary stress energy tensor the on-shell variation of the action must be well defined. This can be achieved by adding appropriate boundary terms (Gibbons-Hawking term and counterterms) to the action\cite{Herzog:2008wg, Brattan:2010bw}
\begin{eqnarray}\label{eq:Sbound}
		S_{boundary} = \frac{1}{16 \pi G_{5}}\int d\xi^{4}\sqrt{-h}\big(2K-6+\mathcal{A}_{a}\mathcal{A}^{a}+3\phi^{2}\big).
		\end{eqnarray}
To find the boundary stress energy tensor associated to this five dimensional effective action we can not directly use Brown-York type analysis\cite{Brown:1992br} as we do not have conformal structure at boundary, because of inhomogeneity in the asymptotic fall off of different metric components. We use the method developed in \cite{Ross:2009ar} to obtain the stress tensors and other currents. We summarise the method in the following subsection. 
   
\subsection{Stress energy complex}\label{stress}
	
To determine the boundary stress energy complex for Schr\'odinger field theory, we consider on-shell variation of action with respect to boundary fields \cite{Hollands:2005ya,Ross:2009ar,Dutta:2018xtr},
\begin{eqnarray}
		\delta S = \frac{1}{16 \pi G_{5}}\int d\xi^{4}\big(s_{\alpha \beta}\delta h^{\alpha\beta}+s^{\alpha}\delta \mathcal{A}_{\alpha}+s_{\phi}\delta \phi+\tilde{s}^{\alpha}\delta\mathcal{C}_{\alpha}\big)
\end{eqnarray}
where,
\begin{eqnarray}
\begin{split}
s_{\alpha\beta} & = \sqrt{-h} (\pi_{\alpha\beta}+3h_{\alpha\beta}+\mathcal{A}_{\alpha}\mathcal{A}_{\beta}-\frac{1}{2}\mathcal{A}_{\gamma}\mathcal{A}^{\gamma}h_{\alpha\beta}-\frac{3}{2}\phi^{2}h_{\alpha\beta})\\
		s_{\alpha} & =  \sqrt{-h}(-n^{a}\mathcal{F}_{a \alpha}e^{\frac{-8\phi}{3}}+2\mathcal{A}_{\alpha}), \qquad
		s_{\phi}=-\sqrt{-h}\Big(\frac{8}{3}n^{a}\partial_{a}\phi-3\phi\Big)\\
		\tilde{s}_{\alpha} & = -\sqrt{-h}\Big(\big(\frac{4}{3}e^{\frac{4\phi}{3}}+\frac{8}{3}e^{-\frac{2\phi}{3}} + 4e^{\frac{-4\phi}{3}}\mathcal{A}_{b}\mathcal{A}^{b}\big)n^{a}\mathcal{G}_{a\alpha}+\frac{8}{\sqrt{3}}e^{-\frac{2\phi}{3}}n^{a}(g_{a}\mathcal{A}_{\alpha}-g_{\alpha}\mathcal{A}_{a})\\ 
  & \qquad -\frac{16 \kappa}{3} n^{a}\epsilon_{a\alpha bcd}\mathcal{C}^{b}\mathcal{G}^{cd}\Big)
  \end{split}
		\end{eqnarray}
here $n^{a}$ is outward directed unit normal to the boundary, and  $\pi_{\alpha\beta}=K_{\alpha\beta}-K h_{\alpha\beta}$ with extrinsic curvature $K_{\alpha\beta}=\nabla_{\alpha}n_{\beta}$.

As we have already mentioned, the TsT transformed asymptotic metric is not conformal to flat space time. Therefore we can not construct the stress tensor by naively varying the action with respect to the boundary data. In \cite{Ross:2009ar} a consistent procedure has been developed to find out the stress energy complex when the boundary is degenerate. In this formalism, we first need to write down the boundary fields (metric and other fields) in terms of tangent space indices
\begin{equation}
    \psi^{\alpha\beta \cdots} = e^\alpha_A e^\beta_B \cdots \psi^{AB \cdots}
\end{equation}
and consider the variation of the boundary action with respect to independent $\psi^{AB\cdots}$. In our case, the set of boundary fields consists of a metric $h^{\alpha\beta}$, a massive gauge field $\cA^a$, a massless gauge field $\cC^a$ and a scalar. We first write these fields in terms of their tangent space indices
\begin{eqnarray}
h^{\alpha\beta} = e^{\alpha}_{A}e^{\beta}_{B}\eta^{A B}, \quad \mathcal{A}^{\alpha} = e^{\alpha}_{A}\mathcal{A}^{A},\quad \mathcal{C}^{\alpha} = e^{\alpha}_{A}\mathcal{C}^{A}, \quad \phi = \phi
\end{eqnarray}
and consider the variation of the action with respect to $e^A_\alpha$, $\cA^A$, $\cC^a$ and $\phi$
\begin{eqnarray}
		\delta S = \frac{1}{16 \pi G_{5}}\int d\xi^{4}\lb T^{\alpha}_{\beta} e^{\beta}_{A}\delta e^{A}_{\alpha} + s_{\alpha}e^{\alpha}_{A} \delta \mathcal{A}^{A} + s_{\phi}\delta \phi + \tilde{s}_{\alpha}e^{\alpha}_{A} \delta \mathcal{C}^{A}\rb 
\end{eqnarray}
where,
\begin{eqnarray}
		T^{\alpha}_{\beta}&=& 2 s^{\alpha}_{\beta}-s^{\alpha}\mathcal{A}_{\beta}-\tilde{s}^{\alpha}\mathcal{C}_{\beta}.
\end{eqnarray}
The components of boundary stress energy complex has been identified as,
		\begin{eqnarray}\label{dic}
		\epsilon=T^{+}_{+}, \hspace{0.5cm}\epsilon^{i}=T^{i}_{+},\hspace{0.5cm}\rho=T^{+}_{-},\hspace{0.5cm}\rho^{i}=T^{i}_{-},\hspace{0.5cm}t^{i}_{j}=-T^{i}_{j},\hspace{0.5cm}Q=\tilde{s}^{+},\hspace{0.5cm}\mathcal{J}^{i}=\tilde{s}^{i}
		\end{eqnarray}
This analysis has been done up to first order in derivative expansion for charged non-relativistic fluids \cite{Dutta:2018xtr,Brattan:2010bw}. Our goal is to use this formulation to obtain the second order stress energy complex for charged non-relativistic fluids.

\subsection{Holographic computation of non-relativistic constitutive relations}\label{sec:holographiccomputation}

On the gravity side, the independent quantities are mass $m$, charge $q$, and four velocity $u^{\mu}$ of the boosted black brane. The other quantities temperature $T$ and chemical potential $\phi$ are related to mass and charge by the following relations \cite{Son:2009tf},
\begin{eqnarray}\label{rel}
		m=\frac{\pi^{4}T^{4}}{16}(\gamma+1)^{3}(3\gamma-1), \qquad q = \frac{\phi}{\sqrt{3}}\frac{\pi^{2}T^{2}}{2}(\gamma+1)^{2}
\end{eqnarray}
		where,
\begin{eqnarray}
		\gamma=\sqrt{1+\frac{8\phi^{2}}{3\pi^{2}T^{2}}}.
\end{eqnarray}
The horizon radius is given by,
\begin{eqnarray}
	R = \frac{\pi}{2}T(\gamma+1).
\end{eqnarray}
In all holographic calculations the unperturbed horizon radius has been set to 1. So, at the leading (zeroth) order these relations become,
\begin{eqnarray}
		m_{0} = 1+q_{0}^{2},\hspace{0.5cm}\phi_{0}=\frac{\sqrt{3}q_{0}}{2},\hspace{0.5cm}T_{0}=\frac{(2-q_{0}^{2})}{2\pi},\hspace{0.5cm}\gamma_{0}=\frac{(2+q_{0}^{2})}{(2-q_{0}^{2})}.
\end{eqnarray}

Following Eqn. (\ref{dic}), the zeroth order non-relativistic quantities are given by,
\begin{eqnarray} 
\rho = 4
  m_{0} (u^{+})^{2},\hspace{0.5cm}p=m_{0},\hspace{0.5cm}\epsilon=m_{0},\hspace{0.5cm}Q=4\sqrt{3}q_{0} u^{+}
\end{eqnarray}
and they satisfy the non-relativistic Euler's equation $\epsilon+p-\rho \rho_{m}-\mu \frac{Q}{4\sqrt{3}}=\tau s $ with mass chemical potential $\rho_{m}$ given by (\ref{eq:rhom}).

\subsubsection{Second order calculations}

We expand the relativistic mass, charge and velocities about their values in local rest frame up to second order in derivative expansion
    \begin{eqnarray}
    \begin{split}
		m & = m_{0}+x^{+}\partial_{+}m+x^{i}\partial_{i}m+\frac{1}{2}((x^{+})^{2}\partial_{+}^{2}m+(x^{i})^{2}\partial_{i}^{2}m+2 x^{+}x^{i}\partial_{+}\partial_{i}m) \\
		q & = q_{0}+x^{+}\partial_{+}q+x^{i}\partial_{i}q+\frac{1}{2}((x^{+})^{2}\partial_{+}^{2}q+(x^{i})^{2}\partial_{i}^{2}q+2 x^{+}x^{i}\partial_{+}\partial_{i}q) \\
		u^{+} & = 1+x^{+}\partial_{+}u^{+}+x^{i}\partial_{i}u^{+}+\frac{1}{2}((x^{+})^{2}\partial_{+}^{2}u^{+}+(x^{i})^{2}\partial_{i}^{2}u^{+}+2 x^{+}x^{i}\partial_{+}\partial_{i}u^{+}) \\
		u^{i} & =  x^{+}\partial_{+}u^{i}+x^{j}\partial_{j}u^{i}+\frac{1}{2}((x^{+})^{2}\partial_{+}^{2}u^{i}+(x^{j})^{2}\partial_{j}^{2}u^{i}+2 x^{+}x^{j}\partial_{+}\partial_{j}u^{i}).
  \end{split}
		\end{eqnarray}
Following the holographic procedure described in this subsection and using the dictionary laid down in Eqn.(\ref{dic}) we calculate the components of stress energy complex corrected up to second order in derivative expansion.  In deriving different components of stress energy complex, we use the constraints obtained from Einstein's and Maxwell's equations. In table \ref{tst constraints} we list the independent first order holographic fluid data coming form equations of motion.
\begin{table}[h!]
\centering
    \begin{tabular}{|c|c|c|c|}
    \hline
     & Data & Constraint & Independent data\\
     \hline
        Scalar & $\partial_{+}m$, $\partial_{+}q$, & $g^{rb}E_{b+} = 0 $,  & $\partial_{i}u^{i}$, $\epsilon^{ij}\omega_{ij}$\\
       & $\epsilon^{ij}\omega_{ij}$, $\partial_{+}u^{+}$, $\partial_{i}u^{i}$ & $g^{rb}E_{b-} = 0 $, $g^{rb}M_{b} = 0 $ & \\
        \hline
         Vector & $\partial_{i}m$, $\partial_{i}q$, $\partial_{i}u^{+}$, $\partial_{+}u^{i}$ & $g^{rb}E_{bi} = 0 $ & $\partial_{i}m$, $\partial_{i}q$, $\partial_{i}u^{+}$ \\
& $\epsilon^{ik}\partial_{k}m$,$ \epsilon^{ik}\partial_{k}q$, $\epsilon^{ik}\partial_{k}u^{+}$ &  &$\epsilon^{ik}\partial_{k}T$, $\epsilon^{ik}\partial_{k}q$, $\epsilon^{ik}\partial_{k}u^{+}$ \\
        \hline 
        Tensor & $\sigma^{ij}$ &  & $\sigma^{ij}$ \\
        \hline 
    \end{tabular}
    \caption{Independent first order data from holography}
    \label{tst constraints}
\end{table}
These holographic constraints (table \ref{tst constraints}) are similar to the constraints obtained by LCR of relativistic conserved currents (table \ref{lcr constraint1}). Using the holographic constraints we write our independent data in terms of derivatives of $m$, $q$ and $u^+$. In addition to these, the holographic fluid also satisfies constraints same as given in table \ref{lcr constraint2}, coming from the zeroth order holographic stress tensor and charge current. Using those relations one can obtain the independent second order scalar, vector and tensor data in terms of derivatives of $m$, $q$ and $u^+$.

We further use the holographic relations (\ref{rel}) to trade derivatives of $m$ and $q$ with the derivatives of non-relativistic fluid variables : mass chemical potential $\mu_m$, temperature $\tau$ and chemical potential $\nu$ defined in (\ref{eq:Ttaumuphi}) and (\ref{eq:numumumup}). At first order the relations are given by 
\begin{eqnarray*}
\begin{split}
    \partial_{i}m & =  4 m_{0} \frac{\partial_{i}\tau}{\tau}+6 q_{0}^{2}\frac{\partial_{i}\nu}{\nu}+4 m_{0}(\partial_{i}\mu_{m})\\ 
    \partial_{i}q & = 3 q_{0} \frac{\partial_{i}\tau}{\tau}+\frac{q_{0}(2+5 q_{0}^{2})}{(2+q_{0}^{2})}\frac{\partial_{i}\nu}{\nu}+3 q_{0}(\partial_{i}\mu_{m})
\end{split}
\end{eqnarray*}
and at second order they are given by
\begin{eqnarray*}
\begin{split}
    \partial_{i}\partial_{j}m &= 4 m_{0}( \partial_{i}\partial_{j}\mu_{m})+4 m_{0}\frac{ \partial_{i}\partial_{j}\tau}{\tau}+6 q_{0}^{2}\frac{ \partial_{i}\partial_{j}\nu}{\nu}  +\frac{6 q_{0}^{2}(2+5 q_{0}^{2})}{(2+q_{0}^{2})}\frac{\partial_{i}\nu\partial_{j}\nu}{\nu^{2}}  \\
  &   +12 m_{0}\frac{\partial_{i}\tau\partial_{j}\tau}{\tau^{2}} + 12m_{0}\partial_{i}\mu_{m}\partial_{j}\mu_{m}+\frac{24 q_{0}^{2}}{\nu \tau}((\partial_{i}\nu)(\partial_{j}\tau)+(\partial_{j}\nu)(\partial_{i}\tau)) \\
  &    +\frac{16 m_{0} }{ \tau}((\partial_{i}\mu_{m})(\partial_{j}\tau) + (\partial_{j}\mu_{m})(\partial_{i}\tau))+\frac{24 q_{0}^{2}}{\nu}((\partial_{i}\mu_{m})(\partial_{j}\nu)+(\partial_{j}\mu_{m})(\partial_{i}\nu))
  \end{split}
\end{eqnarray*}
and
  \begin{eqnarray*}
\begin{split}
   \partial_{i}\partial_{j}q &= 3 q_{0}( \partial_{i}\partial_{j}\mu_{m})+3 q_{0}\frac{ \partial_{i}\partial_{j}\tau}{\tau}+\frac{ q_{0}(2+5 q_{0}^{2})}{(2+q_{0}^{2})}\frac{ \partial_{i}\partial_{j}\nu}{\nu}+\frac{4 q_{0}^{3}(12+8 q_{0}^{2}+5 q_{0}^{4})}{(2+q_{0}^{2})^{3}}\frac{\partial_{i}\nu\partial_{j}\nu}{\nu^{2}}  \\ 
  &   + 6 q_{0}\frac{\partial_{i}\tau\partial_{j}\tau}{\tau^{2}} + 6 q_{0}\partial_{i}\mu_{m}\partial_{j}\mu_{m}+\frac{3 q_{0}(2+5 q_{0}^{2})}{(2+q_{0}^{2})\nu \tau}((\partial_{i}\nu)(\partial_{j}\tau)+(\partial_{j}\nu)(\partial_{i}\tau))  \\
  &   +\frac{9 q_{0} }{ \tau}((\partial_{i}\mu_{m})(\partial_{j}\tau) +(\partial_{j}\mu_{m})(\partial_{i}\tau))+\frac{3q_{0}(2+5q_{0}^{2})}{(2+q_{0}^{2})\nu}((\partial_{i}\mu_{m})(\partial_{j}\nu)+(\partial_{j}\mu_{m})(\partial_{i}\nu)).
    \end{split}
\end{eqnarray*}
Finally we write the holographic constitutive relations in terms of independent non-relativistic fluid data give in table \ref{lcr constraint1} and \ref{lcr constraint2}.

After a dedicated computation we obtain the expressions for different holographic constitutive relations and transports. Here we present the expressions for the stress tensor, charge density and charge current. Other non-relativistic quantities (mass density, energy density, pressure, velocity,energy current) can be found in appendix \ref{charge data}. The stress tensor is given by,
\begin{eqnarray}
\label{eq:nonrelst2ndTsT}
\begin{split}
    t^{i j} & = \rho v^{i} v^{j} + p \delta^{ij}  - n_{1}\tilde{\sigma}^{i j} + n_{a+1} \partial^{\langle i}\partial^{j\rangle} X_a + c_{1} \tilde{\sigma}^{\langle i k}\tilde{\omega}^{k j\rangle}  + \mathbf{c}_{ab} \partial^{\langle i} X_a \partial^{j\rangle} X_b \\ 
    & \quad + g_{a}\epsilon^{\langle i k}\partial^{k}\partial^{j \rangle} X_a + \mathbf{g}_{ab} \epsilon^{\langle i k} \partial_{k} X_a \partial^{j\rangle} X_b+g_{10} \epsilon^{\langle i k}\tilde{\sigma}^{k l}\tilde{\omega}^{l j\rangle}
\end{split}
\end{eqnarray}
where $\hat {\mathbf{c}}_{ab}$ and $\hat {\mathbf{g}}_{ab}$ are holographic counterpart of $\mathbf{c}_{ab}$ and $\mathbf{g}_{ab}$ defined in (\ref{eq:cglcr}). Three sets of holographic transport coefficients are given in table \ref{eq:hatnset}, \ref{eq:hatcset} and \ref{eq:hatgset}.
\begin{table}[h!]
\centering
\begin{tabular}{ | c | c |}
\hline
 $\hat n_{1}$ & $R^{3} u^{+}$  \\
 \hline
 $\hat n_{2}$ & $-\frac{m_{0} \log (\frac{3-\sqrt{4 m_{0}-3}}{\sqrt{4 m_{0}-3}+3})}{2 \sqrt{4 m_{0}-3}}-1$ \\ 
 \hline
 $\hat n_{3}$ & $\frac{-13 m_{0}^3+10 m_{0}^2-3 m_{0}+6}{8 \nu  m_{0}^3+8 \nu  m_{0}^2}-\frac{3 q_{0}^{2} \log (\frac{3-\sqrt{4 m_{0}-3}}{\sqrt{4 m_{0}-3}+3})}{4 \nu  \sqrt{4 m_{0}-3}}$ \\
 \hline
 $\hat n_{4}$ & $\frac{1-4 m_{0}}{4 m_{0} \tau }-\frac{m_{0} \log (\frac{3-\sqrt{4 m_{0}-3}}{\sqrt{4 m_{0}-3}+3})}{2 \sqrt{4 m_{0}-3} \tau }$ \\
\hline
\end{tabular}
\caption{List of $\hat n_i$}
\label{eq:hatnset}
\end{table}
\begin{table}[h!]
\centering
\begin{tabular}{ | c | c |}
\hline
 $\hat c_{1} $ & $-\frac{m_{0} }{4 \sqrt{4 m_{0}-3}}\log (\frac{3-\sqrt{4 m_{0}-3}}{\sqrt{4 m_{0}-3}+3})$\\
 \hline
 $\hat c_{2}$ & $-\frac{24\kappa^{2} q_{0}^{4}}{m_{0}}+\frac{3 m_{0} \log (\frac{3-\sqrt{4 m_{0}-3}}{\sqrt{4 m_{0}-3}+3})}{2 \sqrt{4 m_{0}-3}}+2m_{0}+1$\\
 \hline
$\hat c_{3}$ & $-\frac{36 \kappa^{2} (m_{0}-1)^3}{\nu  m_{0}^2}+\frac{3 (m_{0}-1) \log (\frac{3-\sqrt{4 m_{0}-3}}{\sqrt{4 m_{0}-3}+3})}{2 \nu  \sqrt{4 m_{0}-3}}+\frac{(m_{0}-1) (12 m_{0}^3+19 m_{0}^2-3 m_{0}+6)}{4 \nu  m_{0}^2 (m_{0}+1)}$\\
\hline
$\hat c_{4}$ & $-\frac{24\kappa^{2} q_{0}^{4}}{m_{0} \tau }+\frac{m_{0} \log (\frac{3-\sqrt{4 m_{0}-3}}{\sqrt{4 m_{0}-3}+3})}{\sqrt{4 m_{0}-3} \tau }+\frac{4 m_{0}^2+2 m_{0}-1}{2 m_{0} \tau }$ \\
\hline
$\hat c_{5}$ & $-\frac{27\kappa^{2} (m_{0}-1)^4}{2 \nu ^2 m_{0}^3}+\frac{3 (m_{0}-1) (m_{0}+3) (7 m_{0}-9) \log (\frac{3-\sqrt{4 m_{0}-3}}{\sqrt{4 m_{0}-3}+3})}{16 \nu ^2 m_{0} (m_{0}+1) \sqrt{4 m_{0}-3}}$\\&$+\frac{3 (m_{0}-3)^4 (m_{0}-1) \log (2)}{16 \nu ^2 m_{0}^2 (m_{0}+1)^2}+\frac{(m_{0}-1) (34 m_{0}^6+210 m_{0}^5+245 m_{0}^4-502 m_{0}^3+390 m_{0}^2-396 m_{0}-333)}{32 \nu ^2 m_{0}^3 (m_{0}+1)^3}$ \\
\hline
$\hat c_{6}$ & $ -\frac{18 \kappa ^2 (m_{0}-1)^3}{\nu  m_{0}^2 \tau }+\frac{3 (m_{0}-1) \log (\frac{3-\sqrt{4 m_{0}-3}}{\sqrt{4 m_{0}-3}+3})}{4 \nu  \sqrt{4 m_{0}-3} \tau }-\frac{-3 m_{0}^2-5 m_{0}+8}{2 \nu  \tau +2 \nu  m_{0} \tau }$ \\
\hline
$\hat c_{7}$ & $ -\frac{6 \kappa ^2 q_{0}^{4}}{m_{0} \tau ^2}+\frac{3 m_{0} \log (\frac{3-\sqrt{4 m_{0}-3}}{\sqrt{4 m_{0}-3}+3})}{4 \sqrt{4 m_{0}-3} \tau ^2}+\frac{4 m_{0}^2+14 m_{0}-5}{8 m_{0} \tau ^2}$ \\
\hline
\end{tabular}
\caption{List of $\hat c_i$}
\label{eq:hatcset}
\end{table}

\begin{table}[h!]
\centering
    \begin{tabular}{|c|c|}
    \hline
         $\hat g_1$ & $-\frac{2 \sqrt{3} \kappa  q_{0}^3}{m_{0}}$\\
         \hline
         $\hat g_2$ & $-\frac{3 \sqrt{3} \kappa  q_{0}^5}{2 \nu  m_{0}^2}$\\
         \hline
         $\hat g_3$ & $-\frac{\sqrt{3} \kappa  q_{0}^3}{m_{0} \tau }$\\
         \hline
         $\hat g_4$ & $\frac{6 \sqrt{3} \kappa  q_{0}^3}{m_{0}}$\\
         \hline
         $\hat g_5$ & $\frac{9 \sqrt{3} \kappa  q_{0}^5}{2 \nu  m_{0}^2}$\\
         \hline
         \end{tabular} \quad
         \begin{tabular}{|c|c|}
    \hline
         $\hat g_6$ & $\frac{3 \sqrt{3} \kappa  q_{0}^3}{m_{0} \tau }$\\
         \hline
         $\hat g_7$ & $\frac{3 \sqrt{3} \kappa  q_{0}^5 (9 q_{0}^2 (q_{0}^2+2)-m_{0} (5 q_{0}^2+2))}{2 \nu ^2 m_{0}^3 (q_{0}^2+2)}$\\
         \hline
         $\hat g_8$ & $\frac{3 \sqrt{3} \kappa  q_{0}^5}{\nu  m_{0}^2 \tau }$\\
         \hline
         $\hat g_9$ & $\frac{3 \sqrt{3} \kappa  q_{0}^3}{m_{0} \tau ^2}$\\
         \hline
         $\hat g_{10}$ & $\frac{\sqrt{3} \kappa  q_{0}^3}{2m_{0}}$\\
         \hline
         \end{tabular}
         \caption{List of $\hat g_i$}
         \label{eq:hatgset}
         \end{table}

Non-relativistic charge density $Q$ is given by,
\begin{eqnarray}
    \begin{split}
        Q & = 4\sqrt{3}q u^{+}-\frac{6 q^{2}u^{+}\kappa}{m}\epsilon^{ij}\tilde{\omega_{ij}}-\frac{\sqrt{3} (m_{0}-2)q_{0}}{4 m_{0}^2}\tilde{\sigma}_{ij}\tilde{\sigma}^{ij}+\frac{\sqrt{3}q_{0}^{3}\kappa^{2}}{2m_{0}^{2}}\tilde{\omega}_{ij}\tilde{\omega}^{ij} \\
        & \quad +\hat{n}^{Q}_{a}(\partial^{i}\partial_{i}X_{a})+ {\hat \lambda^Q}_{ab} \partial^{i}X_a \partial_{i} X_b  + \hat {\tilde{{\lambda}}}^Q_{ab} \epsilon^{ij} \partial_{i}X_a \partial_{j} X_b 
    \end{split}
\end{eqnarray}
where,
\begin{equation}
  \hat \lambda^Q =   
  \begin{pmatrix}
\hat \lambda_{1} & \hat \lambda_4/2 & \hat \lambda_6/2\\
\hat \lambda_4/2 & \hat \lambda_2 & \hat \lambda_5/2 \\
\hat \lambda_6/2 & \hat \lambda_5/2 & \hat \lambda_3
\end{pmatrix}, \quad 
\hat {\tilde{{\lambda}}}^Q = \begin{pmatrix}
0 & \hat {\tilde \lambda}_{4}/2 & \hat {\tilde \lambda}_{6}/2\\
\hat {\tilde \lambda}_{4}/2 & 0 & \hat {\tilde \lambda}_{5}/2 \\
\hat {\tilde \lambda}_{6}/2 & \hat {\tilde \lambda}_{5}/2 & 0
\end{pmatrix}.
\end{equation}
Different $\hat{n}^{Q}_i$s,and $\hat \lambda_i$s are given in table \ref{eq:nQ set} and table \ref{eq:hatlambdaset}.
\begin{table}[h!]
    \centering
    \begin{tabular}{|c|c|}
    \hline
       $\hat{n}^{Q}_1$  & -$\frac{2 \sqrt{3} \kappa ^2 q_{0}^3}{m_{0}^2}$ \\
       \hline
       $\hat{n}^{Q}_2$  & $-\frac{\sqrt{3} q_{0} (m_{0}^3+4 m_{0}^2-15 m_{0}+18)}{8 \nu  m_{0}^3}-\frac{3 \sqrt{3} \kappa ^2 q_{0}^{5}}{2 \nu  m_{0}^3}$ \\
       \hline
       $\hat{n}^{Q}_3$ & $-\frac{3 \sqrt{3} q_{0}^{3}}{4 m_{0}^2 \tau }-\frac{\sqrt{3} \kappa ^2 q_{0}^{3}}{m_{0}^2 \tau }$\\
       \hline
    \end{tabular}
    \caption{List of $\hat{n}^{Q}_i$}
    \label{eq:nQ set}
\end{table}
\begin{table}[h!]
\centering
\begin{tabular}{ | c | c |}
\hline
$\hat \lambda_{1}$ & $\frac{2 \sqrt{3} \kappa ^2 q_{0}^{3}}{m_{0}^2}$\\
\hline
$\hat \lambda_{2}$ & $-\frac{3 \sqrt{3} q_{0}^{5}\kappa^{2} (5 m_{0}^2-39 m_{0}+60)}{2 \nu ^2 m_{0}^4 (m_{0}+1)}-\frac{3  \sqrt{3} (m_{0}-3)^2 q_{0}^{3}\log2}{4 \nu ^2 m_{0}^2 (m_{0}+1)}-\frac{\sqrt{3} q_{0}^{3} (m_{0}^4+28 m_{0}^3+24 m_{0}^2+108 m_{0}-369)}{16 \nu ^2 m_{0}^4 (m_{0}+1)}$ \\
\hline
$\hat \lambda_{3}$ & $\frac{\sqrt{3} (5-4 m_{0}) q_{0}}{2 m_{0}^2 \tau ^2}+\frac{\sqrt{3} \kappa ^2 (m_{0}-1)^{3/2}}{m_{0}^2 \tau ^2}$ \\
\hline
$\hat \lambda_{4}$ & $-\frac{\sqrt{3} q_{0}^{3} (5 m_{0}-3)}{\nu  m_{0}^2 (m_{0}+1)}-\frac{12 \sqrt{3} \kappa ^2 (m_{0}-3)^2 q_{0}^{3}}{\nu  m_{0}^3 (m_{0}+1)}$ \\
\hline
 $\hat \lambda_{5}$ & $-\frac{6 \kappa^{2}\sqrt{3} (m_{0}-3)^2 q_{0}^{3}}{\nu  m_{0}^3 (m_{0}+1) \tau }-\frac{\sqrt{3} (m_{0}-3)^2 q_{0}\log2}{2 \nu  m_{0} (m_{0}+1) \tau }-\frac{\sqrt{3} q_{0} (m_{0}^4+55 m_{0}^3-67 m_{0}^2-15 m_{0}+42)}{8 \nu  m_{0}^3 (m_{0}+1) \tau }$\\
 \hline
 $\hat \lambda_{6}$ & $-\frac{3 \sqrt{3} q_{0}^{3}}{m_{0}^2 \tau }$\\
 \hline
$\hat {\tilde \lambda}_{4}$ & $\frac{3 \kappa  q_{0}^{2} (2 m_{0}^2-9 m_{0}+15)}{\nu m_{0}^3}$\\
\hline
$\hat {\tilde \lambda}_{5}$ & $-\frac{3 \kappa  (m_{0}-3) q_{0}^{2} (2 m_{0}^2-7 m_{0}-3)}{2 \nu  m_{0}^3 (m_{0}+1) \tau }$ \\
\hline
$\hat {\tilde \lambda}_{6}$ & $\frac{6 \kappa  q_{0}^{2}}{\tau m_{0}^2}$\\
\hline
\end{tabular}
\caption{List of $\hat \lambda_i$}
\label{eq:hatlambdaset}
\end{table}

Non-relativistic charge current can be written as,
\begin{eqnarray}\label{eq:nonrelJiTsT}
    \begin{split}
       \mathcal{J}^{i} & =  Q v^{i} + \hat{\mathbf{q}}_a \partial^{i} X_a + \hat{\tilde{\mathbf{q}}}_a \epsilon^{ik}\partial_{k}X_a +\hat{n}_{1}^{\cJ}\partial^{k}\tilde{\sigma}^{ik}+\hat{n}_{2}^{\cJ}\partial^{k}\tilde{\omega}^{ik}+\hat{n}_{3}^{\cJ}\epsilon^{ij}\partial^{k}\tilde{\sigma}^{jk}+\hat{n}_{4}^{\cJ}\epsilon^{ij}\partial^{k}\tilde{\omega}^{jk}\\& + \hat{\mathbf{c}}^{\cJ}_a \tilde{\sigma}^{ik} \partial_{k} X_a + \hat{\tilde{\mathbf{c}}}^{\cJ}_a \tilde{\omega}^{ik} \partial_{k} X_a + \hat{\bar{\mathbf{c}}}^{\cJ}_a (\partial_k v^k) \partial^{i} X_a + \hat{\mathbf{g}}^{\cJ}_a \epsilon^{il}\tilde{\sigma}^{lk}\partial_{k} X_a\\
       &  + \hat{\tilde{\mathbf{g}}}^{\cJ}_a \epsilon^{il}\tilde{\omega}^{lk}\partial_{k} X_a + \hat{\bar{\mathbf{g}}}^{\cJ}_a (\partial_k v^k) \epsilon^{il} \partial_{l} X_a.
    \end{split}
\end{eqnarray}
Again all the hatted transports are the holographic counterparts of the non-relativistic transports defined in (\ref{eq:non-relarray}). Holographic values of different charge transports are given in table \ref{eq:hatnJset}, \ref{eq:hatqJset}, \ref{eq:hatcJset} and \ref{eq:hatgJset}.
\begin{table}[h!]
\centering
    \begin{tabular}{|c|c|}
    \hline
      $\hat{n}_{1}^{\cJ}$   &  $\Big(\frac{q_{0}\sqrt{3}}{2\sqrt{4m_{0}-3}}\log\big(\frac{3+\sqrt{4m_{0}-3}}{3-\sqrt{4m_{0}-3}}\big)+\frac{\sqrt{3}q_{0}(m_{0}-2)}{2m_{0}^{2}}\Big)$\\
      \hline
       $\hat{n}_{2}^{\cJ}$  & $\Big(-\frac{\sqrt{3}q_{0}(m_{0}+2)}{4m_{0}^{2}}+\frac{\sqrt{3} \kappa ^2 q_{0}^3}{m_{0}^2}-\frac{q_{0}\sqrt{3}}{2\sqrt{4m_{0}-3}}\log\big(\frac{3+\sqrt{4m_{0}-3}}{3-\sqrt{4m_{0}-3}}\big)\Big)$\\
       \hline
       $\hat{n}_{3}^{\cJ}$ & $\frac{6q_{0}^{2}\kappa}{m_{0}^{2}}$\\
       \hline
       $\hat{n}_{4}^{\cJ}$ &$-\frac{3(m_{0}-2)q_{0}^{2}\kappa}{m_{0}^{2}}$\\
       \hline
    \end{tabular}
\caption{List of $\hat n^{\cJ}_i$.}
    \label{eq:hatnJset}
\end{table}
\begin{table}[h!]
\centering
    \begin{tabular}{|c|c|}
    \hline
       $\hat \lambda_q$  &  $\frac{\sqrt{3}q R^{3}}{m u^{+}}+\frac{2 q\tau\sqrt{3}}{m R}\Big(\frac{(2 R^{4}+3m)}{T(3\gamma-1)}-\frac{(m+R^{4})}{\gamma T}\Big)+\frac{q\nu \tau\sqrt{3}}{m R}\Big(\frac{\sqrt{3}q(2 R^{4}+3m)}{m}-\frac{(m+R^{4})(3\gamma-2)}{\gamma \phi}\Big)$\\
       \hline
        $\hat \sigma_q $
         & $\frac{q\tau u^{+}\sqrt{3}}{m R}\Big(\frac{\sqrt{3}q(2 R^{4}+3m)}{m}-\frac{(m+R^{4})(3\gamma-2)}{\gamma \phi}\Big)$\\
         \hline
         $\kappa_q$ & $\frac{2 q u^{+}\sqrt{3}}{m R}\Big(\frac{(2 R^{4}+3m)}{T(3\gamma-1)}-\frac{(m+R^{4})}{\gamma T}\Big)+\frac{q\nu u^{+}\sqrt{3}}{m R}\Big(\frac{\sqrt{3}q(2 R^{4}+3m)}{m}-\frac{(m+R^{4})(3\gamma-2)}{\gamma \phi}\Big)$\\
         \hline
         $\hat{\tilde{ \lambda}}_q$ & $-\frac{12\kappa q^{2}}{m}\Big(1+\frac{2u^{+}}{3\gamma-1}+\frac{\sqrt{3}q\nu \tau}{m}\Big)$\\
         \hline 
         $\hat{\tilde{\sigma}}_q$ & $-\frac{12\kappa q^{2}}{m}\Big(\tau u^{+}\frac{\sqrt{3}q}{m}\Big)$\\
         \hline
         $\hat{\tilde{\kappa}}_q$ & $-\frac{12\kappa q^{2}}{m}\Big(\frac{2}{(3\gamma-1)\tau}+\frac{\sqrt{3}q\nu u^{+}}{m}\Big)$\\
         \hline
         \end{tabular}
         \caption{List of $\hat {\mathbf{q}}^{\cJ}_a$ and $\hat {\tilde{ \mathbf{q}}}^{\cJ}_a$.}
    \label{eq:hatqJset}
\end{table}
\begin{table}[h!]
\centering
    \begin{tabular}{|c|c|}
    \hline
         $\hat c_1^{\cJ}$ & $\frac{2 \sqrt{3} q_{0}^{3}}{m_{0}^2}-\frac{2 \sqrt{3} \kappa ^2 q_0^3}{m_0^2}$ \\
         \hline
         $\hat c_2^{\cJ}$ & $\frac{2 \sqrt{3} q_0^3}{m_0^{2}}-\frac{2 \sqrt{3} \kappa ^2 (12 m_0-13) q_0^3}{m_0^2}$\\
         \hline
         $\hat c_3^{\cJ}$ & $0$\\
         \hline
         $\hat c_4^{\cJ}$ & $\frac{\sqrt{3} (5 m_0^2+24 m_0-9) q_0^3}{8 \nu  m_0^3 (m_0+1)}+\frac{2\sqrt{3}q_{0}^{3}}{4 m_{0}\nu\sqrt{4m_{0}-3}}\log(\frac{3+\sqrt{4m_{0}-3}}{3-\sqrt{4m_{0}-3}})-\frac{3 \sqrt{3} \kappa ^2 q_0^5}{2 \nu  m_0^3}$\\
         \hline
         $\hat c_5^{\cJ}$ & $\frac{\sqrt{3} (11 m_0^4-4 m_0^3-23 m_0^2+12 m_0-12) q_0}{8 \nu  m_0^3 (m_0+1)}-\frac{3 \sqrt{3} \kappa ^2 q_{0}^{3} (12 m_0^3-17 m_0^2+12 m_0-23)}{2 \nu  m_0^3 (m_0+1)}$\\
         \hline
         $\hat c_6^{\cJ}$ & $0$\\
         \hline
         $\hat c_7^{\cJ}$ & $\frac{\sqrt{3} (5 m_0-6) q_0}{4 m_0^2 \tau }+\frac{\sqrt{3}q_{0}}{2\tau\sqrt{4m_{0}-3}}\log(\frac{3+\sqrt{4m_{0}-3}}{3-\sqrt{4m_{0}-3}})-\frac{\sqrt{3} \kappa ^2 q_0^3}{m_0^2 \tau }$\\
         \hline
         $\hat c_8^{\cJ}$ & $\frac{\sqrt{3} (4 m_0^2-3 m_0-2) q_0}{4 m_0^2 \tau }-\frac{\sqrt{3} \kappa ^2 q_{0}^{3} (12 m_0-13)}{m_0^2 \tau }$\\
         \hline
         $\hat c_9^{\cJ}$ & $0$\\
         \hline
    \end{tabular}
    \caption{List of $\hat {\mathbf{c}}^{\cJ}_a$, $\hat {\tilde{ \mathbf{c}}}^{\cJ}_a$ and $\hat {\bar{ \mathbf{c}}}^{\cJ}_a$.}
    \label{eq:hatcJset}
    \end{table}
\begin{table}[h!]
\centering
\begin{tabular}{|c|c|}
    \hline
         $\hat g_1^{\cJ}$ & $0$\\
         \hline
         $\hat g_2^{\cJ}$ & $0$\\
         \hline
         $\hat g_3^{\cJ}$ & $0$\\
         \hline
         \end{tabular} \quad
         \begin{tabular}{|c|c|}
    \hline
         $\hat g_4^{\cJ}$ & $\frac{3 q_{0}^{2}\kappa}{m_{0}^{2}\tau}$\\
         \hline
         $\hat g_5^{\cJ}$ & $\frac{6q_{0}^{2}\kappa}{m_{0}\tau}$\\
         \hline
         $\hat g_6^{\cJ}$ & $0$\\
         \hline
         \end{tabular} \quad
         \begin{tabular}{|c|c|}
    \hline
         $\hat g_7^{\cJ}$ & $\frac{9 q_{0}^{6}\kappa}{m_{0}^{3}(m_{0}+1)\nu}$\\
         \hline
         $\hat g_8^{\cJ}$ & $\frac{9q_{0}^{4}\kappa}{m_{0}^{2}\nu}$\\
         \hline
         $\hat g_9^{\cJ}$ & $0$\\
         \hline
    \end{tabular}
    \caption{List of $\hat {\mathbf{g}}^{\cJ}_a$, $\hat {\tilde{ \mathbf{g}}}^{\cJ}_a$ and $\hat {\bar{ \mathbf{g}}}^{\cJ}_a$.}
    \label{eq:hatgJset}
\end{table}

\section{Discussion : Comparison between LCR and TsT} \label{dis}

In order to compare the two approaches to obtain the non-relativistic constitutive relations we first note that the stress tensor (\ref{eq:nonrelst2nd}) and charge current (\ref{eq:nonrelJi}) obtained via LCR have exactly same form as stress tensor (\ref{eq:nonrelst2ndTsT}) and charge current (\ref{eq:nonrelJiTsT}) obtained via TsT transformations. Not only the stress tensor and charge current, one can check that other constitutive relations like mass density, charge density, energy density, energy current etc. have the same structure of terms. The expressions can be found in appendix \ref{lcrgen} and \ref{charge data}. The transport coefficients appearing in light-cone reduced stress tensor and charge current depend on relativistic data (i.e. relativistic transports and fluid variables). The same is true for other light-cone reduced constitutive relations also. These relations are very generic and depend on the parent relativistic system (\ref{eq:st2ndorder}) and (\ref{eq:Jrel}) and do not depend on any holographic model. In appendix \ref{lcrgen} we have listed all these transports. However, if we use the holographic values of the relativistic transports (\ref{eq:reltransports}) then it turns out that the values of the transports appearing in light-cone reduced constitutive relations exactly match those obtained via TsT. For example let us look at the holographic (TsT) value of $\hat{n}_{3}$ in table \ref{eq:hatnset}. The value of $n_{3}$ for light-cone reduced fluid depends on the relativistic transports and other fluid variables (see eqn. (\ref{eq:app-n}) and (\ref{eq:app-nt})). However, if one uses the holographic values of relativistic transports (\ref{eq:reltransports}) then it turns out that $\hat{n}_{3} = n_{3}$. The same is true for other transports. Thus we find that the non-relativistic fluid obtained by LCR or by TsT transformation are identical order by order in derivative expansion. Motivated by this we speculate that the observation is true for any order in derivative expansions.  

The matching of TsT and LCR results, we think, is non-trivial for the following reason. It was studied in \cite{Maldacena:2008wh,Adams:2008wt} that the TsT transformation of asymptotically AdS space generates a new solution which has Schr\"odinger isometry at the boundary. On the other hand Schr\"odinger in the boundary theory can be obtained by LCR of conformal algebra. In this paper we tried to understand this connection in the context of hydrodynamics. LCR reduction of relativistic constitutive equations (in particular conservation equations) renders non-relativistic constitutive relations, namely continuity equation, Euler equations, energy current equations and charge conservation equation at different orders of derivative expansion. These equations are consistent with Schr\"odinger isometry. Different non-relativistic fluid data i.e. mass density, charge density, fluid velocity, energy current, energy density and different transports can be obtained in terms of relativistic fluid data and transports. Obtaining the non-relativistic fluid via TsT transformation is rather non-trivial. First of all we need to obtain the correct holographic dual of a relativistic fluid. We then uplift the solution to 10 dimensions and perform the TsT transformation after identifying two isometry directions. The transformation mixes different components of the metric non trivially. In order to find the boundary currents we first reduced the TsT transformed 10 dimensional solution to 5 dimensions and wrote an effective action (following \cite{Ross:2009ar}). The boundary currents are obtained from the variation of this effective action following the dictionary \cite{Hollands:2005ya,Ross:2009ar,Dutta:2018xtr}. Since the variation is done with respect to the tangent space variables of the asymptotic fields, there is a further mixing in different components of boundary currents. Therefore, it is not very easy to see that the non-relativistic constitutive relations obtained in this way match with those obtained via LCR. After very dedicated calculations we see that the two sides in deed match at every order of derivative expansion exactly.

\section*{Acknowledgement}
We would like to that N. Banerjee, N. Ganguli, P. Ramadevi and A. Rudra for useful discussion. The work of TM is supported by a Simons Foundation Grant Award ID 509116 and by the South African Research Chairs initiative of the Department of Science and Technology and the National Research Foundation. The work of SD is supported by the MATRICS (grant no. MTR/2019/ 000390, the Department of Science and Technology, Government of India).  Finally, we are indebted to people of India for their unconditional support toward the researches in basic science.

\begin{appendices}
\section{Relativistic bulk metric}\label{bulkg}
Here we have written down the leading order pieces of the components of bulk charged relativistic metric which appear in the second order correction.

Scalars:
\begin{eqnarray}
\begin{split}
h(r)&=-\frac{1}{12 r^{2}}\Big(2(\partial_{i}u^{+})^{2}+2(\partial_{+}u^{i})^{2}+2(\partial_{x}u^{y})^{2}+2(\partial_{y}u^{x})^{2}+(\partial_{x}u^{x}-\partial_{y}u^{y})^{2}\Big)+\mathcal{O}\big(\frac{1}{r^{5}}\big)\\
k(r)&=\frac{r^{2}}{6m_{0}}\Big(2\partial_{+}^{2}m+2\partial^{2}_{i}m+6m_{0}(\partial_{i}u^{+})^{2}+16m_{0}(\partial_{i}u^{+})(\partial_{+}u^{i})-22m_{0}(\partial_{+}u^{i})^{2}\\&-11m_{0}(\partial_{+}u^{x})^{2}-3m_{0}(\partial_{x}u^{x})^{2}-3m_{0}(\partial_{y}u^{y})^{2}+6m_{0}(\partial_{y}u^{x})^{2}+16 m_{0}(\partial_{y}u^{x})(\partial_{x}u^{y})\\&-34m_{0}(\partial_{x}u^{x})(\partial_{y}u^{y})\Big)+\mathcal{O}\big(\frac{1}{r}\big)\\
 w(r)&=\frac{1}{24m_{0}^{2}r}\Big((1+m_{0})(-3q_{0}\partial_{+}^{2}m-3q_{0}\partial_{i}^{2}m-+4m_{0}\partial_{+}^{2}q+4m_{0}\partial_{i}^{2}q+20m_{0}(\partial_{+}q)(\partial_{+}u^{+})+\\&20m_{0}(\partial_{i}q)(\partial_{+}u^{i})+72m_{0}q_{0}(\partial_{+}u^{+})^{2}+72m_{0}q_{0}(\partial_{+}u^{i})^{2}+4q_{0}m_{0}((\partial_{i}q)^{2}+(\partial_{+}q)^{2}+9 q_{0}^{2}(\partial_{+}u^{+})^{2}\\&+9 q_{0}^{2}(\partial_{+}u^{i})^{2}+6 q_{0}(\partial_{+}q)(\partial_{+}u^{+})+6 q_{0}(\partial_{i}q)(\partial_{+}u^{i}))-32\sqrt{3}q_{0}m_{0}\kappa\epsilon^{ij}((\partial_{i}q)(\partial_{+}u_{j})\\&-(\partial_{i}q)(\partial_{j}u^{+})-3q_{0}(\partial_{+}u^{+})(\partial_{i}u_{j})+3q_{0}(\partial_{i}u^{+})(\partial_{j}u^{+})-(\partial_{+}q)(\partial_{i}u_{j}))\Big)+\mathcal{O}\big(\frac{1}{r^{2}}\big)
\end{split}
\end{eqnarray}
Vectors:
\begin{eqnarray}
j_{\mu}(r)&=&-\frac{1}{12 r^{2}}(P^{\nu}_{\mu}\mathcal{D}_{\lambda}\sigma^{\lambda}_{\nu}+P^{\nu}_{\mu}\mathcal{D}_{\lambda}\omega^{\lambda}_{\nu})+\mathcal{O}\big(\frac{1}{r^{5}}\big)\nonumber\\
g_{\mu}(r)&=&-\frac{2}{\sqrt{3}q_{0}r^{7}}\Big(\frac{3q^{3}\kappa}{2m^{2}}(l^{\lambda}\sigma_{\lambda\mu})-\frac{3\sqrt{3}q^{2}}{8m_{0}}P^{\nu}_{\mu}\mathcal{D}_{\lambda}\sigma^{\lambda}_{\nu}-\frac{\sqrt{3}q^{4}\kappa^{2}}{4m^{2}}P^{\nu}_{\mu}\mathcal{D}_{\lambda}\omega^{\lambda}_{\nu}\nonumber\\&&-\frac{\sqrt{3}q\log2}{4}\sigma_{\mu}^{\lambda}\mathcal{D}_{\lambda}q+\frac{\sqrt{3}q_{0}}{16m_{0}^{2}}(m_{0}^{2}-48 q_{0}^{2}\kappa^{2}+3)\omega_{\mu}^{\lambda}\mathcal{D}_{\lambda}q\Big)
    \end{eqnarray}

    Tensors:
    \begin{eqnarray}
    \begin{split}
    \alpha_{\mu\nu}&=\frac{1}{r^{2}}\Big(\omega_{\mu\lambda}\sigma^{\lambda}_{\nu}+\omega_{\nu\lambda}\sigma^{\lambda}_{\mu}+\sigma_{\mu\lambda}\sigma^{\lambda}_{\nu}-\frac{P_{\mu\nu}}{3}\sigma^{\alpha\beta}\sigma_{\alpha\beta}-\omega_{\mu\lambda}\omega^{\lambda}_{\nu}-\frac{P_{\mu\nu}}{3}\omega^{\alpha\beta}\omega_{\alpha\beta}\Big)\\&+\frac{1}{4 r^{4}}\Big(\mathcal{N}_{1} (u^{\lambda}\mathcal{D}_{\lambda}\sigma^{\mu\nu})+\mathcal{N}_{2}(\omega^{\mu}\hspace{0.2pt}_{\lambda}\sigma^{\lambda\nu}+\omega^{\nu}\hspace{0.2pt}_{\lambda}\sigma^{\lambda\mu})+\mathcal{N}_{3}(\sigma^{\mu}\hspace{0.2pt}_{\lambda}\sigma^{\lambda\nu}-\frac{1}{3}P^{\mu\nu}\sigma^{\alpha\beta}\sigma_{\alpha\beta})\\&+\mathcal{N}_{4}(\omega^{\mu}\hspace{0.2pt}_{\lambda}\omega^{\lambda\nu}+\frac{1}{3}P^{\mu\nu}\omega^{\alpha\beta}\omega_{\alpha\beta})+\mathcal{N}_{5}(\Pi^{\mu\nu\alpha\beta}\mathcal{D}_{\alpha}\mathcal{D}_{\beta}n)+\mathcal{N}_{6}(\Pi^{\mu\nu\alpha\beta}\mathcal{D}_{\alpha}n\mathcal{D}_{\beta}n)\\&+\mathcal{N}_{7}\big(\Pi^{\mu\nu\alpha\beta}(\mathcal{D}_{\alpha}l_{\beta}+\mathcal{D}_{\beta}l_{\alpha})-\epsilon_{\alpha\beta\gamma\mu}u^{\alpha}a^{\beta}(\partial_{\nu}u^{\gamma})-\epsilon_{\alpha\beta\gamma\nu}u^{\alpha}a^{\beta}(\partial_{\mu}u^{\gamma})+\frac{2}{3}(l^{\alpha}a_{\alpha})P_{\mu\nu}\big)\Big)
       \end{split}
    \end{eqnarray}
Here $i,j$ runs over $(x,y)$  and $\mu,\nu$ over $(+,-,x,y)$.
\section{Non relativistic charged fluid from LCR}\label{lcrgen}
    The transport coefficient involved in stress tensor are given by,
\begin{eqnarray}\label{eq:app-n}
\begin{split}
n_{1} & = \eta_{r}u^{+},\ \
n_{2} =\nu _0 \tilde{n}_3 \tau _0+\tilde{n}_4 \tau _0+\tilde{n}_2, \ \
n_{3} =  \tilde{n}_3 \tau _0, \ \
n_{4} = \nu _0 \tilde{n}_3 + \tilde{n}_4, \\
c_{1} & = \frac{\mathcal{N}_{2}}{2},  \ \
c_{2} = \tilde{c}_5 \nu _0^2 \tau _0^2+ \tilde{c}_6 \nu _0 \tau _0^2+ \tilde{c}_3 \nu _0 \tau _0+\tilde{c}_7 \tau _0^2+ \tilde{c}_4 \tau _0+\tilde{c}_2,  \\
c_{3} &=2\tilde{c}_5 \nu _0 \tau _0^2+\tilde{c}_6 \tau _0^2+\tilde{c}_3 \tau _0+2\tilde{n}_3 \tau _0, \\
c_{4} &= 2 \tilde{c}_5 \nu _0^2 \tau _0+ 2\tilde{c}_6 \nu _0 \tau _0+\tilde{c}_3 \nu _0+2\tilde{c}_7 \tau _0+\tilde{c}_4+2\nu _0 \tilde{n}_3+2\tilde{n}_4, \ \
c_{5} = \tilde{c}_5 \tau _0^2, \\
c_{6} & =2 \tilde{c}_5 \nu _0 \tau _0+\tilde{c}_6 \tau _0+2\tilde{n}_3,  \ \
c_{7} = \tilde{c}_5 \nu _0^2+ \tilde{c}_6 \nu _0+\tilde{c}_7 \\
g_{1} & =\tau _0 (\tilde{g}_2 \nu _0+\tilde{g}_3)+\tilde{g}_1, \ \
g_{2} = \tilde{g}_2 \tau _0,  \ \
g_{3} = \tilde{g}_2 \nu _0+\tilde{g}_3,  \\
g_{4} & = \tau _0 (\tilde{g}_7 \nu _0^2\tau _0+2 \tilde{g}_8 \nu _0 \tau _0+2 \tilde{g}_5 \nu _0+\tilde{g}_9 \tau _0+2\tilde{g}_6)+\tilde{g}_4, \ \
g_{5} = \tau _0 (\tilde{g}_8  \tau _0+\tilde{g}_7 \nu \tau+\tilde{g}_2+\tilde{g}_5), \\
g_{6} &= \tilde{g}_7 \nu _0^2 \tau _0+2 \tilde{g}_8 \nu _0 \tau _0+\tilde{g}_2 \nu _0+\tilde{g}_5 \nu _0+\tilde{g}_9 \tau _0+\tilde{g}_3+\tilde{g}_6, \ \
g_{7} = \tilde{g}_7 \tau _0^2,\ \
g_{8} =\tilde{g}_8  \tau _0+\tilde{g}_7 \tau \nu +\tilde{g}_2,\\ 
g_{9}& =\tilde{g}_8 \nu _0 +\tilde{g}_7 \nu_{0}^{2}+\tilde{g}_9,\ \ g_{10}= -\frac{\mathcal{N}_{7}}{2}
\end{split}
\end{eqnarray}
where,
\begin{eqnarray}\label{eq:app-nt}
\begin{split}
    \tilde{n}_{2} &= -\frac{\eta_{r}{}^2}{4 p_0}, \ \
    \tilde{n}_{3} =-\frac{1}{16 p_0^2 q_0}\Big(2 p_0 (\mathcal{N}_{1} q_0+3 \mathcal{N}_{5} q_0) (\partial_{\phi}p)-8 \mathcal{N}_{5} p_0^2 (\partial_{\phi}q)-q_0 (\partial_{\phi}p) \eta_{r}{}^2\Big), \\
   \tilde{n}_4 &=  -\frac{1}{16 p_0^2 q_0}\Big(2 p_0 (\mathcal{N}_{1} q_0+3 \mathcal{N}_{5} q_0) (\partial_{T}p)-8 \mathcal{N}_{5} p_0^2 (\partial_{T}q)-q_0 (\partial_{T}p) \eta_{r}{}^2\Big)
   \end{split}
 \end{eqnarray}
 \begin{eqnarray}
 \begin{split}
   \tilde{c}_{2} &=-\frac{1}{8 p_0}\Big((4 \mathcal{N}_{2}-\mathcal{N}_{3}+4 \mathcal{N}_{4}) p_0-3 \eta_{r}{}^2\Big),  \\  
  \tilde{c}_{3} & =\frac{1}{16 p_0^2}\Big(p_0 ((2 \mathcal{N}_{1}-\mathcal{N}_{3}-4 \mathcal{N}_{4}) (\partial_{\phi}p)-4 \eta_{r}
  (\partial_{\phi}\eta_{r}))+4 (\partial_{\phi}p) \eta_{r}{}^2\Big), \\  
  \tilde{c}_{4} & =\frac{1}{16 p_0^2}\Big(p_0 ((2 \mathcal{N}_{1}-\mathcal{N}_{3}-4 \mathcal{N}_{4}) (\partial_{T}p)-4 \eta_{r}
  (\partial_{T}\eta_{r}))+4 (\partial_{T}p) \eta_{r}{}^2\Big), \\
  \tilde{c}_{5} & = -\frac{1}{128 p_0^3 q_0^2}\Big(16 p_0^2 (2 (4 \mathcal{N}_{5} q_0+3 \mathcal{N}_{6} q_0) (\partial_{\phi}p) (\partial_{\phi}q)+q_0 (\mathcal{N}_{1} q_0+3 \mathcal{N}_{5} q_0)(\partial_{\phi}^{2}p))\\&-p_0 (((20 \mathcal{N}_{1}+4 \mathcal{N}_{2}+\mathcal{N}_{3}-4 \mathcal{N}_{4}) q_0^2+108 \mathcal{N}_{5} q_0^2+36 \mathcal{N}_{6} q_0^2) (\partial_{\phi}p){}^2+8 q_0^2
   (\partial_{\phi}p) \eta_{r}(\partial_{\phi}\eta_{r})\\&+8 q_0^2(\partial_{\phi}^{2}p) \eta_{r}{}^2)-64 p_0^3
   (\mathcal{N}_{6} (\partial_{\phi}q){}^2+\mathcal{N}_{5} q_0(\partial_{\phi}^{2}q))+17 q_0^2 (\partial_{\phi}p){}^2 \eta_{r}{}^2\Big),\\
   \tilde{c}_{6}& = -\frac{1}{64 p_0^3 q_0^2}\Big(16 p_0^2 ((4 \mathcal{N}_{5} q_0+3 \mathcal{N}_{6} q_0) (\partial_{T}p) (\partial_{\phi}q)+(4 \mathcal{N}_{5} q_0+3 \mathcal{N}_{6} q_0) (\partial_{\phi}p) (\partial_{T}q)\\&+q_0
   (\mathcal{N}_{1} q_0+3 \mathcal{N}_{5} q_0) (\partial_{\phi}\partial_{T}p))-p_0 ((\partial_{\phi}p) (((20 \mathcal{N}_{1}+4 \mathcal{N}_{2}+\mathcal{N}_{3}-4 \mathcal{N}_{4}) q_0^2\\&+108 \mathcal{N}_{5} q_0^2+36 \mathcal{N}_{6} q_0^2) (\partial_{T}p)+4 q_0^2 \eta_{r}(\partial_{T}\eta_{r}))+4
   q_0^2 \eta_{r} ((\partial_{T}p)(\partial_{\phi}\eta_{r})+2 (\partial_{\phi}\partial_{T}p) \eta_{r}))\\&-64 p_0^3 (\mathcal{N}_{6} (\partial_{\phi}q) (\partial_{T}q)+\mathcal{N}_{5} q_0
   (\partial_{\phi}\partial_{T}q))+17 q_0^2 (\partial_{\phi}p) (\partial_{T}p)\eta_{r}{}^2\Big),\\
   \tilde{c}_{7} & = -\frac{1}{128 p_0^3 q_0^2}\Big(16 p_0^2 (2 (4 \mathcal{N}_{5} q_0+3 \mathcal{N}_{6} q_0) (\partial_{T}p) (\partial_{T}q)+q_0 (\mathcal{N}_{1} q_0+3 \mathcal{N}_{5} q_0)(\partial_{T}^{2}p))\\&-p_0 (((20 \mathcal{N}_{1}+4
   \mathcal{N}_{2}+\mathcal{N}_{3}-4 \mathcal{N}_{4}) q_0^2+108 \mathcal{N}_{5} q_0^2+36 \mathcal{N}_{6} q_0^2) (\partial_{T}p){}^2+8 q_0^2 (\partial_{T}p) \eta_{r}(\partial_{T}\eta_{r})\\&+8 q_0^2(\partial_{T}^{2}p) \eta_{r}{}^2)-64 p_0^3 (\mathcal{N}_{6} (\partial_{T}q){}^2+\mathcal{N}_{5} q_0(\partial_{T}^{2}q))+17 q_0^2 (\partial_{T}p){}^2 \eta_{r}{}^2\Big)\\
   \tilde{g}_{1} & =- \mathcal{N}_{7}, \ \
    \tilde{g}_{2} = -\frac{1}{4 p_0}(\mathcal{N}_{7} (\partial_{\phi}p)), \ \
    \tilde{g}_{3}  = -\frac{1}{4 p_0}(\mathcal{N}_{7} (\partial_{T}p)) \ \
    \tilde{g}_{4} = \mathcal{N}_{7}\\
    \tilde{g}_{5} & =\frac{1}{32 p_0 q_0}\Big((8 \mathcal{N}_{7} q_0+(6 \mathcal{N}_{8}-3 \mathcal{N}_{9}) q_0) (\partial_{\phi}p)+4 (\mathcal{N}_{9}-2 \mathcal{N}_{8}) p_0 (\partial_{\phi}q)\Big)\\
    \tilde{g}_{6} & = \frac{1}{32 p_0 q_0}\Big((8 \mathcal{N}_{7} q_0+(6 \mathcal{N}_{8}-3 \mathcal{N}_{9}) q_0) (\partial_{T}p)+4 (\mathcal{N}_{9}-2 \mathcal{N}_{8}) p_0
   (\partial_{T}q)\Big)\\
    \tilde{g}_{7} & = -\frac{1}{64 p_0^2 q_0}\Big(4 (2 \mathcal{N}_{8}+\mathcal{N}_{9}) p_0 (\partial_{\phi}p) (\partial_{\phi}q)-3 (8 \mathcal{N}_{7}
   q_0+(2 \mathcal{N}_{8}+\mathcal{N}_{9}) q_0) (\partial_{\phi}p){}^2+16 \mathcal{N}_{7} p_0 q_0 (\partial_{\phi}^{2}p)\Big)\\
    \tilde{g}_{8} & = -\frac{1}{64 p_0^2 q_0}\Big((\partial_{\phi}p) (2 (2 \mathcal{N}_{8}+\mathcal{N}_{9}) p_0 (\partial_{T}q)-3 (8
   \mathcal{N}_{7} q_0+(2 \mathcal{N}_{8}+\mathcal{N}_{9}) q_0) (\partial_{T}p))\\&+2 p_0 ((2 \mathcal{N}_{8}+\mathcal{N}_{9})
   (\partial_{T}p) (\partial_{\phi}q)+8 \mathcal{N}_{7} q_0 (\partial_{\phi}\partial_{T}p))\Big)\\
    \tilde{g}_{9} & = -\frac{1}{64 p_0^2 q_0}\Big(4 (2 \mathcal{N}_{8}+\mathcal{N}_{9}) p_0 (\partial_{T}p) (\partial_{T}q)-3 (8 \mathcal{N}_{7}
   q_0+(2 \mathcal{N}_{8}+\mathcal{N}_{9}) q_0) (\partial_{T}p){}^2+16 \mathcal{N}_{7} p_0 q_0 (\partial_{T}^{2}p)\Big)\\
    \end{split}
\end{eqnarray}

The transports appearing in the charge current are given by,
\begin{eqnarray}
\begin{split}
    \lambda_{q} & =\eta_{r}\Big(\frac{\sqrt{3}q }{P u^{+}}-\frac{\sqrt{3}q T u^{+}(\partial_{T}p)}{4 P^{2}}-\frac{3q^{2}(\partial_{\phi}p)}{8 u^{+} P^{2}}\Big) \\
    & \quad + \lambda_{r} \Big(\frac{\sqrt{3}(3q\partial_{T}p-4P\partial_{T}q)T u^{+}}{P}+\frac{3q(3q \partial_{\phi}p-4P\partial_{\phi}q)}{2u^{+}P}\Big) \\
     \kappa_{q} & =\frac{\eta_{r}}{\tau}\Big(-\frac{\sqrt{3}q T (u^{+})^{2}(\partial_{T}p)}{4 P^{2}}-\frac{3q^{2}(\partial_{\phi}p)}{8 u^{+} P^{2}}\Big) \\ 
     & \quad +\frac{\lambda_{r}}{\tau}\Big(\frac{\sqrt{3}(3q\partial_{T}p-4P\partial_{T}q) T(u^{+})^{2}}{P}+\frac{3q(3q \partial_{\phi}p-4P\partial_{\phi}q)}{2u^{+}P}\Big) \\
     \sigma_{q} & =\eta_{r}\Big(-\frac{\sqrt{3}q(\partial_{\phi}p)T}{4 P^{2}}\Big)+\lambda_{r}\Big(\frac{\sqrt{3}(3q \partial_{\phi}p-4P\partial_{\phi}q)T}{P}\Big),\\
     \tilde{\lambda}_{q} & = -\xi_{r}\Big(\frac{1}{ u^{+}}+\frac{(\partial_{T}p)T u^{+}}{4 P}+\frac{(\sqrt{3}q\partial_{\phi}p)}{8 u^{+} P}\Big)\\
     \tilde{\kappa}_{q} & =-\frac{\xi_{r}}{\tau}\Big(\frac{(\partial_{T}p)T(u^{+})^{2}}{4 P}+\frac{\sqrt{3}q(\partial_{\phi}p)}{8 u^{+} P}\Big),\ \
     \tilde{\sigma}_{q}=-\xi_{r}\Big(\frac{(\partial_{\phi}p)T}{4 P}\Big)\\
     \end{split}
\end{eqnarray}

\begin{eqnarray}
\begin{split}
    c^{\mathcal{J}}_{1}&=\tilde{c}^{\mathcal{J}}_{1}+\tau u^{+}\tilde{c}^{\mathcal{J}}_{4}+\tau\tilde{c}^{\mathcal{J}}_{7},\ \
    c^{\mathcal{J}}_{2}=\tilde{c}^{\mathcal{J}}_{2}+\tau u^{+}\tilde{c}^{\mathcal{J}}_{5}+\tau\tilde{c}^{\mathcal{J}}_{8},\ \
    c^{\mathcal{J}}_{3}=\tilde{c}^{\mathcal{J}}_{3}+\tau u^{+}\tilde{c}^{\mathcal{J}}_{6}+\tau\tilde{c}^{\mathcal{J}}_{9}\\
    c^{\mathcal{J}}_{4}&=\nu u^{+}\tilde{c}^{\mathcal{J}}_{4}+\tau\tilde{c}^{\mathcal{J}}_{7},\ \
    c^{\mathcal{J}}_{5}=\nu u^{+}\tilde{c}^{\mathcal{J}}_{5}+\tau\tilde{c}^{\mathcal{J}}_{8},\ \
    c^{\mathcal{J}}_{6}=\nu u^{+}\tilde{c}^{\mathcal{J}}_{6}+\tau\tilde{c}^{\mathcal{J}}_{9}\\
    c^{\mathcal{J}}_{7}&=u^{+}\tau\tilde{c}^{\mathcal{J}}_{7},\ \
    c^{\mathcal{J}}_{8}=u^{+}\tau\tilde{c}^{\mathcal{J}}_{8},\ \
    c^{\mathcal{J}}_{9}=u^{+}\tau\tilde{c}^{\mathcal{J}}_{9}\\
    g^{\mathcal{J}}_{1}&=\tilde{g}^{\mathcal{J}}_{1}+\tau u^{+}\tilde{g}^{\mathcal{J}}_{4}+\tau\tilde{g}^{\mathcal{J}}_{7},\ \
    g^{\mathcal{J}}_{2}=\tilde{g}^{\mathcal{J}}_{2}+\tau u^{+}\tilde{g}^{\mathcal{J}}_{5}+\tau\tilde{g}^{\mathcal{J}}_{8},\\
    g^{\mathcal{J}}_{3}&=\tilde{g}^{\mathcal{J}}_{3}+\tau u^{+}\tilde{g}^{\mathcal{J}}_{6}+\tau\tilde{g}^{\mathcal{J}}_{9},\ \
    g^{\mathcal{J}}_{4}=\nu u^{+}\tilde{g}^{\mathcal{J}}_{4}+\tau\tilde{g}^{\mathcal{J}}_{7},\ \
    g^{\mathcal{J}}_{5}=\nu u^{+}\tilde{g}^{\mathcal{J}}_{5}+\tau\tilde{g}^{\mathcal{J}}_{8},\\
    g^{\mathcal{J}}_{6}&=\nu u^{+}\tilde{g}^{\mathcal{J}}_{6}+\tau\tilde{g}^{\mathcal{J}}_{9}\ \
    g^{\mathcal{J}}_{7}=u^{+}\tau\tilde{g}^{\mathcal{J}}_{7},\ \
    g^{\mathcal{J}}_{8}=u^{+}\tau\tilde{g}^{\mathcal{J}}_{8},\ \
    g^{\mathcal{J}}_{9}=u^{+}\tau\tilde{g}^{\mathcal{J}}_{9}
    \end{split}
\end{eqnarray}
where,
\begin{eqnarray}
\begin{split}
{n}^{\mathcal{J}}_{1}&=\frac{-\sqrt{3} p_0 q_0 (\mathcal{N}_{1}+12 \eta _r \lambda _r)+2 \gamma _1 p_0^2+\sqrt{3} q_0 \eta _r^2}{2 p_0^2}\\  
 {n}^{\mathcal{J}}_{2}&=\frac{-2 \sqrt{3} p_0 q_0 (\mathcal{N}_{1}+6 \eta _r \lambda _r)+2 (\gamma _1+\gamma _2) p_0^2+\sqrt{3} q_0 \eta _r^2}{4 p_0^2}\\
 {n}^{\mathcal{J}}_{3}&=\frac{\eta_{r}\xi_{r}}{2 p_{0}},\ \
 {n}^{\mathcal{J}}_{4}=\frac{\eta_{r}\xi_{r}-4\sqrt{3}q \mathcal{N}_{7}}{4 p_{0}}\\
 \tilde{c}^{\mathcal{J}}_{1}&=\frac{\sqrt{3} p_0 q_0 (\mathcal{N}_{1}-2 \mathcal{N}_{2}-\mathcal{N}_{3}-12 \eta _r \lambda _r)+2 (\gamma _1-\gamma _2) p_0^2+\sqrt{3} q_0 \eta _r^2}{4 p_0^2}\\
 \tilde{c}^{\mathcal{J}}_{2}&=\frac{\sqrt{3} p_0 q_0 (\mathcal{N}_{1}+2 \mathcal{N}_{2}-4 \mathcal{N}_{4}+12 \eta _r \lambda _r)-2 (\gamma _1-\gamma _2) p_0^2-\sqrt{3} q_0 \eta _r^2}{4 p_0^2}\\
 \tilde{c}^{\mathcal{J}}_{3}&=\frac{1}{8 p_0^2 ((\partial_{T}p) (\partial_{\phi}q)-(\partial_{\phi}p)(\partial_{T}q))}\Big(\sqrt{3} q_0 (\eta _r-12 p_0 \lambda _r) (3 \eta _r ((\partial_{T}p) (\partial_{\phi}q)-(\partial_{\phi}p)
   (\partial_{T}q))\\&+q_0 (3 (\partial_{\phi}p) (\partial_{T}\eta_{r})-3 (\partial_{T}p) (\partial_{\phi}\eta_{r}))+4 p_0 ((\partial_{T}q) (\partial_{\phi}\eta_{r})-(\partial_{\phi}q) (\partial_{T}\eta_{r})))\Big)
  \end{split}
\end{eqnarray}
\begin{eqnarray}
\begin{split}
\tilde{c}^{\mathcal{J}}_{4}&=\frac{1}{16 p_0^2 q_0}\Big(p_0 (8 \sqrt{3} q_0 (\mathcal{N}_{5} (\partial_{\phi}q)-6 q_0 \lambda _r (\partial_{\phi}\eta_{r}))+(6 \gamma _1 q_0-2 \gamma _2 q_0-6 \gamma _4 q_0) (\partial_{\phi}p))\\&+\sqrt{3} q_0 (((-3 \mathcal{N}_{1}-2 \mathcal{N}_{2}+\mathcal{N}_{3}) q_0-6 \mathcal{N}_{5} q_0) (\partial_{\phi}p)+4 q_0 \eta _r (\partial_{\phi}\eta_{r}))+8 \gamma _4 p_0^2 (\partial_{\phi}q)\Big)\\
 \tilde{c}^{\mathcal{J}}_{5}&=\frac{1}{16 p_0^3 q_0}\Big(2 p_0^2 (4 \sqrt{3} \mathcal{N}_{5} q_0 (\partial_{\phi}q)+((\gamma _1+\gamma _2) q_0-3 \gamma _5 q_0) (\partial_{\phi}p))\\&-\sqrt{3}p_0 q_0 (\partial_{\phi}p) (q_0 (\mathcal{N}_{1}+2 \mathcal{N}_{2}+4 \mathcal{N}_{4}+12 \eta _r \lambda _r)+6 \mathcal{N}_{5} q_0)+\sqrt{3} q_0^2 \eta _r^2 (\partial_{\phi}p)+8 \gamma_5 p_0^3 (\partial_{\phi}q)\Big)\\
 \tilde{c}^{\mathcal{J}}_{6}&=-\frac{1}{32 p_0^3 ((\partial_{T}p) (\partial_{\phi}q)-(\partial_{\phi}p(\partial_{T}q))}\Big(\sqrt{3} q_0 (\partial_{\phi}p) (\eta _r-12 p_0 \lambda _r) (3 \eta _r ((\partial_{T}p) (\partial_{\phi}q)-(\partial_{\phi}p) (\partial_{T}q))\\&+q_0 (3 (\partial_{\phi}p) (\partial_{T}\eta_{r})-3
   (\partial_{T}p) (\partial_{\phi}\eta_{r}))+4 p_0 ((\partial_{T}q) (\partial_{\phi}\eta_{r})-(\partial_{\phi}q) (\partial_{T}\eta_{r})))\Big)\\
 \tilde{c}^{\mathcal{J}}_{7}&=\frac{1}{16 p_0^2 q_0}\Big(p_0 (8 \sqrt{3} q_0 (\mathcal{N}_{5} (\partial_{T}q)-6 q_0 \lambda _r (\partial_{T}\eta_{r}))+(6 \gamma _1 q_0-2 \gamma _2 q_0-6\gamma _4 q_0) (\partial_{T}p))\\&+\sqrt{3} q_0 (((-3 \mathcal{N}_{1}-2 \mathcal{N}_{2}+\mathcal{N}_{3}) q_0-6 \mathcal{N}_{5} q_0) (\partial_{T}p)+q_0 \eta _r(\partial_{T}\eta_{r}))+8 \gamma _4 p_0^2 (\partial_{T}q)\Big)\\
 \tilde{c}^{\mathcal{J}}_{8}&=\frac{1}{16 p_0^3 q_0}\Big(2 p_0^2 (4 \sqrt{3} \mathcal{N}_{5} q_0 (\partial_{T}q)+((\gamma _1+\gamma _2) q_0-3 \gamma _5 q_0) (\partial_{T}p))\\&-\sqrt{3}p_0 q_0 (\partial_{T}p) (q_0 (\mathcal{N}_{1}+2 \mathcal{N}_{2}+4 \mathcal{N}_{4}+12 \eta _r \lambda _r)+6 \mathcal{N}_{5} q_0)+\sqrt{3} q_0^2 \eta _r^2 (\partial_{T}p)+8 \gamma_5 p_0^3 (\partial_{T}q)\Big)\\
   \tilde{c}^{\mathcal{J}}_{9}&=\frac{1}{32 p_0^3 ((\partial_{T}p) (\partial_{\phi}q)-(\partial_{\phi}p) (\partial_{T}q))}\Big(\sqrt{3} q_0 (\partial_{T}p) (\eta _r-12 p_0 \lambda _r) (\eta _r (3 (\partial_{\phi}p) (\partial_{T}q)-3(\partial_{T}p) (\partial_{\phi}q))\\&+3 q_0 ((\partial_{T}p) (\partial_{\phi}\eta_{r})-(\partial_{\phi}p) (\partial_{T}\eta_{r}))+4 p_0 ((\partial_{\phi}q) (\partial_{T}\eta_{r})-(\partial_{T}q) (\partial_{\phi}\eta_{r})))\Big)\\
 \tilde{g}^{\mathcal{J}}_{1}&=\frac{1}{4} (2 \gamma _3+\frac{\eta _r \xi _r}{p_0}),\ \
 \tilde{g}^{\mathcal{J}}_{2}=\frac{\gamma _3}{2}-\frac{2 \sqrt{3} \mathcal{N}_{7} q_0}{p_0}\\
 \tilde{g}^{\mathcal{J}}_{3}&=\frac{1}{8p_0 ((\partial_{T}p) (\partial_{\phi}q)-(\partial_{\phi}p) (\partial_{T}q))}\Big(\xi _r (3 \eta _r ((\partial_{T}p) (\partial_{\phi}q)-(\partial_{\phi}p) (\partial_{T}q))+q_0 (3(\partial_{\phi}p) (\partial_{T}\eta_{r})\\&-3 (\partial_{T}p) (\partial_{\phi}\eta_{r}))+4 p_0((\partial_{T}q) (\partial_{\phi}\eta_{r})-(\partial_{\phi}q) (\partial_{T}\eta_{r})))\Big)\\
 \tilde{g}^{\mathcal{J}}_{4}&=\frac{1}{16 p_0^2 q_0}\Big(2 p_0 (-2 \sqrt{3} \mathcal{N}_{9} q_0 (\partial_{\phi}q)+\gamma _3 q_0 (\partial_{\phi}p)+2 q_0 \xi _r (\partial_{\phi}\eta_{r}))+3\sqrt{3} \mathcal{N}_{9} q_0^2 (\partial_{\phi}p)\Big)\\
 \tilde{g}^{\mathcal{J}}_{5}&=\frac{1}{8 p_0^2 q_0}\Big(\sqrt{3} q_0 (4 \mathcal{N}_{7} q_0+3 \mathcal{N}_{8} q_0) (\partial_{\phi}p)-p_0 (4 \sqrt{3} \mathcal{N}_{8} q_0 (\partial_{\phi}q)+\gamma _3 q_0 (\partial_{\phi}p))\Big)\\
 \tilde{g}^{\mathcal{J}}_{6}&=\frac{1}{32 p_0^2 ((\partial_{T}p) (\partial_{\phi}q)-(\partial_{\phi}p)(\partial_{T}q))}\Big(\xi _r (\partial_{\phi}p) (\eta _r (3 (\partial_{\phi}p) (\partial_{T}q)-3 (\partial_{T}p)
   (\partial_{\phi}q))+3 q_0 ((\partial_{T}p) (\partial_{\phi}\eta_{r})\\&-(\partial_{\phi}p) (\partial_{T}\eta_{r}))+4 p_0 ((\partial_{\phi}q) (\partial_{T}\eta_{r})-(\partial_{T}q) (\partial_{\phi}\eta_{r})))\Big)\\
 \tilde{g}^{\mathcal{J}}_{7}&=\frac{1}{16 p_0^2 q_0}\Big(2 p_0 (-2 \sqrt{3} \mathcal{N}_{9} q_0 (\partial_{T}q)+\gamma _3 q_0 (\partial_{T}p)+2 q_0 \xi _r (\partial_{T}\eta_{r}))+3\sqrt{3} \mathcal{N}_{9} q_0^2 (\partial_{T}p)\Big)\\
 \tilde{g}^{\mathcal{J}}_{8}&=\frac{1}{8 p_0^2 q_0}\Big(\sqrt{3} q_0 (4 \mathcal{N}_{7} q_0+3 \mathcal{N}_{8} q_0) (\partial_{T}p)-p_0 (4 \sqrt{3} \mathcal{N}_{8} q_0 (\partial_{T}q)+\gamma _3 q_0 (\partial_{T}p))\Big)\nonumber
 \end{split}
\end{eqnarray}
\begin{eqnarray}
    \begin{split}
        \tilde{g}^{\mathcal{J}}_{9}&=\frac{1}{32 p_0^2 ((\partial_{T}p) (\partial_{\phi}q)-(\partial_{\phi}p)(\partial_{T}q))}\Big(\xi _r (\partial_{T}p) (\eta _r (3 (\partial_{\phi}p) (\partial_{T}q)-3 (\partial_{T}p)(\partial_{\phi}q))+3 q_0 ((\partial_{T}p) (\partial_{\phi}\eta_{r})\\&-(\partial_{\phi}p) (\partial_{T}\eta_{r}))+4 p_0 ((\partial_{\phi}q) (\partial_{T}\eta_{r})-(\partial_{T}q) (\partial_{\phi}\eta_{r})))\Big)\nonumber
    \end{split}
\end{eqnarray}

    Density,
    \begin{eqnarray}
    \begin{split}
    \rho =&(E+P)(u^{+})^{2}+\tilde{n}^{(\rho)}_{1}(\partial_{k}\partial_{k}\mu_{m})+\tilde{n}^{(\rho)}_{2}(\partial_{k}\partial_{k}\phi)+\tilde{n}^{(\rho)}_{1}(\partial_{k}\partial_{k}T)+\tilde{c}^{(\rho)}_{1}(\partial_{k}\mu_{m})(\partial^{k}\mu_{m})\\&+\tilde{c}^{(\rho)}_{2}(\partial_{k}\phi)(\partial^{k}\phi)+\tilde{c}^{(\rho)}_{3}(\partial_{k}T)(\partial^{k}T)+\tilde{c}^{(\rho)}_{4}(\partial_{k}\mu_{m})(\partial^{k}\phi)+\tilde{c}^{(\rho)}_{5}(\partial_{k}T)(\partial^{k}\phi)\\&+\tilde{c}^{(\rho)}_{6}(\partial_{k}\mu_{m})(\partial^{k}T)+\tilde{c}^{(\rho)}_{7}\sigma^{ij}\sigma_{ij}+\tilde{c}^{(\rho)}_{8}\omega^{ij}\omega_{ij}+\tilde{g}^{(\rho)}_{1}\epsilon^{ij}(\partial_{i}\mu_{m})(\partial_{j}\phi)\\&+\tilde{g}^{(\rho)}_{2}\epsilon^{ij}(\partial_{i}\mu_{m})(\partial_{j}T)+\tilde{g}^{(\rho)}_{3}\epsilon^{ij}(\partial_{i}T)(\partial_{j}\phi)
    \end{split}
\end{eqnarray}
where,
\begin{eqnarray}
\begin{split}
\tilde{n}^{(\rho)}_{1}&=-\frac{\eta _r^2}{2 p_0},\ \
\tilde{n}^{(\rho)}_{2}=\frac{1}{24 p_0^2 q_0}\Big((6 \mathcal{N}_{5} p_0 q_0 (\partial_{\phi}p)+3 q_0 \eta _r^2 (\partial_{\phi}p)-8 \mathcal{N}_{5} p_0^2 (\partial_{\phi}q))\Big)\\
\tilde{n}^{(\rho)}_{3}&=\frac{1}{24 p_0^2 q_0}\Big((6 \mathcal{N}_{5} p_0 q_0 (\partial_{T}p)+3 q_0 \eta _r^2 (\partial_{T}p)-8 \mathcal{N}_{5} p_0^2 (\partial_{T}q))\Big)\\
\tilde{c}^{(\rho)}_{1}&=\frac{1}{12 p_0}\Big(((12 \mathcal{N}_{2}+\mathcal{N}_{3}-4 \mathcal{N}_{4}) p_0+8 \eta _r^2)\Big)\\
\tilde{c}^{(\rho)}_{2}&=-\frac{1}{192 p_0^3 q_0^2}\Big((16 q_0^2 \eta _r^2 (\partial_{\phi}p){}^2-16 p_0^2 (2 (5 \mathcal{N}_{5} q_0+3 \mathcal{N}_{6} q_0) (\partial_{\phi}p) (\partial_{\phi}q)+3 \mathcal{N}_{5} q_0^2 (\partial_{\phi}^{2}p))\\&+p_0 (((12 \mathcal{N}_{1}+12 \mathcal{N}_{2}-\mathcal{N}_{3}+4 \mathcal{N}_{4}) q_0^2+132 \mathcal{N}_{5} q_0^2+36 \mathcal{N}_{6} q_0^2) (\partial_{\phi}p){}^2\\&-24 q_0^2 \eta _r (\partial_{\phi}p) (\partial_{\phi}\eta_{r})-24 q_0^2 \eta _r^2 (\partial_{\phi}^{2}p))+64 p_0^3(\mathcal{N}_{6} (\partial_{\phi}q){}^2+\mathcal{N}_{5} q_0 (\partial_{\phi}^{2}q)))\Big)\\
\tilde{c}^{(\rho)}_{3}&=-\frac{1}{192 p_0^3 q_0^2}\Big((16 q_0^2 \eta _r^2 (\partial_{T}p){}^2-16 p_0^2 (2 (5 \mathcal{N}_{5} q_0+3 \mathcal{N}_{6} q_0) (\partial_{T}p) (\partial_{T}q)+3 \mathcal{N}_{5} q_0^2 (\partial_{T}^{2}p))\\&+p_0 (((12 \mathcal{N}_{1}+12 \mathcal{N}_{2}-\mathcal{N}_{3}+4 \mathcal{N}_{4}) q_0^2+132 \mathcal{N}_{5} q_0^2+36 \mathcal{N}_{6} q_0^2) (\partial_{T}p){}^2\\&-24 q_0^2 \eta _r (\partial_{T}p) (\partial_{T}\eta_{r})-24 q_0^2 \eta _r^2 (\partial_{T}^{2}p))+64 p_0^3(\mathcal{N}_{6} (\partial_{T}q){}^2+\mathcal{N}_{5} q_0 (\partial_{T}^{2}q)))\Big)\\
\tilde{c}^{(\rho)}_{4}&=-\frac{1}{24 p_0^2}\Big((5 \eta _r^2 (\partial_{\phi}p)+p_0 ((-6 \mathcal{N}_{1}+\mathcal{N}_{3}+4 \mathcal{N}_{4}) (\partial_{\phi}p)+12 \eta _r (\partial_{\phi}\eta_{r})))\Big)\\
\tilde{c}^{(\rho)}_{5}&=\frac{1}{96 p_0^3 q_0^2}\Big((-16 q_0^2 \eta _r^2 (\partial_{\phi}p) (\partial_{T}p)+16 p_0^2 ((5 \mathcal{N}_{5} q_0+3 \mathcal{N}_{6} q_0) (\partial_{\phi}q) (\partial_{T}p)\\&+(5 \mathcal{N}_{5} q_0+3 \mathcal{N}_{6} q_0) (\partial_{\phi}p) (\partial_{T}q)+3 \mathcal{N}_{5} q_0^2 (\partial_{\phi}\partial_{T}p))\\&+p_0 (-(\partial_{\phi}p) (((12 \mathcal{N}_{1}+12 \mathcal{N}_{2}-\mathcal{N}_{3}+4 \mathcal{N}_{4}) q_0^2+132 \mathcal{N}_{5} q_0^2+36 \mathcal{N}_{6} q_0^2) (\partial_{T}p)\\&-12q_0^2 \eta _r (\partial_{T}\eta_{r}))+12 q_0^2 \eta _r ((\partial_{\phi}\eta_{r}) (\partial_{T}p)+2 \eta _r (\partial_{T}\partial_{\phi}p)))-64 p_0^3 (\mathcal{N}_{6} (\partial_{\phi}q) (\partial_{T}q)+\mathcal{N}_{5} q_0 (\partial_{\phi}\partial_{T}q)))\Big)\\
\tilde{c}^{(\rho)}_{6}&=-\frac{1}{24 p_0^2}\Big((5 \eta _r^2 (\partial_{T}p)+p_0 ((-6 \mathcal{N}_{1}+\mathcal{N}_{3}+4 \mathcal{N}_{4}) (\partial_{T}p)+12 \eta _r (\partial_{T}\eta_{r})))\Big)\\
\tilde{c}^{(\rho)}_{7}&=-\frac{p_0 \mathcal{N}_{3} -4 \eta _r^2}{12 p_0},\ \
\tilde{c}^{(\rho)}_{8}=\frac{\mathcal{N}_{4}}{3}
\end{split}
\end{eqnarray}
\begin{eqnarray}
\begin{split}
\tilde{g}^{(\rho)}_{1}&=\frac{1}{24 p_0 q_0}\Big((2 \mathcal{N}_{8}+3 \mathcal{N}_{9}) (3 q_0 (\partial_{\phi}p)-4 p_0 (\partial_{\phi}q))\Big)\\
\tilde{g}^{(\rho)}_{2}&=\frac{1}{24 p_0 q_0}\Big((2 \mathcal{N}_{8}+3 \mathcal{N}_{9}) (3 q_0 (\partial_{T}p)-4 p_0 (\partial_{T}q))\Big)\\
\tilde{g}^{(\rho)}_{3}&=-\frac{1}{24 p_0 q_0}\Big((2 \mathcal{N}_{8}-3 \mathcal{N}_{9}) ((\partial_{\phi}q) (\partial_{T}p)-(\partial_{\phi}p) (\partial_{T}q))\Big)\nonumber
\end{split}
\end{eqnarray}
Velocity,
\begin{eqnarray}
\begin{split}
v^{i}&=\frac{u^{i}}{u^{+}}+\frac{\eta_{r}}{16 P^{2}u^{+}}\Big(\partial_{i}P-4 P\frac{\partial_{i}u^{+}}{u^{+}}\Big)+\tilde{n}^{(v)}_{1}\partial_{k}\sigma^{ik}+\tilde{n}^{(v)}_{2}\partial_{k}\omega^{ik}+\tilde{n}^{(v)}_{3}\epsilon^{il}\partial_{k}\sigma^{lk}+\tilde{n}^{(v)}_{4}\epsilon^{il}\partial_{k}\omega^{lk}\\&+\tilde{c}^{(v)}_{1}\sigma^{ik}(\partial_{k}\mu_{m})+\tilde{c}^{(v)}_{2}\omega^{ik}(\partial_{k}\mu_{m})+\tilde{c}^{(v)}_{3}(\partial_{k}v^{k})(\partial_{i}\mu_{m})+\tilde{c}^{(v)}_{4}\sigma^{ik}(\partial_{k}\phi)+\tilde{c}^{(v)}_{5}\omega^{ik}(\partial_{k}\phi)\\&+\tilde{c}^{(v)}_{6}(\partial_{k}v^{k})(\partial_{i}\phi)+\tilde{c}^{(v)}_{7}\sigma^{ik}(\partial_{k}T)+\tilde{c}^{(v)}_{8}\omega^{ik}(\partial_{k}T)+\tilde{c}^{(v)}_{9}(\partial_{k}v^{k})(\partial^{i}T)+\tilde{g}^{(v)}_{1}\epsilon^{il}\sigma^{lk}(\partial_{k}\mu_{m})\\&+\tilde{g}^{(v)}_{2}\epsilon^{il}\omega^{lk}(\partial_{k}\mu_{m})+\tilde{g}^{(v)}_{3}(\partial_{k}v^{k})\epsilon^{il}(\partial_{l}\mu_{m})+\tilde{g}^{(v)}_{4}\epsilon^{il}\sigma^{lk}(\partial_{k}\phi)+\tilde{g}^{(v)}_{5}\epsilon^{il}\omega^{lk}(\partial_{k}\phi)\\&+\tilde{g}^{(v)}_{6}(\partial_{k}v^{k})\epsilon^{il}(\partial_{l}\phi)+\tilde{g}^{(v)}_{7}\epsilon^{il}\sigma^{lk}(\partial_{k}T)+\tilde{g}^{(v)}_{8}\epsilon^{il}\omega^{lk}(\partial_{k}T)+\tilde{g}^{(v)}_{9}(\partial_{k}v^{k})\epsilon^{il}(\partial^{l}T)
\end{split}
\end{eqnarray}
where,
\begin{eqnarray}
\begin{split}
 \tilde{n}^{(v)}_{1}& =\frac{\mathcal{N}_{1} p_0-\eta _r^2}{8 p_0^2},\ \   
 \tilde{n}^{(v)}_{2} =\frac{2 \mathcal{N}_{1} p_0-\eta _r^2}{16 p_0^2},\ \
 \tilde{n}^{(v)}_{3} =0,\ \
 \tilde{n}^{(v)}_{4} =\frac{\mathcal{N}_{7}}{4 p_0},\\
 \tilde{c}^{(v)}_{1}& =\frac{(-\mathcal{N}_{1}+2 \mathcal{N}_{2}+\mathcal{N}_{3}) p_0-\eta _r^2}{16 p_0^2},\ \
 \tilde{c}^{(v)}_{2} =\frac{(\mathcal{N}_{1}+2 \mathcal{N}_{2}-4 \mathcal{N}_{4}) p_0-\eta _r^2}{16 p_0^2}\\
 \tilde{c}^{(v)}_{3}& =\frac{1}{32 p_0^2 ((\partial_{T}p)(\partial_{\phi}q)-(\partial_{\phi}p) (\partial_{T}q))}\Big(\eta _r (\eta _r (3 (\partial_{\phi}p) (\partial_{T}q)-3 (\partial_{T}p) (\partial_{\phi}q))\\&+(\partial_{\phi}\eta_{r}) (3 q_0 (\partial_{T}p)-4 p_0 (\partial_{T}q))+(\partial_{T}\eta_{r}) (4 p_0 (\partial_{\phi}q)-3 q_0 (\partial_{\phi}p)))\Big)\\
\tilde{c}^{(v)}_{4}& =-\frac{1}{64 p_0^2 q_0}\Big(-3 \mathcal{N}_{1} q_0 (\partial_{\phi}p)-2 \mathcal{N}_{2} q_0 (\partial_{\phi}p)+\mathcal{N}_{3} q_0 (\partial_{\phi}p)-6 \mathcal{N}_{5} q_0 (\partial_{\phi}p)\\&+8 \mathcal{N}_{5}
   p_0 (\partial_{\phi}q)+4 q_0 \eta _r (\partial_{\phi}\eta_{r})\Big)\\
 \tilde{c}^{(v)}_{5}& =-\frac{1}{64 p_0^3 q_0}\Big(\mathcal{N}_{1} p_0 q_0 (\partial_{\phi}p)+2 \mathcal{N}_{2} p_0 q_0 (\partial_{\phi}p)+4 \mathcal{N}_{4} p_0 q_0 (\partial_{\phi}p)+6 \mathcal{N}_{5} p_0 q_0  (\partial_{\phi}p)\\&-8 \mathcal{N}_{5} p_0^2 (\partial_{\phi}q)-q_0 \eta _r^2 (\partial_{\phi}p)\Big)\\
 \tilde{c}^{(v)}_{6}& =\frac{1}{128 p_0^3((\partial_{T}p) (\partial_{\phi}q)-(\partial_{\phi}p) (\partial_{T}q))}\Big(4 p_0 \eta _r (\partial_{\phi}p) (\partial_{T}q) (\partial_{\phi}\eta_{r})-4 p_0 \eta _r (\partial_{\phi}p)(\partial_{\phi}q) (\partial_{T}\eta_{r})\\&+3 \eta _r^2 (\partial_{\phi}p) (\partial_{T}p) (\partial_{\phi}q)-3 \eta _r^2 (\partial_{\phi}p){}^2 (\partial_{T}q)-3 q_0 \eta _r (\partial_{\phi}p) (\partial_{T}p)
(\partial_{\phi}\eta_{r})+3 q_0 \eta _r (\partial_{\phi}p){}^2 (\partial_{T}\eta_{r})\Big)\\
 \tilde{c}^{(v)}_{7}& =\frac{1}{64 p_0^2 q_0}\Big(((3 \mathcal{N}_{1}+2 \mathcal{N}_{2}-\mathcal{N}_{3}) q_0+6 \mathcal{N}_{5} q_0) (\partial_{T}p)-4 (2 \mathcal{N}_{5} p_0 (\partial_{T}q)+q_0 \eta _r (\partial_{T}\eta_{r}))\Big)\\
 \tilde{c}^{(v)}_{8}& =\frac{1}{64 p_0^3 q_0}\Big(-p_0 ((\mathcal{N}_{1}+2 \mathcal{N}_{2}+4 \mathcal{N}_{4}) q_0+6 \mathcal{N}_{5} q_0) (\partial_{T}p)+8 \mathcal{N}_{5} p_0^2 (\partial_{T}q)+q_0 \eta _r^2  (\partial_{T}p)\Big)
 \end{split}
\end{eqnarray}
\begin{eqnarray}
\begin{split}
 \tilde{c}^{(v)}_{9}& =\frac{1}{128 p_0^3 ((\partial_{T}p) (\partial_{\phi}q)-(\partial_{\phi}p)
   (\partial_{T}q))}\Big(\eta _r (\partial_{T}p) (3 \eta _r ((\partial_{T}p) (\partial_{\phi}q)-(\partial_{\phi}p)
   (\partial_{T}q))+q_0 (3 (\partial_{\phi}p) (\partial_{T}\eta_{r})\\&-3 (\partial_{T}p)(\partial_{\phi}\eta_{r}))+4 p_0 ((\partial_{T}q) (\partial_{\phi}\eta_{r})-(\partial_{\phi}q) (\partial_{T}\eta_{r})))\Big)\\  
 \tilde{g}^{(v)}_{1}& =0,\ \
 \tilde{g}^{(v)}_{2} =-\frac{\mathcal{N}_{7}}{2 p_0},\ \
 \tilde{g}^{(v)}_{3} =0, \ \ \tilde{g}^{(v)}_{4} =\frac{1}{64 p_0^2 q_0}\Big(\mathcal{N}_{9} (4 p_0 (\partial_{\phi}q)-3 q_0 (\partial_{\phi}p))\Big),\\
 \tilde{g}^{(v)}_{5}& =-\frac{1}{32 p_0^2 q_0}\Big(-4 \mathcal{N}_{7} q_0 (\partial_{\phi}p)-3 \mathcal{N}_{8} q_0 (\partial_{\phi}p)+4 \mathcal{N}_{8} p_0 (\partial_{\phi}q)\Big),\ \
 \tilde{g}^{(v)}_{6} =0,\\
 \tilde{g}^{(v)}_{7} &=\frac{1}{64 p_0^2 q_0}\Big(\mathcal{N}_{9} (4 p_0 (\partial_{T}q)-3 q_0 (\partial_{T}p))\Big),\\
 \tilde{g}^{(v)}_{8} &=\frac{1}{32 p_0^2 q_0}\Big((4 \mathcal{N}_{7} q_0+3 \mathcal{N}_{8} q_0) (\partial_{T}p)-4 \mathcal{N}_{8} p_0 (\partial_{T}q)\Big),\ \
 \tilde{g}^{(v)}_{9} =0\nonumber
 \end{split}
\end{eqnarray}
Pressure,
\begin{eqnarray}
\begin{split}
    p&=P+\tilde{n}^{(p)}_{1}(\partial_{k}\partial_{k}\mu_{m})+\tilde{n}^{(p)}_{2}(\partial_{k}\partial_{k}\phi)+\tilde{n}^{(p)}_{1}(\partial_{k}\partial_{k}T)+\tilde{c}^{(p)}_{1}(\partial_{k}\mu_{m})(\partial^{k}\mu_{m})\\&+\tilde{c}^{(p)}_{2}(\partial_{k}\phi)(\partial^{k}\phi)+\tilde{c}^{(p)}_{3}(\partial_{k}T)(\partial^{k}T)+\tilde{c}^{(p)}_{4}(\partial_{k}\mu_{m})(\partial^{k}\phi)+\tilde{c}^{(p)}_{5}(\partial_{k}T)(\partial^{k}\phi)\\&+\tilde{c}^{(p)}_{6}(\partial_{k}\mu_{m})(\partial^{k}T)+\tilde{c}^{(p)}_{7}\sigma^{ij}\sigma_{ij}+\tilde{c}^{(p)}_{8}\omega^{ij}\omega_{ij}+\tilde{g}^{(p)}_{1}\epsilon^{ij}(\partial_{i}\mu_{m})(\partial_{j}\phi)\\&+\tilde{g}^{(p)}_{2}\epsilon^{ij}(\partial_{i}\mu_{m})(\partial_{j}T)+\tilde{g}^{(p)}_{3}\epsilon^{ij}(\partial_{i}T)(\partial_{j}\phi)
    \end{split}
\end{eqnarray}
where,
\begin{eqnarray}
\begin{split}
\tilde{n}^{(p)}_{1}&=\frac{1}{4 p_0}\Big(\eta_{r}{}^2\Big),\ \
\tilde{n}^{(p)}_{2}=\frac{1}{48 p_0^2 q_0}\Big(-6 N_5 p_0 q_0 (\partial_{\phi}p)+8 N_5 p_0^2 (\partial_{\phi}q)-3 q_0 (\partial_{\phi}p)
   \eta_{r}{}^2\Big)\\
\tilde{n}^{(p)}_{3}&=\frac{1}{48 p_0^2 q_0}\Big(-6 N_5 p_0 q_0 (\partial_{T}p)+8 N_5 p_0^2 (\partial_{T}q)-3 q_0 (\partial_{T}p)
   \eta_{r}{}^2\Big)\\
\tilde{c}^{(p)}_{1}&=\frac{1}{24 p_0}\Big(-11 \eta_{r}{}^2-12 N_2 p_0-N_3 p_0+4 N_4 p_0\Big)\\
\tilde{c}^{(p)}_{2}&=\frac{1}{384 p_0^3 q_0^2}\Big(-160 N_5 p_0^2 q_0 (\partial_{\phi}p) (\partial_{\phi}q)-96 N_6 p_0^2 q_0 (\partial_{\phi}p) (\partial_{\phi}q)-48 N_5 p_0^2 q_0^2 (\partial_{\phi}^{2}p)\\&+12 N_1 p_0 q_0^2 (\partial_{\phi}p){}^2+12 N_2 p_0 q_0^2 (\partial_{\phi}p){}^2-N_3 p_0 q_0^2 (\partial_{\phi}p){}^2+4 N_4 p_0 q_0^2
   (\partial_{\phi}p){}^2\\&+132 N_5 p_0 q_0^2 (\partial_{\phi}p){}^2+36 N_6 p_0 q_0^2 (\partial_{\phi}p){}^2+64 N_6 p_0^3 (\partial_{\phi}q){}^2+64 N_5 p_0^3 q_0 (\partial_{\phi}^{2}q)\\&-24 p_0 q_0^2
   (\partial_{\phi}p) \eta_{r} (\partial_{\phi}\eta_{r})-24 p_0 q_0^2
   (\partial_{\phi}^{2}p) \eta_{r}{}^2+13 q_0^2 (\partial_{\phi}p){}^2 \eta_{r}{}^2\Big)\\
\tilde{c}^{(p)}_{3}&=\frac{1}{384 p_0^3 q_0^2}\Big(-160 N_5 p_0^2 q_0 (\partial_{T}p) (\partial_{T}q)-96 N_6 p_0^2 q_0 (\partial_{T}p) (\partial_{T}q)-48 N_5 p_0^2 q_0^2 (\partial_{T}^{2}p)\\&+12 N_1 p_0 q_0^2 (\partial_{T}p){}^2+12 N_2 p_0 q_0^2 (\partial_{T}p){}^2-N_3 p_0 q_0^2 (\partial_{T}p){}^2+4 N_4 p_0 q_0^2
   (\partial_{T}p){}^2\\&+132 N_5 p_0 q_0^2 (\partial_{T}p){}^2+36 N_6 p_0 q_0^2 (\partial_{T}p){}^2+64 N_6 p_0^3 (\partial_{T}q){}^2+64 N_5 p_0^3 q_0 (\partial_{T}^{2}q)\\&-24 p_0 q_0^2
   (\partial_{T}p) \eta_{r} (\partial_{T}\eta_{r})-24 p_0 q_0^2
   (\partial_{T}^{2}p) \eta_{r}{}^2+13 q_0^2 (\partial_{T}p){}^2 \eta_{r}{}^2\Big)
   \end{split}
\end{eqnarray}
\begin{eqnarray}
\begin{split}
\tilde{c}^{(p)}_{4}&=\frac{1}{48 p_0^2}\Big(-6 N_1 p_0 (\partial_{\phi}p)+N_3 p_0 (\partial_{\phi}p)+4 N_4 p_0 (\partial_{\phi}p)+8
   (\partial_{\phi}p) \eta_{r}{}^2+12 p_0 (\partial_{\phi}\eta_{r})\eta_{r}\Big)\\
\tilde{c}^{(p)}_{5}&=\frac{1}{48 p_0^2}\Big(-6 N_1 p_0 (\partial_{T}p)+N_3 p_0 (\partial_{T}p)+4 N_4 p_0 (\partial_{T}p)+8
   (\partial_{T}p) \eta_{r}{}^2+12 p_0 (\partial_{T}\eta_{r}) \eta_{r}\Big)\\
\tilde{c}^{(p)}_{6}&=\frac{1}{384 p_0^3 q_0^2}\Big(-32 p_0^2 (5 N_5 q_0 (\partial_{T}p) (\partial_{\phi}q)\\&+3 N_6 q_0 (\partial_{T}p) (\partial_{\phi}q)+5 N_5 q_0 (\partial_{\phi}p) (\partial_{T}q)+3 N_6 q_0 (\partial_{\phi}p) (\partial_{T}q)+3 N_5 q_0^2 (\partial_{\phi}\partial_{T}p))\\&+2 p_0 (12
   N_1 q_0^2 (\partial_{\phi}p) (\partial_{T}p)+12 N_2 q_0^2 (\partial_{\phi}p)
   (\partial_{T}p)-N_3 q_0^2 (\partial_{\phi}p) (\partial_{T}p)+4 N_4 q_0^2
   (\partial_{\phi}p) (\partial_{T}p)\\&+132 N_5 q_0^2 (\partial_{\phi}p)
   (\partial_{T}p)+36 N_6 q_0^2 (\partial_{\phi}p) (\partial_{T}p)-12 q_0^2
   (\partial_{T}p) \eta_{r} (\partial_{\phi}\eta_{r})-12 q_0^2
   (\partial_{\phi}p) \eta_{r} (\partial_{T}\eta_{r})\\&-24 q_0^2
   (\partial_{\phi}\partial_{T}p) \eta_{r}{}^2)+128 p_0^3 (N_6 (\partial_{\phi}q)
   (\partial_{T}q)+N_5 q_0 (\partial_{\phi}\partial_{T}q))+26 q_0^2 (\partial_{\phi}p)
   (\partial_{T}p) \eta_{r}{}^2\Big)\\
\tilde{c}^{(p)}_{7}&=\frac{1}{24 p_0}\Big(N_3 p_0-4 \eta_{r}{}^2\Big),\ \
\tilde{c}^{(p)}_{8}= -\frac{N_4}{6},\\
\tilde{g}^{(p)}_{1}&=\frac{1}{48 p_0 q_0}\Big(-6 N_8 q_0 (\partial_{\phi}p)-9 N_9 q_0 (\partial_{\phi}p)+8 N_8 p_0 (\partial_{\phi}q)+12
   N_9 p_0 (\partial_{\phi}q)\Big),\\
\tilde{g}^{(p)}_{2}&=\frac{1}{48 p_0 q_0}\Big(-6 N_8 q_0 (\partial_{T}p)-9 N_9 q_0 (\partial_{T}p)+8 N_8 p_0 (\partial_{T}q)+12
   N_9 p_0 (\partial_{T}q)\Big)\\
\tilde{g}^{(p)}_{3}&=-\frac{1}{48 p_0 q_0}\Big(-2 N_8 (\partial_{T}p) (\partial_{\phi}q)+3 N_9 (\partial_{T}p)
   (\partial_{\phi}q)+2 N_8 (\partial_{\phi}p) (\partial_{T}q)-3 N_9 (\partial_{\phi}p) (\partial_{T}q)\Big)\nonumber
\end{split}
\end{eqnarray}
Energy,
\begin{eqnarray}
\begin{split}
\epsilon&=\frac{E-P}{2}+\frac{1}{2}\rho v^{2}+\tilde{n}^{(\epsilon)}_{1}(\partial_{k}\partial_{k}\mu_{m})+\tilde{n}^{(\epsilon)}_{2}(\partial_{k}\partial_{k}\phi)+\tilde{n}^{(\epsilon)}_{1}(\partial_{k}\partial_{k}T)+\tilde{c}^{(\epsilon)}_{1}(\partial_{k}\mu_{m})(\partial^{k}\mu_{m})\\&+\tilde{c}^{(\epsilon)}_{2}(\partial_{k}\phi)(\partial^{k}\phi)+\tilde{c}^{(\epsilon)}_{3}(\partial_{k}T)(\partial^{k}T)+\tilde{c}^{(\epsilon)}_{4}(\partial_{k}\mu_{m})(\partial^{k}\phi)+\tilde{c}^{(\epsilon)}_{5}(\partial_{k}T)(\partial^{k}\phi)\\&+\tilde{c}^{(\epsilon)}_{6}(\partial_{k}\mu_{m})(\partial^{k}T)+\tilde{c}^{(\epsilon)}_{7}\sigma^{ij}\sigma_{ij}+\tilde{c}^{(\epsilon)}_{8}\omega^{ij}\omega_{ij}+\tilde{g}^{(\epsilon)}_{1}\epsilon^{ij}(\partial_{i}\mu_{m})(\partial_{j}\phi)\\&+\tilde{g}^{(\epsilon)}_{2}\epsilon^{ij}(\partial_{i}\mu_{m})(\partial_{j}T)+\tilde{g}^{(\epsilon)}_{3}\epsilon^{ij}(\partial_{i}T)(\partial_{j}\phi)
\end{split}
\end{eqnarray}
where,
\begin{eqnarray}
\begin{split}
\tilde{n}^{(\epsilon)}_{1}&=\frac{\eta _r^2}{4 p_0},\ \
\tilde{n}^{(\epsilon)}_{2}=-\frac{1}{48 p_0^2 q_0}\Big(6 \mathcal{N}_{5} p_0 q_0 (\partial_{\phi}p)-8 \mathcal{N}_{5} p_0^2 (\partial_{\phi}q)+3 q_0 \eta _r^2 (\partial_{\phi}p))\\
\tilde{n}^{(\epsilon)}_{3}&=-\frac{1}{48 p_0^2 q_0}\Big(6 \mathcal{N}_{5} p_0 q_0 (\partial_{T}p)-8 \mathcal{N}_{5} p_0^2 (\partial_{T}q)+3 q_0 \eta _r^2 (\partial_{T}p)\Big),\\
\tilde{c}^{(\epsilon)}_{1}&=-\frac{1}{24 p_0}\Big((12 \mathcal{N}_{2}+\mathcal{N}_{3}-4 \mathcal{N}_{4}) p_0+11 \eta _r^2\Big)\\
\tilde{c}^{(\epsilon)}_{2}&=\frac{1}{384 p_0^3 q_0^2}\Big(-16 p_0^2 (2 (5 \mathcal{N}_{5} q_0+3 \mathcal{N}_{6} q_0) (\partial_{\phi}p) (\partial_{\phi}q)+3 \mathcal{N}_{5} q_0^2 (\partial_{\phi}^{2}p))\\&+p_0
   (((12 \mathcal{N}_{1}+12 \mathcal{N}_{2}-\mathcal{N}_{3}+4 \mathcal{N}_{4}) q_0^2+132 \mathcal{N}_{5} q_0^2+36 \mathcal{N}_{6} q_0^2) (\partial_{\phi}p){}^2\\&-24 q_0^2 \eta _r (\partial_{\phi}p)
   (\partial_{\phi}\eta_{r})-24 q_0^2 \eta _r^2 (\partial_{\phi}^{2}p))+64 p_0^3 (\mathcal{N}_{6} (\partial_{\phi}q){}^2+\mathcal{N}_{5} q_0
   (\partial_{\phi}^{2}q))+13 q_0^2 \eta _r^2 (\partial_{\phi}p){}^2\Big)\nonumber
   \end{split}
\end{eqnarray}
\begin{eqnarray}
\begin{split}
\tilde{c}^{(\epsilon)}_{3}&=\frac{1}{384 p_0^3 q_0^2}\Big(-16 p_0^2 (2 (5 \mathcal{N}_{5} q_0+3 \mathcal{N}_{6} q_0) (\partial_{T}p) (\partial_{T}q)+3 \mathcal{N}_{5} q_0^2 (\partial_{T}^{2}p))\\&+p_0
   (((12 \mathcal{N}_{1}+12 \mathcal{N}_{2}-\mathcal{N}_{3}+4 \mathcal{N}_{4}) q_0^2+132 \mathcal{N}_{5} q_0^2+36 \mathcal{N}_{6} q_0^2) (\partial_{T}p){}^2\\&-24 q_0^2 \eta _r (\partial_{T}p)
   (\partial_{T}\eta_{r})-24 q_0^2 \eta _r^2 (\partial_{T}^{2}p))+64 p_0^3 (\mathcal{N}_{6} (\partial_{T}q){}^2+\mathcal{N}_{5} q_0
   (\partial_{T}^{2}q))+13 q_0^2 \eta _r^2 (\partial_{T}p){}^2\Big)\\
\tilde{c}^{(\epsilon)}_{4}&=\frac{1}{48 p_0^2}\Big(p_0 ((-6 \mathcal{N}_{1}+\mathcal{N}_{3}+4 \mathcal{N}_{4}) (\partial_{\phi}p)+12 (\partial_{\phi}\eta_{r}))+8 \eta _r^2(\partial_{\phi}p)\Big)\\
\tilde{c}^{(\epsilon)}_{5}&=\frac{1}{48 p_0^2}\Big(p_0 ((-6 \mathcal{N}_{1}+\mathcal{N}_{3}+4 \mathcal{N}_{4}) (\partial_{T}p)+12 (\partial_{T}\eta_{r}))+8 \eta _r^2 (\partial_{T}p)\Big)\\
\tilde{c}^{(\epsilon)}_{6}&=\frac{1}{192 p_0^3 q_0^2}\Big(-16 p_0^2 ((5 \mathcal{N}_{5} q_0+3 \mathcal{N}_{6} q_0) (\partial_{T}p) (\partial_{\phi}q)+(5 \mathcal{N}_{5} q_0+3 \mathcal{N}_{6} q_0) (\partial_{\phi}p)
   (\partial_{T}q)\\&+3 \mathcal{N}_{5} q_0^2 (\partial_{T}\partial_{\phi}p))+p_0 ((\partial_{\phi}p) (((12 \mathcal{N}_{1}+12 \mathcal{N}_{2}-\mathcal{N}_{3}+4 \mathcal{N}_{4})
   q_0^2+132 \mathcal{N}_{5} q_0^2\\&+36 \mathcal{N}_{6} q_0^2) (\partial_{T}p)-12 q_0^2 (\partial_{T}\eta_{r}))-12 q_0^2 \eta _r (2 \eta _r
   (\partial_{T}\partial_{\phi}p)+(\partial_{T}p) (\partial_{\phi}\eta_{r})))+64 p_0^3 (\mathcal{N}_{6} (\partial_{\phi}q)
   (\partial_{T}q)\\&+\mathcal{N}_{5} q_0 (\partial_{T}\partial_{\phi}q))+13 q_0^2 \eta _r^2 (\partial_{\phi}p) (\partial_{T}p)\Big)\\
   \tilde{c}^{(\epsilon)}_{7}&=\frac{1}{24} (\mathcal{N}_{3}-\frac{4 \eta _r^2}{p_0}),\ \
\tilde{c}^{(\epsilon)}_{8}=-\frac{\mathcal{N}_{4}}{6}
\end{split}
\end{eqnarray}
\begin{eqnarray}
\begin{split}
\tilde{g}^{(\epsilon)}_{1}&=\frac{1}{48 p_0 q_0}\Big((2 \mathcal{N}_{8}+3 \mathcal{N}_{9}) (4 p_0 (\partial_{\phi}q)-3 q_0 (\partial_{\phi}p))\Big),\ \ \tilde{g}^{(\epsilon)}_{2}=\frac{1}{48 p_0 q_0}\Big((2 \mathcal{N}_{8}+3 \mathcal{N}_{9}) (4 p_0 (\partial_{T}q)-3 q_0 (\partial_{T}p))\Big)\\
\tilde{g}^{(\epsilon)}_{3}&=-\frac{1}{48 p_0 q_0}\Big((2 \mathcal{N}_{8}-3 \mathcal{N}_{9}) ((\partial_{T}p) (\partial_{\phi}q)-(\partial_{\phi}p) (\partial_{T}q))\Big)\nonumber
\end{split}
\end{eqnarray}
Energy current,
\begin{eqnarray}
\begin{split}
e^{i}&=(e+P+\frac{1}{2}\rho v^{2})v^{i}+\eta_{r}\Big(\frac{\partial_{i}u^{+}}{(u^{+})^{2}}-\frac{\partial_{i}P}{4 Pu^{+}}\Big)-\eta_{r}\sigma_{ij}v^{j}+\tilde{n}^{(ec)}_{1}\partial_{k}\sigma^{ik}+\tilde{n}^{(ec)}_{2}\partial_{k}\omega^{ik}+\tilde{n}^{(ec)}_{3}\epsilon^{il}\partial_{k}\sigma^{lk}\\&+\tilde{n}^{(ec)}_{4}\epsilon^{il}\partial_{k}\omega^{lk}+\tilde{c}^{(ec)}_{1}\sigma^{ik}(\partial_{k}\mu_{m})+\tilde{c}^{(ec)}_{2}\omega^{ik}(\partial_{k}\mu_{m})+\tilde{c}^{(ec)}_{3}(\partial_{k}v^{k})(\partial_{i}\mu_{m})+\tilde{c}^{(ec)}_{4}\sigma^{ik}(\partial_{k}\phi)\\&+\tilde{c}^{(ec)}_{5}\omega^{ik}(\partial_{k}\phi)+\tilde{c}^{(ec)}_{6}(\partial_{k}v^{k})(\partial_{i}\phi)+\tilde{c}^{(ec)}_{7}\sigma^{ik}(\partial_{k}T)+\tilde{c}^{(ec)}_{8}\omega^{ik}(\partial_{k}T)+\tilde{c}^{(ec)}_{9}(\partial_{k}v^{k})(\partial^{i}T)\\&+\tilde{g}^{(ec)}_{1}\epsilon^{il}\sigma^{lk}(\partial_{k}\mu_{m})+\tilde{g}^{(ec)}_{2}\epsilon^{il}\omega^{lk}(\partial_{k}\mu_{m})+\tilde{g}^{(ec)}_{3}(\partial_{k}v^{k})\epsilon^{il}(\partial_{l}\mu_{m})+\tilde{g}^{(ec)}_{4}\epsilon^{il}\sigma^{lk}(\partial_{k}\phi)+\tilde{g}^{(ec)}_{5}\epsilon^{il}\omega^{lk}(\partial_{k}\phi)\\&+\tilde{g}^{(ec)}_{6}(\partial_{k}v^{k})\epsilon^{il}(\partial_{l}\phi)+\tilde{g}^{(ec)}_{7}\epsilon^{il}\sigma^{lk}(\partial_{k}T)+\tilde{g}^{(ec)}_{8}\epsilon^{il}\omega^{lk}(\partial_{k}T)+\tilde{g}^{(ec)}_{9}(\partial_{k}v^{k})\epsilon^{il}(\partial^{l}T)
\end{split}
\end{eqnarray}
where,
\begin{eqnarray}
\begin{split}
  \tilde{n}^{(ec)}_{1}&=\frac{1}{2} (\frac{\eta _r^2}{p_0}-\mathcal{N}_{1}),\ \ 
   \tilde{n}^{(ec)}_{2}=\frac{\eta _r^2-2 \mathcal{N}_{1} p_0}{4 p_0},\ \ 
   \tilde{n}^{(ec)}_{3}=0,\ \ 
   \tilde{n}^{(ec)}_{4}=-\mathcal{N}_{7},\\
   \tilde{c}^{(ec)}_{1}&=\frac{1}{4} (\mathcal{N}_{1}-2 \mathcal{N}_{2}-\mathcal{N}_{3}),\ \
   \tilde{c}^{(ec)}_{2}=\frac{1}{4} (\mathcal{N}_{1}+2 \mathcal{N}_{2}-4 \mathcal{N}_{4}-\frac{\eta _r^2}{p_0})\nonumber
   \end{split}
\end{eqnarray}
\begin{eqnarray}
\begin{split}
   \tilde{c}^{(ec)}_{3}&=\frac{1}{8 p_0 ((\partial_{T}p) (\partial_{\phi}q)-(\partial_{\phi}p) (\partial_{T}q))}\Big(\eta _r (3 \eta _r ((\partial_{T}p) (\partial_{\phi}q)-(\partial_{\phi}p) (\partial_{T}q))+4 p_0
   ((\partial_{T}q) (\partial_{\phi}\eta_{r})\\&-(\partial_{\phi}q) (\partial_{T}\eta_{r}))+q_0
   (3 (\partial_{\phi}p) (\partial_{T}\eta_{r})-3 (\partial_{T}p) (\partial_{\phi}\eta_{r})))\Big)\\
   \tilde{c}^{(ec)}_{4}&=\frac{1}{16 p_0^2 q_0}\Big(p_0 (((-3 \mathcal{N}_{1}-2 \mathcal{N}_{2}+\mathcal{N}_{3}) q_0-6 \mathcal{N}_{5} q_0) (\partial_{\phi}p)+4 q_0 \eta _r (\partial_{\phi}\eta_{r}))\\&+8 \mathcal{N}_{5} p_0^2
   (\partial_{\phi}q)+q_0 \eta _r^2 (\partial_{\phi}p)\Big)\\
   \tilde{c}^{(ec)}_{5}&=\frac{1}{16 p_0^2 q_0}\Big(-p_0 ((\mathcal{N}_{1}+2 \mathcal{N}_{2}+4 \mathcal{N}_{4}) q_0+6 \mathcal{N}_{5} q_0) (\partial_{\phi}p)+8 \mathcal{N}_{5} p_0^2 (\partial_{\phi}q)+q_0 \eta _r^2 (\partial_{\phi}p)\Big)\\
   \tilde{c}^{(ec)}_{6}&=\frac{1}{32 p_0^2 ((\partial_{T}p) (\partial_{\phi}q)-(\partial_{\phi}p)
   (\partial_{T}q))}\Big(\eta _r (\partial_{\phi}p) (\eta _r (3 (\partial_{\phi}p) (\partial_{T}q)-3 (\partial_{T}p)
   (\partial_{\phi}q))+3 q_0 ((\partial_{T}p) (\partial_{\phi}\eta_{r})\\&-(\partial_{\phi}p) (\partial_{T}\eta_{r}))+4 p_0 ((\partial_{\phi}q) (\partial_{T}\eta_{r})-(\partial_{T}q) (\partial_{\phi}\eta_{r})))\Big)\\
   \tilde{c}^{(ec)}_{7}&=\frac{1}{16 p_0^2 q_0}\Big(p_0 (((-3 \mathcal{N}_{1}-2 \mathcal{N}_{2}+\mathcal{N}_{3}) q_0-6 \mathcal{N}_{5} q_0) (\partial_{T}p)+4 q_0 \eta _r (\partial_{T}\eta_{r}))\\&+8 \mathcal{N}_{5} p_0^2
   (\partial_{T}q)+q_0 \eta _r^2 (\partial_{T}p)\Big)\\
   \tilde{c}^{(ec)}_{8}&=\frac{1}{16 p_0^2 q_0}\Big(-p_0 ((\mathcal{N}_{1}+2 \mathcal{N}_{2}+4 \mathcal{N}_{4}) q_0+6 \mathcal{N}_{5} q_0) (\partial_{T}p)+8 \mathcal{N}_{5} p_0^2 (\partial_{T}q)+q_0 \eta _r^2 (\partial_{T}p)\Big)\\
   \tilde{c}^{(ec)}_{9}&=\frac{1}{32 p_0^2 ((\partial_{T}p) (\partial_{\phi}q)-(\partial_{\phi}p)
   (\partial_{T}q))}\Big(\eta _r (\partial_{T}p) (\eta _r (3 (\partial_{\phi}p) (\partial_{T}q)-3 (\partial_{T}p)
   (\partial_{\phi}q))+3 q_0 ((\partial_{T}p) (\partial_{\phi}\eta_{r})\\&-(\partial_{\phi}p) (\partial_{T}\eta_{r}))+4 p_0 ((\partial_{\phi}q) (\partial_{T}\eta_{r})-(\partial_{T}q) (\partial_{\phi}\eta_{r})))\Big)
   \end{split}
   \end{eqnarray}
   \begin{eqnarray}
   \begin{split}
  \tilde{g}^{(ec)}_{1}&=0,\ \
   \tilde{g}^{(ec)}_{2}=-2\mathcal{N}_{7},\ \
   \tilde{g}^{(ec)}_{3}=0,\ \
   \tilde{g}^{(ec)}_{4}=\frac{1}{16 p_0 q_0}\Big(\mathcal{N}_{9} (3 q_0 (\partial_{\phi}p)-4 p_0 (\partial_{\phi}q))\Big)\\
   \tilde{g}^{(ec)}_{5}&=\frac{1}{8 p_0 q_0}\Big((4 \mathcal{N}_{7} q_0+3 \mathcal{N}_{8} q_0) (\partial_{\phi}p)-4 \mathcal{N}_{8} p_0 (\partial_{\phi}q)\Big),\ \
   \tilde{g}^{(ec)}_{6}=0,\\
   \tilde{g}^{(ec)}_{7}&=\frac{1}{16 p_0 q_0}\Big(\mathcal{N}_{9} (3 q_0 (\partial_{T}p)-4 p_0 (\partial_{T}q))\Big)\\
   \tilde{g}^{(ec)}_{8}&=\frac{1}{8 p_0 q_0}\Big((4 \mathcal{N}_{7} q_0+3 \mathcal{N}_{8} q_0) (\partial_{T}p)-4 \mathcal{N}_{8} p_0 (\partial_{T}q)\Big),\ \
   \tilde{g}^{(ec)}_{9}=0\nonumber
   \end{split}
   \end{eqnarray}
   Charge,
   \begin{eqnarray}
   \begin{split}
 Q&=4\sqrt{3}q u^{+}-\xi_{r}(u^{+})^{2}\epsilon^{ij}(\partial_{i}v_{j})+\tilde{n}^{(q)}_{1}(\partial_{k}\partial_{k}\mu_{m})+\tilde{n}^{(q)}_{2}(\partial_{k}\partial_{k}\phi)+\tilde{n}^{(q)}_{1}(\partial_{k}\partial_{k}T)+\tilde{c}^{(q)}_{1}(\partial_{k}\mu_{m})(\partial^{k}\mu_{m})\\&+\tilde{c}^{(q)}_{2}(\partial_{k}\phi)(\partial^{k}\phi)+\tilde{c}^{(q)}_{3}(\partial_{k}T)(\partial^{k}T)+\tilde{c}^{(q)}_{4}(\partial_{k}\mu_{m})(\partial^{k}\phi)+\tilde{c}^{(q)}_{5}(\partial_{k}T)(\partial^{k}\phi)+\tilde{c}^{(q)}_{6}(\partial_{k}\mu_{m})(\partial^{k}T)\\&+\tilde{c}^{(q)}_{7}\sigma^{ij}\sigma_{ij}+\tilde{c}^{(q)}_{8}\omega^{ij}\omega_{ij}+\tilde{g}^{(q)}_{1}\epsilon^{ij}(\partial_{i}\mu_{m})(\partial_{j}\phi)+\tilde{g}^{(q)}_{2}\epsilon^{ij}(\partial_{i}\mu_{m})(\partial_{j}T)+\tilde{g}^{(q)}_{3}\epsilon^{ij}(\partial_{i}T)(\partial_{j}\phi)
   \end{split}
\end{eqnarray}
\begin{eqnarray}
\begin{split}
\tilde{n}^{(q)}_{1}&=\frac{1}{2 p_0}\Big((\gamma _1-\gamma _2) p_0-6 \sqrt{3} q_0 \eta _r \lambda _r\Big)\\
\tilde{n}^{(q)}_{2}&=-\frac{1}{8 p_0^2}\Big(p_0 (\partial_{\phi}p) (\gamma _1+\gamma _2-24 \sqrt{3} q_0 \lambda _r^2)-6 \sqrt{3} q_0 \eta _r \lambda _r (\partial_{\phi}p)+32 \sqrt{3}p_0^2 \lambda _r^2 (\partial_{\phi}q)\Big)\\
\tilde{n}^{(q)}_{3}&=-\frac{1}{8 p_0^2}\Big(p_0 (\partial_{T}p) (\gamma _1+\gamma _2-24 \sqrt{3} q_0 \lambda _r^2)-6 \sqrt{3} q_0 \eta _r \lambda _r (\partial_{T}p)+32 \sqrt{3} p_0^2 \lambda _r^2 (\partial_{T}q)\Big)\\
\tilde{c}^{(q)}_{1}&=\frac{1}{2 p_0}\Big((\gamma _2-\gamma _1) p_0+8 \sqrt{3} q_0 \eta _r \lambda _r\Big)\\
\tilde{c}^{(q)}_{2}&=\frac{1}{32 p_0^3 q_0}\Big(4 p_0 q_0 (-p_0 (\partial_{\phi}^{2}p) (\gamma _1+\gamma _2-24 \sqrt{3} q_0 \lambda _r^2)+6 \sqrt{3} q_0 \eta _r \lambda _r (\partial_{\phi}^{2}p)-32 \sqrt{3} p_0^2 \lambda _r (\lambda _r (\partial_{\phi}^{2}q)\\&+(\partial_{\phi}q) (\partial_{\phi}\lambda_{r})))+4 p_0 (\partial_{\phi}p) (6 \sqrt{3} q_0^2 \lambda _r (4 p_0 (\partial_{\phi}\lambda_{r})+(\partial_{\phi}\eta_{r}))-p_0 (\partial_{\phi}q) (\gamma _4+\gamma _5-24 \sqrt{3} q_0 \lambda _r^2))\\&+(\partial_{\phi}p){}^2 (4 \gamma _2 p_0 q_0+q_0 (3 (\gamma _4+\gamma _5) p_0-16 \sqrt{3} q_0 \lambda _r (6 p_0 \lambda _r+\eta _r)))\Big)\\
\tilde{c}^{(q)}_{3}&=\frac{1}{32 p_0^3 q_0}\Big(4 p_0 q_0 (-p_0 (\partial_{T}^{2}p) (\gamma _1+\gamma _2-24 \sqrt{3} q_0 \lambda _r^2)+6 \sqrt{3} q_0 \eta _r \lambda _r (\partial_{T}^{2}p)-32 \sqrt{3} p_0^2 \lambda _r (\lambda _r (\partial_{T}^{2}q)\\&+(\partial_{T}q) (\partial_{T}\lambda_{r})))+4 p_0 (\partial_{T}p) (6 \sqrt{3} q_0^2 \lambda _r (4 p_0 (\partial_{T}\lambda_{r})+(\partial_{T}\eta_{r}))-p_0 (\partial_{T}q) (\gamma _4+\gamma _5-24 \sqrt{3} q_0 \lambda _r^2))\\&+(\partial_{T}p){}^2 (4 \gamma _2 p_0 q_0+q_0 (3 (\gamma _4+\gamma _5) p_0-16 \sqrt{3} q_0 \lambda _r (6 p_0 \lambda _r+\eta _r)))\Big)\\
\tilde{c}^{(q)}_{4}&=\frac{1}{8 p_0^2 q_0}\Big((\partial_{\phi}p) (p_0 (4 \gamma _1 q_0+3 (\gamma _5-\gamma _4) q_0)-10 \sqrt{3} q_0^2 \eta _r \lambda _r)+4 p_0 ((\gamma _4-\gamma _5) p_0 (\partial_{\phi}q)\\&-6 \sqrt{3} q_0^2 \lambda_r (\partial_{\phi}\eta_{r}))\Big)\\\tilde{c}^{(q)}_{5}&=\frac{1}{8 p_0^2 q_0}\Big((\partial_{T}p) (p_0 (4 \gamma _1 q_0+3 (\gamma _5-\gamma _4) q_0)-10 \sqrt{3} q_0^2 \eta _r \lambda _r)+4 p_0 ((\gamma_4-\gamma _5) p_0 (\partial_{T}q)\\&-6 \sqrt{3} q_0^2 \lambda _r (\partial_{T}\eta_{r}))\Big)\\
\tilde{c}^{(q)}_{6}&=\frac{1}{16 p_0^3 q_0}\Big((\partial_{\phi}p) ((\partial_{T}p) (4 \gamma _2 p_0 q_0+q_0 (3 p_0 (\gamma_{4}+\gamma_{5})-16 \sqrt{3} q_0 \lambda _r
   (6 p_0 \lambda _r+\eta _r)))\\&+2 p_0 (6 \sqrt{3} q_0^2 \lambda _r (4 p_0 (\partial_{T}\lambda_{r})+(\partial_{T}\eta_{r}))-p_0 (\partial_{T}q) (\gamma_{4}+\gamma_{5}-24 \sqrt{3} q_0 \lambda_r^2)))\\&+2 p_0 (2q_0 (p_0 ((-\gamma _1-\gamma _2) (\partial_{T}\partial_{\phi}p)-16 \sqrt{3} p_0 \lambda _r (2 \lambda _r (\partial_{T}\partial_{\phi}q)+(\partial_{T}q) (\partial_{\phi}\lambda_{r})))\\&+3 \sqrt{3} q_0 \lambda _r (2 (\partial_{T}\partial_{\phi}p) (4p_0 \lambda _r+\eta _r)+(\partial_{T}p) (4 p_0 (\partial_{\phi}\lambda_{r})+(\partial_{\phi}\eta_{r}))))\\&+p_0 (\partial_{\phi}q) ((\partial_{T}p) (-\gamma_{4}-\gamma_{5}+24 \sqrt{3} q_0 \lambda_r^2)-32 \sqrt{3} p_0 q_0 \lambda _r (\partial_{T}\lambda_{r})))\Big)\\
\tilde{c}^{(q)}_{7}&=-\frac{1}{4 p_0}\Big(\gamma _1 p_0-8 \sqrt{3} q_0 \eta _r \lambda _r\Big),\ \
\tilde{c}^{(q)}_{8}=\frac{\gamma _2}{4}\\
\tilde{g}^{(q)}_{1}&=\frac{1}{16 p_0^2}\Big(4 p_0 ((\partial_{\phi}p) (-\gamma_{3}-2 \lambda _r \xi _r)+\xi _r (\partial_{\phi}\eta_{r}))-5 \eta _r \xi _r
   (\partial_{\phi}p)+16 p_0^2 \lambda _r (\partial_{\phi}\xi)\Big)
   \end{split}
\end{eqnarray}
\begin{eqnarray}
\begin{split}
\tilde{g}^{(q)}_{2}&=\frac{1}{16 p_0^2}\Big(4 p_0 ((\partial_{T}p) (-\gamma_{3}-2 \lambda _r \xi _r)+\xi _r (\partial_{T}\eta_{r}))-5 \eta _r \xi _r
   (\partial_{T}p)+16 p_0^2 \lambda _r (\partial_{T}\xi)\Big)\\
\tilde{g}^{(q)}_{3}&=\frac{1}{16 p_0^2}\Big(\xi _r ((\partial_{T}p) (\partial_{\phi}\eta_{r})-(\partial_{\phi}p) (\partial_{T}\eta_{r}))+4 p_0 \lambda _r ((\partial_{\phi}p) (\partial_{T}\xi)-(\partial_{T}p) (\partial_{\phi}\xi))\Big)\nonumber
\end{split}
\end{eqnarray}

\section{Holographic Non-relativistic charged fluid}\label{charge data}

Here we are presenting all the quantities (not presented in the main text) describing non-relativistic charged holographic fluid, in terms of mass, charge and local velocities of black brane.

The pressure of the holographic fluid is given by,
  \begin{eqnarray}
  \begin{split}
     p&=m+\frac{1}{4m_{0}}(\partial_{i}\partial^{i}u^{+})+\frac{(m_{0}-2)}{16m_{0}^{2}}(\partial_{i}\partial^{i}m)-\frac{q_{0}}{12m_{0}}(\partial_{i}\partial^{i}q)\\&+\frac{m_{0}}{2\sqrt{4m_{0}-3}}\log\Big(\frac{3+\sqrt{4m_{0}-3}}{3-\sqrt{4m_{0}-3}}\Big)\Big(\frac{1}{16m_{0}^{2}}(\partial^{i}m)(\partial_{i}m)-\frac{1}{2}(\partial_{i}m)(\partial^{i}u^{+})+(\partial_{i}u^{+})(\partial^{i}u^{+})\Big)\\&-\frac{\log(2)}{4}\Big((\partial_{i}q)(\partial^{i}q)+\frac{9q_{0}^{2}}{16m_{0}^{2}}(\partial_{i}m)(\partial^{i}m)-\frac{3q_{0}}{2m_{0}}(\partial_{i}q)(\partial^{i}m)\Big)-\frac{2q_{0}^{4}\kappa^{2}}{m_{0}}\Big(\frac{\partial_{i}m}{4 m_{0}}+\partial_{i}u^{+}\Big)^{2}\\&+\frac{(22 m_{0}^3-44 m_{0}^2-181 m_{0}+237)}{24 (m_{0}-3) m_{0} (16 m_{0}^2)}(\partial^{i}m)(\partial_{i}m)-\frac{(6 m_{0}^2-8 m_{0}-39) q_{0}}{48 (m_{0}-3) m_{0}^2}(\partial^{i}m)(\partial_{i}q)\\&-\frac{(2 m_{0}^3-3 m_{0}^2-4 m_{0}+12)}{24 (m_{0}-3) m_{0}^2}(\partial^{i}m)(\partial_{i}u^{+})-\frac{1}{12}(\partial^{i}q)(\partial_{i}q)+\frac{3 q_{0}}{2 (m_{0}-3) m_{0}}(\partial^{i}q)(\partial_{i}u^{+})\\&-\frac{(4 m_{0}^2-2 m_{0}+11)}{24 m_{0}}(\partial^{i}u^{+})(\partial_{i}u^{+})-\frac{(2-m_{0})}{12 m_{0}}\sigma^{ij}\sigma_{ij}+\big(\frac{2 q_{0}^{4}\kappa^{2}}{m_{0}}-\frac{q_{0}^{2}}{6}\big)\omega^{ij}\omega_{ij}
     \end{split}
   \end{eqnarray}
Mass density is given by,
	\begin{eqnarray}
        \begin{split}
	\rho&= 4 m(u^{+})^{2}+\frac{2-m_{0}}{6m_{0}}\sigma^{ij}\sigma_{ij}+\frac{q_{0}^{2}}{3}\omega^{ij}\omega_{ij}+\frac{4 q_{0}^{4}\kappa^{2}}{m_{0}}(l^{i}l_{i}-\omega^{ij}\omega_{ij})+\frac{q_{0}}{6m_{0}}\partial_{i}\partial^{i}q-\frac{1}{2m_{0}}\partial_{i}\partial^{i}u^{+}\\&+\frac{(2-m_{0})}{8m_{0}^{2}}\partial_{i}\partial^{i}m-\frac{\log2}{2}\partial_{i}q\partial_{i}q+\frac{3 q_{0}\log2}{4 m_{0}}\partial_{i}q\partial_{i}m-\frac{9 q_{0}^{2}\log2}{32 m_{0}^{2}}\partial_{i}m\partial_{i}m\\&-\frac{m_{0}}{2\sqrt{4 m_{0}-3}}\log(\frac{3-\sqrt{4m_{0}-3}}{3+\sqrt{4m_{0}-3}})(\frac{1}{4m_{0}}(\partial_{i}m)-(\partial_{i}u^{+}))^{2}-\frac{3 q_{0}}{2(m_{0}-3)m_{0}}\partial_{i}q\partial_{i}u^{+}\\&+\frac{q_{0}}{4m_{0}}\frac{(-39-8m_{0}+6 m_{0}^{2})}{6 m_{0}(m_{0}-3)}	\partial_{i}q\partial_{i}m+\frac{1}{6}\partial_{i}q\partial_{i}q+\frac{(4-m_{0}+2m_{0}^{2})}{6 m_{0}}(\partial_{i}u^{+})(\partial_{i}u^{+})\\&+\frac{(123-92 m_{0}-22 m_{0}^{2}+11 m_{0}^{3})}{96m_{0}^{3}(m_{0}-3)}\partial_{i}m\partial_{i}m+\frac{(15-5 m_{0}-6 m_{0}^{2}+4 m_{0}^{3})}{24 m_{0}^{2}(m_{0}-3)}\partial_{i}u^{+}\partial_{i}m
 \end{split}
   \end{eqnarray}
Non-relativistic velocity has the following expression,
   \begin{eqnarray}
        \begin{split}
   v_{i}&=\frac{u_{i}}{u^{+}}-\frac{1}{\rho}(\frac{3u^{+}(q^{2}-m)+(2m_{0}-3)u^{+}}{2(m_{0}-3)})(\frac{\partial_{i}u^{+}}{u^{+}}-\frac{1}{4 m}\partial_{i}m)+\frac{\sqrt{3}q_{0}^{3}\kappa}{4 m_{0}^{2}}\epsilon_{ij}\partial_{k}\omega_{jk}\\&+\frac{\sqrt{3}q_{0}^{3}\kappa}{8 m_{0}^{3}}\epsilon_{ij}(\partial_{k}m)\omega_{jk}+\frac{\sqrt{3}q_{0}^{3}\kappa}{2 m_{0}^{2}}\epsilon_{ij}(\partial_{k} u^{+})\omega_{jk}+\frac{(4m_{0}-1)}{16 m_{0}^{2}}\partial_{k}\omega_{ik}+\frac{(2m_{0}-1)}{8 m_{0}^{2}}\partial_{k}\sigma_{ik}\\&-\frac{1}{8\sqrt{4m_{0}-3}}\log\Big(\frac{3+\sqrt{4 m_{0}-3}}{3-\sqrt{4 m_{0}-3}}\Big)\Big(\partial_{k}\sigma_{ik}+\partial_{k}\omega_{ik}-(\partial_{k}u^{+})\sigma_{ik}+\frac{1}{4m_{0}}(\partial_{k}m)\sigma_{ik}\Big)\\&+\frac{1}{16m_{0}^{2}}\sigma_{ik}\Big(\frac{q_{0}(m_{0}-6)}{(m_{0}-3)}(\partial_{k}q)-(\partial_{k}u^{+})+\frac{(m_{0}^{2}+6m_{0}-9)}{4 m_{0}(m_{0}-3)}(\partial_{k}m)\Big)\\&-\frac{1}{16m_{0}^{2}}\omega_{ik}\Big(q_{0}(\partial_{k}q)+(1+2m_{0}-4 m_{0}^{2})(\partial_{k}u^{+})-\frac{(4m_{0}^{2}-3m_{0}-2)}{4 m_{0}}(\partial_{k}m)\Big)
   \end{split}
   \end{eqnarray}
Energy density is given by
   \begin{eqnarray}
        \begin{split}
      e&=m+\frac{1}{2}\rho v^{2}-\frac{2 q_{0}^{4}\kappa^{2}}{m_{0}}\omega^{ij}\omega_{ij}+\log(2)\Big(\frac{1}{4}\partial_{i}q\partial_{i}q-\frac{3q_{0}}{8m_{0}}\partial_{i}q\partial_{i}m+\frac{9q_{0}^{2}}{64m_{0}^{2}}\partial_{i}m\partial_{i}m\Big)\\&-\frac{m_{0}}{4\sqrt{4m_{0}-3}}\log(\frac{3+\sqrt{4m_{0}-3}}{3-\sqrt{4m_{0}-3}})\Big(\partial_{i}u^{+}-\frac{\partial_{i}m}{4m_{0}}\Big)^{2}-\frac{q_{0}}{12m_{0}}\partial_{i}\partial_{i}q+\frac{1}{4m_{0}}\partial_{i}\partial_{i}u^{+}\\&-\frac{1}{12}\partial_{i}q\partial_{i}q+\frac{3q_{0}}{4m_{0}(m_{0}-3)}\partial_{i}q\partial_{i}u^{+}+\frac{q_{0}(6 m_{0}^{2}-8 m_{0}-39)}{48m_{0}^{2}(m_{0}-3)}\partial_{i}q\partial_{i}m\\&-\frac{(4m_{0}^{2}-2m_{0}+11)}{24m_{0}}\partial_{i}u^{+}\partial_{i}u^{+}+\frac{q_{0}^{4}\kappa^{2}}{8m_{0}^{3}}\partial_{i}m\partial_{i}m+\frac{q_{0}^{4}\kappa^{2}}{m_{0}^{2}}\partial_{i}m\partial_{i}u^{+}\\&+\frac{(-12 + 4 m_{0} + 3 m_{0}^2 - 2 m_{0}^3)}{(24 (-3 + m_{0}) m_{0}^2)}\partial_{i}m\partial_{i}u^{+}+\frac{(m_{0}-2)}{12m_{0}}\sigma_{ij}\sigma^{ij}+\frac{q_{0}^{2}}{6}\omega_{ij}\omega^{ij}\\&+\frac{(-237 + 181 m_{0} + 44 m_{0}^{2} - 22 m_{0}^{3})}{(384 (-3 + m_{0}) m_{0}^{3})}\partial_{i}m\partial_{i}m+\frac{2 q_{0}^{4}\kappa^{2}}{m_{0}}(\partial_{i}u^{+})(\partial^{i}u^{+})
   \end{split}
   \end{eqnarray}
Finally, the energy current is given by,
  \begin{eqnarray}
        \begin{split}
     e^{i}&=(e+p+\frac{1}{2}\rho v^{2})v^{i}+\frac{R^{3}}{u^{+}}(\frac{\partial_{i}u^{+}}{u^{+}}-\frac{\partial_{i}m}{4m})-R^{3}\sigma^{ik}v^{k}-\frac{\sqrt{3}q_{0}^{3}\kappa}{m_{0}}\epsilon^{ik}\partial^{j}\omega^{kj}\\&+\frac{q_{0}^{3}\kappa\sqrt{3}}{2m_{0}^{2}}\epsilon^{ij}\omega^{jk}\partial^{k}m-\frac{12 q_{0}^{4}\kappa^{2}}{m_{0}}\omega^{ij}(\partial^{j}u^{+}+\frac{1}{4m_{0}}\partial^{j}m)+\frac{(1-2m_{0})}{2m_{0}}\partial^{k}\sigma^{ik}\\&+\frac{m_{0}}{2\sqrt{4m_{0}-3}}\log\big(\frac{3-\sqrt{4m_{0}-3}}{3+\sqrt{4m_{0}-3}}\big)(\sigma^{ik}(\partial^{k}u^{+})-\partial^{k}\sigma^{ik}-\partial^{k}\omega^{ik}-\frac{1}{4m_{0}}\sigma^{ik}(\partial^{k}m))\\&+\frac{(1-4m_{0})}{4m-{0}}\partial^{k}\omega^{ik}-\frac{(m_{0}-6)q_{0}}{4m_{0}(m_{0}-3)}\sigma^{ik}(\partial^{k}q)-\frac{q_{0}}{4m_{0}}\omega^{ik}(\partial^{k}q)-\frac{2\sqrt{3}q_{0}^{3}\kappa}{m_{0}}\epsilon^{ij}\omega^{jk}\partial^{k}u^{+}\\&+\frac{(4m_{0}^{2}-2m_{0}-1)}{4m_{0}}\sigma^{ik}(\partial^{k}u^{+})-\big(\frac{(m_{0}^{2}+5m_{0}-6)}{16 m_{0}^{2}(m_{0}-3)}\sigma^{ik}(\partial^{k}m)\\&-\frac{(4m_{0}^{2}-2-3m_{0})}{16m_{0}^{2}}\omega^{ik}(\partial^{k}m)\big)
  \end{split}
   \end{eqnarray}
  \end{appendices}

\bibliographystyle{hieeetr}
\bibliography{tst}{}

\end{document}

%% file: gdefn.tex
\usepackage{epsfig,graphicx,bm,amsmath,amsfonts}
\usepackage{epstopdf}
\usepackage{amssymb,xcolor,float}
\usepackage{amsthm}
\usepackage{hyperref}
\usepackage{cite}
\input epsf.sty
\usepackage{cite}
\usepackage{comment}
\usepackage{epstopdf}
\usepackage{float}
\usepackage{caption}
\usepackage{subcaption}
\usepackage{enumitem}
\usepackage{bbm}
\usepackage{bm}
\usepackage{longtable}
\usepackage{parskip}
\usepackage{enumitem}
\usepackage{graphicx,epsfig}
\usepackage{ae,aecompl}
\usepackage{dcolumn}
\usepackage{latexsym}
\usepackage{psfrag}
\usepackage{ytableau}
\usepackage{tabularx}
\usepackage{float}
\usepackage{wrapfig}

\usepackage{xcolor}
\usepackage[english]{babel}
\usepackage[T1]{fontenc}
\usepackage[utf8]{inputenc}
\usepackage{authblk}
\usepackage{mathtools}
\usepackage{slashed}
\usepackage{mattens,amssymb}
\usepackage{amsmath,amssymb}
\usepackage{amsfonts}
\usepackage{graphicx,color}
\usepackage{cite}
\usepackage{hyperref}
\hypersetup{
	colorlinks=true,
	linkcolor=blue,
	filecolor=magenta,      
	urlcolor=cyan,
}
\usepackage[capitalize]{cleveref}
\numberwithin{equation}{section} 

\definecolor{refcol}{rgb}{0.9,0.1,0.1}
\hypersetup{colorlinks=true,linkcolor=blue,citecolor=refcol,urlcolor=cyan,linktocpage}

\usepackage{wrapfig}

\newcommand{\ben}{\begin{eqnarray}\displaystyle}
\newcommand{\een}{\end{eqnarray}}

\newcommand{\be}{\begin{equation}}
\newcommand{\ee}{\end{equation}}

\newcommand{\bc}{\begin{center}}
\newcommand{\ec}{\end{center}}

\newcommand{\eesp}{\end{split}}
\newcommand{\bsp}{\begin{split}}

\newcommand{\Rmnum}[1]{\expandafter\@slowromancap\romannumeral #1@}
\newcommand{\cA}{\mathcal{A}}

\newcommand{\cC}{\mathcal{C}}
\newcommand{\cD}{\mathcal{D}}

\newcommand{\cG}{\mathcal{G}}

\newcommand{\cJ}{\mathcal{J}}

\newcommand{\ra}{\rightarrow}

\newcommand{\lB}{\left [}
\newcommand{\rB}{\right ]}
\newcommand{\lb}{\left (}
\newcommand{\rb}{\right )}

\newcommand{\bensp}{\begin{eqnarray}\begin{split}}
\newcommand{\eensp}{\end{eqnarray}\end{split}}

\newcommand{\bnm}{\begin{matrix}}
\newcommand{\enm}{\end{matrix}}
\def\XXint#1#2#3{{\setbox0=\hbox{$#1{#2#3}{\int}$ }
\vcenter{\hbox{$#2#3$ }}\kern-.6\wd0}}

%% file: main.bbl
\begin{thebibliography}{10}

\bibitem{Maldacena:1997re}
J.~M. Maldacena, ``{The Large N limit of superconformal field theories and
  supergravity},'' {\em Adv. Theor. Math. Phys.}, vol.~2, pp.~231--252, 1998,
  hep-th/9711200.

\bibitem{Gubser:1998bc}
S.~S. Gubser, I.~R. Klebanov, and A.~M. Polyakov, ``{Gauge theory correlators
  from noncritical string theory},'' {\em Phys. Lett. B}, vol.~428,
  pp.~105--114, 1998, hep-th/9802109.

\bibitem{Witten:1998qj}
E.~Witten, ``{Anti-de Sitter space and holography},'' {\em Adv. Theor. Math.
  Phys.}, vol.~2, pp.~253--291, 1998, hep-th/9802150.

\bibitem{PhysRevD.5.377}
C.~R. Hagen, ``Scale and conformal transformations in galilean-covariant field
  theory,'' {\em Phys. Rev. D}, vol.~5, pp.~377--388, Jan 1972.

\bibitem{Mehen:1999nd}
T.~Mehen, I.~W. Stewart, and M.~B. Wise, ``{Conformal invariance for
  nonrelativistic field theory},'' {\em Phys. Lett. B}, vol.~474, pp.~145--152,
  2000, hep-th/9910025.

\bibitem{Nishida:2007pj}
Y.~Nishida and D.~T. Son, ``{Nonrelativistic conformal field theories},'' {\em
  Phys. Rev. D}, vol.~76, p.~086004, 2007, 0706.3746.

\bibitem{Sakaguchi:2008ku}
M.~Sakaguchi and K.~Yoshida, ``{More super Schrodinger algebras from
  psu(2,2|4)},'' {\em JHEP}, vol.~08, p.~049, 2008, 0806.3612.

\bibitem{Kovtun:2008qy}
P.~Kovtun and D.~Nickel, ``{Black holes and non-relativistic quantum
  systems},'' {\em Phys. Rev. Lett.}, vol.~102, p.~011602, 2009, 0809.2020.

\bibitem{Duval:2008jg}
C.~Duval, M.~Hassaine, and P.~A. Horvathy, ``{The Geometry of Schrodinger
  symmetry in gravity background/non-relativistic CFT},'' {\em Annals Phys.},
  vol.~324, pp.~1158--1167, 2009, 0809.3128.

\bibitem{Rangamani:2009zz}
M.~Rangamani, ``{Holography for non-relativistic CFTs},'' {\em Acta Phys.
  Polon. B}, vol.~40, pp.~3745--3770, 2009.

\bibitem{Son:2008ye}
D.~T. Son, ``{Toward an AdS/cold atoms correspondence: A Geometric realization
  of the Schrodinger symmetry},'' {\em Phys. Rev. D}, vol.~78, p.~046003, 2008,
  0804.3972.

\bibitem{Balasubramanian:2008dm}
K.~Balasubramanian and J.~McGreevy, ``{Gravity duals for non-relativistic
  CFTs},'' {\em Phys. Rev. Lett.}, vol.~101, p.~061601, 2008, 0804.4053.

\bibitem{Goldberger:2008vg}
W.~D. Goldberger, ``{AdS/CFT duality for non-relativistic field theory},'' {\em
  JHEP}, vol.~03, p.~069, 2009, 0806.2867.

\bibitem{Barbon:2008bg}
J.~L.~F. Barbon and C.~A. Fuertes, ``{On the spectrum of nonrelativistic
  AdS/CFT},'' {\em JHEP}, vol.~09, p.~030, 2008, 0806.3244.

\bibitem{Adams:2008zk}
A.~Adams, A.~Maloney, A.~Sinha, and S.~E. Vazquez, ``{1/N Effects in
  Non-Relativistic Gauge-Gravity Duality},'' {\em JHEP}, vol.~03, p.~097, 2009,
  0812.0166.

\bibitem{Schafer-Nameki:2009dsc}
S.~Schafer-Nameki, M.~Yamazaki, and K.~Yoshida, ``{Coset Construction for Duals
  of Non-relativistic CFTs},'' {\em JHEP}, vol.~05, p.~038, 2009, 0903.4245.

\bibitem{Herzog:2008wg}
C.~P. Herzog, M.~Rangamani, and S.~F. Ross, ``{Heating up Galilean
  holography},'' {\em JHEP}, vol.~11, p.~080, 2008, 0807.1099.

\bibitem{Rangamani:2008gi}
M.~Rangamani, S.~F. Ross, D.~T. Son, and E.~G. Thompson, ``{Conformal
  non-relativistic hydrodynamics from gravity},'' {\em JHEP}, vol.~01, p.~075,
  2009, 0811.2049.

\bibitem{Adams:2009dm}
A.~Adams, C.~M. Brown, O.~DeWolfe, and C.~Rosen, ``{Charged Schrodinger Black
  Holes},'' {\em Phys. Rev. D}, vol.~80, p.~125018, 2009, 0907.1920.

\bibitem{Brattan:2010bw}
D.~K. Brattan, ``{Charged, conformal non-relativistic hydrodynamics},'' {\em
  JHEP}, vol.~10, p.~015, 2010, 1003.0797.

\bibitem{Lunin:2005jy}
O.~Lunin and J.~M. Maldacena, ``{Deforming field theories with U(1) x U(1)
  global symmetry and their gravity duals},'' {\em JHEP}, vol.~05, p.~033,
  2005, hep-th/0502086.

\bibitem{Adams:2008wt}
A.~Adams, K.~Balasubramanian, and J.~McGreevy, ``{Hot Spacetimes for Cold
  Atoms},'' {\em JHEP}, vol.~11, p.~059, 2008, 0807.1111.

\bibitem{Alishahiha:2003ru}
M.~Alishahiha and O.~J. Ganor, ``{Twisted backgrounds, PP waves and nonlocal
  field theories},'' {\em JHEP}, vol.~03, p.~006, 2003, hep-th/0301080.

\bibitem{Gimon:2003xk}
E.~G. Gimon, A.~Hashimoto, V.~E. Hubeny, O.~Lunin, and M.~Rangamani, ``{Black
  strings in asymptotically plane wave geometries},'' {\em JHEP}, vol.~08,
  p.~035, 2003, hep-th/0306131.

\bibitem{Bhattacharyya:2007vjd}
S.~Bhattacharyya, V.~E. Hubeny, S.~Minwalla, and M.~Rangamani, ``{Nonlinear
  Fluid Dynamics from Gravity},'' {\em JHEP}, vol.~02, p.~045, 2008, 0712.2456.

\bibitem{Ross:2009ar}
S.~F. Ross and O.~Saremi, ``{Holographic stress tensor for non-relativistic
  theories},'' {\em JHEP}, vol.~09, p.~009, 2009, 0907.1846.

\bibitem{Dutta:2018xtr}
S.~Dutta and H.~Krishna, ``{Light-Cone Reduction vs. TsT transformations : A
  Fluid Dynamics Perspective},'' {\em JHEP}, vol.~05, p.~029, 2018, 1803.03948.

\bibitem{Banerjee:2008th}
N.~Banerjee, J.~Bhattacharya, S.~Bhattacharyya, S.~Dutta, R.~Loganayagam, and
  P.~Surowka, ``{Hydrodynamics from charged black branes},'' {\em JHEP},
  vol.~01, p.~094, 2011, 0809.2596.

\bibitem{Erdmenger:2008rm}
J.~Erdmenger, M.~Haack, M.~Kaminski, and A.~Yarom, ``{Fluid dynamics of
  R-charged black holes},'' {\em JHEP}, vol.~01, p.~055, 2009, 0809.2488.

\bibitem{Banerjee:2014mka}
N.~Banerjee, S.~Dutta, A.~Jain, and D.~Roychowdhury, ``{Entropy current for
  non-relativistic fluid},'' {\em JHEP}, vol.~08, p.~037, 2014, 1405.5687.

\bibitem{Dutta:2020gcg}
S.~Dutta, T.~Mandal, and S.~Parihar, ``{Holographic constraints on generalized
  Rivlin-Ericksen fluid},'' {\em Phys. Rev. D}, vol.~103, no.~2, p.~026014,
  2021, 2009.11335.

\bibitem{RE}
R.~S. RIVLIN and J.~L. ERICKSEN, ``Stress-deformation relations for isotropic
  materials,'' {\em Journal of Rational Mechanics and Analysis}, vol.~4,
  pp.~323--425, 1955.

\bibitem{Maldacena:2008wh}
J.~Maldacena, D.~Martelli, and Y.~Tachikawa, ``{Comments on string theory
  backgrounds with non-relativistic conformal symmetry},'' {\em JHEP}, vol.~10,
  p.~072, 2008, 0807.1100.

\bibitem{Yamada:2008if}
D.~Yamada, ``{Thermodynamics of Black Holes in Schrodinger Space},'' {\em
  Class. Quant. Grav.}, vol.~26, p.~075006, 2009, 0809.4928.

\bibitem{Bobev:2009zf}
N.~Bobev and A.~Kundu, ``{Deformations of Holographic Duals to Non-Relativistic
  CFTs},'' {\em JHEP}, vol.~07, p.~098, 2009, 0904.2873.

\bibitem{Bobev:2009mw}
N.~Bobev, A.~Kundu, and K.~Pilch, ``{Supersymmetric IIB Solutions with
  Schr\"odinger Symmetry},'' {\em JHEP}, vol.~07, p.~107, 2009, 0905.0673.

\bibitem{Imeroni:2009cs}
E.~Imeroni and A.~Sinha, ``{Non-relativistic metrics with extremal limits},''
  {\em JHEP}, vol.~09, p.~096, 2009, 0907.1892.

\bibitem{Kim:2010tf}
B.~S. Kim and D.~Yamada, ``{Properties of Schroedinger Black Holes from AdS
  Space},'' {\em JHEP}, vol.~07, p.~120, 2011, 1008.3286.

\bibitem{Banerjee:2011jb}
N.~Banerjee, S.~Dutta, and D.~P. Jatkar, ``{Geometry and Phase Structure of
  Non-Relativistic Branes},'' {\em Class. Quant. Grav.}, vol.~28, p.~165002,
  2011, 1102.0298.

\bibitem{Mazzucato:2008tr}
L.~Mazzucato, Y.~Oz, and S.~Theisen, ``{Non-relativistic Branes},'' {\em JHEP},
  vol.~04, p.~073, 2009, 0810.3673.

\bibitem{Brown:1992br}
J.~D. Brown and J.~W. York, Jr., ``{Quasilocal energy and conserved charges
  derived from the gravitational action},'' {\em Phys. Rev. D}, vol.~47,
  pp.~1407--1419, 1993, gr-qc/9209012.

\bibitem{Hollands:2005ya}
S.~Hollands, A.~Ishibashi, and D.~Marolf, ``{Counter-term charges generate bulk
  symmetries},'' {\em Phys. Rev. D}, vol.~72, p.~104025, 2005, hep-th/0503105.

\bibitem{Son:2009tf}
D.~T. Son and P.~Surowka, ``{Hydrodynamics with Triangle Anomalies},'' {\em
  Phys. Rev. Lett.}, vol.~103, p.~191601, 2009, 0906.5044.

\end{thebibliography}
